\documentclass[12pt]{article}
\usepackage{a4wide}
\usepackage{epsfig}

\newfont{\xLfont}{pzcmi8r at 12pt}
\newcommand{\xloops}{{\xLfont xloops}}

\renewcommand{\arraystretch}{1.3}
\newcommand{\pfrac}[2]{\left(\frac{#1}{#2}\right)}

\newcommand{\eps}{\varepsilon}
\newcommand{\real}{\mathop{\rm Re}\nolimits}
\newcommand{\imag}{\mathop{\rm Im}\nolimits}
\newcommand{\GeV}{\mbox{\rm\,GeV}}
\newcommand{\Li}{\mathop{\rm Li}\nolimits}
\newcommand{\sqrts}{\sqrt[\star]}
\begin{document}
\thispagestyle{empty}
\begin{flushright}
MZ-TH/05-17\\
hep-ph/0508173\\
August 2005\\
\end{flushright}

\begin{center}
{\Large\bf Evaluating massive planar\\[.3truecm]
  two-loop tensor vertex integrals}\\[1truecm]
{\large Stefan Groote$^{1,2}$ and Markus M.\ Knodel$^1$}\\[.7cm]
$^1$ Institut f\"ur Physik der Johannes-Gutenberg-Universit\"at,\\
  Staudinger Weg 7, 55099 Mainz, Germany\\[.5truecm]
$^2$ F\"u\"usika-Keemiateaduskond, Tartu \"Ulikool,
  T\"ahe 4, 51010 Tartu, Estonia\\[.3truecm] 
\vspace{1truecm}
\end{center}

\begin{abstract}
Using the parallel/orthogonal space method, we calculate the planar two-loop
three-point diagram and two rotated reduced planar two-loop three-point
diagrams. Together with the crossed topology, these diagrams are the most
complicated ones in the two-loop corrections necessary, for instance, for the
decay of the $Z^0$ boson. Instead of calculating particular decay processes,
we present the new algorithm which allows one to calculate arbitrary NNLO
calculations for massive planar two-loop vertex functions in the general mass
case. All integration steps up to the last two ones are performed analytically
and will be implemented under \xloops\ as part of the Mainz \xloops-GiNaC
project. The last two integrations are done numerically using methods like
VEGAS and Divonne. Thresholds originating from Landau singularities are found
and discussed in detail. In order to demonstrate the numeric stability of our
methods we consider particular Feynman integrals which contribute to different
physical processes. Our results can be generalized to the case of the crossed
topology.
\end{abstract}

\newpage

\section{Introduction}
Precision measurements at the LEP collider at CERN and other colliders like
e.g.\ SLC at SLAC, TEVATRON at Fermilab and HERA at DESY have reached a
precision which has exceeded all expectations. This is true especially for
electron colliders~\cite{Hollik:2000,Bardin:2001sk,Hollik:2004dz,LEP:2005em}. 
At the moment the precision of measurements related to the parameters of the
Standard Model of electroweak interactions reaches values up to
$O(10^{-4})$~\cite{LEP:2005em}. At future colliders like the LHC or the ILC
(including GigaZ), further improvements are expected~\cite{Hollik:2004dz,
Aguilar-Saavedra:2001rg,Heinemeyer:2000jd,Weiglein:2003tr,Weiglein:2004hn}.

Compared to this, theoretical predictions accomplish this precision only in
very few cases. To check the validity of the Standard Model and to be able to
draw conclusions about ``new physics'', progress in theoretical methods and
their application is necessary~\cite{Bardin:2001sk,Hollik:2004dz}. The
complexity of the calculations and the number of the graphs which have to be
calculated within perturbation theory grows considerably order by order. At
second order only a few observables are calculated~\cite{Heinemeyer:2004an}.
Still there is no method available which allows for the (semi)automatic
calculation of arbitrary processes at this order. Much work has been done on
NNLO calculations~\cite{Kreimer:1992zv,Usyukina:1994eg,Davydychev:1995nq,
Fujimoto:1995ev,Pivovarov:1999mr,Chetyrkin:1996my,Groote:1998ic,
Ghinculov:1997pd,Davydychev:2002hy,Aglietti:2003yc,Bernreuther:2004ih,
Bonciani:2003te,Caffo:2002wm,Mastrolia:2002gt,Bauberger:1994by,
Post:1997,Post:1996gg,Do:2003}. While for mixing QCD and electroweak
$O(\alpha\alpha_s)$ NNLO vertex corrections a few calculations were done
several years ago~\cite{Ghinculov:1997pd}, for the evaluation of two-loop
three-point diagrams in the general mass case only the methods proposed in
Refs.~\cite{Passarino:2001wv,Ferroglia:2003yj,Actis:2004bp} have been used
successfully in order to calculate electroweak NNLO
corrections~\cite{Hollik:2005va}. With increasing energy as it will be used at
the ILC, radiative corrections will become increasingly important. Therefore,
there is still need for independent methods to calculate general massive
two-loop vertex diagrams.

In this paper we present a new algorithm which allows one to calculate
NNLO-correc\-tions for general massive planar two-loop vertex functions as
they arise in the Standard Model of electroweak interactions using Feynman
gauge, e.g.\ for the effective weak mixing angle
$\sin^2\theta_{\rm eff}^{\rm lept}$~\cite{Hollik:2005va,Awramik:2004ge}
occuring in $Z^0\to l^+l^-$ and for other processes like $Z^*\to t\bar t$.
Even though the calculation of particular physical processes is not subject of
this paper, the new algorithm enables one to perform such calculations. For
our algorithm we use the parallel/orthogonal space
method~\cite{Kreimer:1992hs} which allows one to separate effects coming from
inner momenta from those of momenta of the outer particles of the process.

The benefit of the parallel/orthogonal space method is the fact that the
calculation stays close to the physical process. The introduction of Feynman
parameters and the application of the Wick rotation which estrange the
calculation from the physical process for other methods are not necessary
for the parallel/orthogonal space method. Instead, a decay process can be
calculated in the rest frame of the decaying particle. The introduction of
Gram determinants which might cause artificial divergences is not necessary.
The analytical integration leads to still rather simple basic functions like
logarithms and dilogarithms. For the remaining numerical integration the
integrand can be analyzed at physical thresholds. Landau singularities are
mirrored directly onto the parameters of the integrands. Finally, the method
is totally independent from other methods and therefore allows for an
independent check.

Up to now, the parallel/orthogonal space method was used successfully to
evaluate the general massive scalar integrals in the case of the planar
topology~\cite{Czarnecki:1994td} and in the case of the crossed
topology~\cite{Frink:1996ya}. However, it was not possible to evaluate
two-loop three-point functions which contain loop momenta in the numerator.
The method proposed in this paper allows one to calculate also tensor
integrals for the planar vertex topologies.

We present a tensor reduction which reduces the planar and rotated reduced
planar two-loop three-point topologies containing an arbitrary tensor
structure to a set of master integrals with strongly restricted numerator
structure~\cite{Kreimer:1993tx}. As for most of the master integrals which
remain after tensor reduction there are already established methods
available~\cite{Bauberger:1994by,Post:1997,Post:1996gg,Do:2003}, we do not
give results in these cases. The same is valid for
particular mass cases~\cite{Kreimer:1992zv,Kotikov:1991hm,Fleischer:1997bw,
Smirnov:1998vk} and for the UV-singular parts~\cite{Kreimer:1993tx}. Efficient
methods for several cases including few masses have been developed in
Refs.~\cite{Aglietti:2003yc,Bernreuther:2004ih,Bonciani:2003te}. We also
do not dwell on the calculation of massless diagrams for which other methods
are applied successfully (see e.g.\ \cite{Birthwright:2004kk}). As IR
divergences occur in this case, there is no systematic method available within
the parallel/orthogonal space method because for each special case the
subtraction has to be found seperately ``by hand''~\cite{Fleischer:1997bq,
Frink:2000}. Instead, we concentrate on the most complicated part,
namely the calculation of the UV-finite parts of those master integrals which
keep the planar or rotated reduced planar topology after tensor reduction.

Two-loop integration in four-dimensional space-time needs eight integration
steps. Using the parallel/orthogonal space method, we can do six of them
analytically using different techniques. The last two integration steps 
are done numerically. Because most of the integrations can be performed
analytically, the Landau singularities of the Feynman diagrams can be related
graphically to the integrands in the remaining two-dimensional integration
region. Using different algorithms, we demonstrate the numerical stability.

The algorithm for the numerical calculation of the last two integrations is
developed and its reliability is demonstrated. In addition, further exhaustive
numerical tests are done. While for the numerical integration programs like
the Monte Carlo integration routine VEGAS~\cite{Lepage:1980dq,Ohl:1998jn} and
the program Divonne of the library CUBA~\cite{Hahn:2004fe} are used, the
implementation of the algorithms for the semi-automatical calculation in
\xloops~\cite{Kreimer:1996qy,Brucher:1996ww,Franzkowski:1996wx} is part of the
work described in this paper. In the \xloops-GiNaC project developed by the
members of the ThEP working group in Mainz~\cite{Frink:1997sg,Bauer:2001ig},
work on the automatic generation of Feynman diagrams and their evaluation with
analytical and numerical means is in progress. For this purpose the language
for algebraic calculation GiNaC was developed~\cite{Bauer:2000cp}. While most
of the master two-loop integrals not considered in this paper are already
implemented in \xloops, the missing planar topologies will be added in the
course of the work introduced here. The complete set of algorithms described
in this paper is implemented under \xloops-GiNaC. However, the implementation
of parts of the algorithms for known topologies as mentioned in
Sec.~\ref{sumredpro} still has to be done. Because in this spirit the
implementation is not yet finished, the calculation of physical processes is
not subject of this paper but will be presented in a future publication.

The paper is organized as follows. In Sec.~2 we introduce the main tools of
the analytical calculation associated with the parallel/orthogonal space
method. The tensor reduction procedure is explained in detail in Sec.~3 for
the non-reduced planar two-loop three-point diagram, and modifications in case
of the rotated reduced planar diagrams are mentioned and explained. The tensor
reduction leads to master integrals which are integrated analytically in Sec.~4
up to the last two integrations. In Sec.~5 we deal with the analysis of Landau
singularities and thresholds. Sec.~6 is devoted to the numerical integration.
In both sections we present examples to show the reliability of our procedure.
In Sec.~7 we give our conclusions. In Appendix~A we deal with the integral
basis.

\section{Tools for the calculation}
In this section we provide the reader with the tools necessary for the
calculation of massive two-loop tensor vertex integrals. Most of the tools
were already introduced in the literature~\cite{Kreimer:1992hs,
Czarnecki:1994td,Frink:1996ya,Frink:2000,Kreimer:1991wj,Kreimer:1992ps,
Frink:1996}. Therefore, we can be brief in presenting these tools.

\subsection{The parallel/orthogonal space method}
The integrals we have to calculate are determined by the two momenta of the
produced particles, $p_1$ and $p_2$. Because the integrals are expressed in
terms of covariant quantities, we are free to decompose any loop momentum
$k$ into two covariant vectors, $k^\mu=k_\parallel^\mu+k_\perp^\mu$, where
$k_\parallel$ has components in the {\em parallel space\/} which is the linear
span of the external momenta $p_i$, while $k_\perp$ is the orthogonal
complement with $\sum_\mu k_\perp^\mu p_{i\mu}=0$ with components in the
{\em orthogonal space}~\cite{Kreimer:1992zv,Ghinculov:1997pd}. For three-point
functions we can consider the process in the rest frame of the decaying
particle. In this frame the two emerging particles are produced back-to-back,
defining the $z$-axis.\footnote{The $z$-component will be written as the
second component of the four-vector in the following.} In this frame the
representation
\begin{equation}
p_1=(E_1;q_z,0,0),\qquad p_2=(E_2;-q_z,0,0),\qquad
p=p_1+p_2=(E;0,0,0)
\end{equation}
can be used where $p$ is the momentum of the decaying particle with
$E=E_1+E_2$ and $p^2=E^2$. Accordingly, the loop momenta $k$ and $l$ necessary
for the description of the two-loop integral are parametrized
by~\cite{Kreimer:1992ps}
\begin{equation}
k=(k_0;k_1,\vec k_\perp),\qquad l=(l_0;l_1,\vec l_\perp).
\end{equation}
The two-dimensional vectors $\vec k_\perp$ and $\vec l_\perp$ are represented
in polar coordinates, using the squared absolute values
$s=\vec k_\perp^2=k_\perp^2$ and $t=\vec l_\perp^2=l_\perp^2$ and two angles,
the angle $\alpha$ of $\vec k_\perp$ with the $x$-axis and the relative angle
$\gamma$. We can write
\begin{equation}
\vec k_\perp\cdot\vec l_\perp=k_\perp l_\perp z=\sqrt{st}\,z,\qquad
  z=\cos\gamma=\cos(\vec k_\perp,\vec l_\perp).
\end{equation}
The choice of parallel and orthogonal subspaces itself is Lorentz invariant,
so that the calculation can still be done in any Lorentz frame. The benefit of
the parallel/orthogonal space method (P/O-space method) is that the
contributions belonging to the orthogonal space can be integrated out and we
are left with the contributions in parallel space only. The integration
measure is accordingly written as
\begin{equation}
\int d^4k\int d^4l=\pi\int_{-\infty}^{+\infty}
  dk_0\,dl_0\,dk_1\,dl_1\int_0^\infty ds\,dt
  \int_{-1}^{+1}\frac{dz}{\sqrt{1-z^2}}
\end{equation}
where the trivial integration over $\alpha$ has already been performed.

\subsection{The linearization}
Using Feynman gauge, the denominators of integrals occurring in two-loop
vertex calculations contain up to six propagator factors. A typical factor is
given by
\begin{equation}
P_1=(k+p_1)^2-m_1^2+i\eta
\end{equation}
where $\eta>0$. In the P/O-space representation this propagator factor reads
\begin{equation}
P_1=(k_0+E_1)^2-(k_1+q_z)^2-k_\perp^2-m_1^2+i\eta.
\end{equation}
If we replace $k_0=k'_0\pm k_1$, we obtain 
\begin{equation}
P_1=(k'_0+E_1)^2\pm2k_1(k'_0+E_1\mp q_z)-q_z^2-k_\perp^2-m_1^2+i\eta
\end{equation}
where $k_1$ no longer appears quadratically. This replacement, also known as
{\em linearization\/}, is allowed because the integral for $k_0$ ranges from
$-\infty$ to $+\infty$. However, because of the occurrence of the mixing
propagator factor
\begin{equation}
P_3=(k+l)^2-m_3^2+i\eta,
\end{equation}
the signs for the linearizations in $k$ and $l$ are coupled. In applications
of this linearization we use the sign which is the most appropriate for our
aims. The benefit of the linearization is that the integrals over $k_1$ and
$l_1$ can then be calculated by using the residue theorem.

\subsection{The integration over $z$}
The quantity $z$ occurs only in the just mentioned propagator factor $P_3$.
After linearization this propagator factor can be written as $P_3=A+Bz+i\eta$
where
\begin{equation}\label{defAB}
A=A(k_1,l_1)=(k'_0+l'_0)^2\pm2(k_1+l_1)(k'_0+l'_0)-(s+t)-m_3^2,\qquad
B=-2\sqrt{st}.
\end{equation}
$P_3$ occurs only once in the denominator (if at all). The resulting integral
\begin{equation}\label{sqrts}
\int_{-1}^{+1}\frac{dz}{\sqrt{1-z^2}}\ \frac1{A+Bz+i\eta}
  =\frac\pi{\sqrts{(A+i\eta)^2-B^2}}
  =\frac\pi{\sqrts{(A+i\eta)^2-4st}}
\end{equation}
can be written in this closed form only if we make a convention different from
the usual one~\cite{Kreimer:1992zv}. We stipulate that the cut of the square
root is located on the positive real axis instead of the negative one. For
this reason we have added the star in $\sqrt[\star]{(A+i\eta)^2-B^2}$ which
reminds one of the different cut. The occurrence of the modified square root
has consequences for the subsequent integrations because in closing paths for
the residue theorem we have to avoid crossing this cut. In order to find
conditions for this we split the square root up into a product of two square
roots $\sqrts{A\pm B+i\eta}$. If one of the radicands is equal to a positive
real number $x$, we are definitely crossing the cut. Written in terms of $k_1$
and $l_1$, we obtain
\begin{equation}\label{k1l1cut}
(k_1+l_1)^{\rm cut}
  =\pm\frac{x-(k'_0+l'_0)^2+m_3^2+(\sqrt s\mp\sqrt t)^2-i\eta}{2(k'_0+l'_0)}.
\end{equation}
Note that the two undetermined signs are not correlated. While the sign
between $\sqrt s$ and $\sqrt t$ depends on which of the two square roots is
taken, the global sign of the right hand side is due to the sign of the
linearization. Considering only the linearizations $k_0\rightarrow k'_0+k_1$,
$l_0\rightarrow l'_0+l_1$, the signs of the real and imaginary part of the cut
still depend on the sign of $(k'_0+l'_0)$. If $(k'_0+l'_0)>0$, the cut is
located in the lower complex half plane. The real part starts (for $x=0$) at
some finite value and runs to $+\infty$, independent of which square root we
selected. In order to avoid the cut, we therefore have to close the
integration path for $k_1$ resp.\ $l_1$ in the upper complex half plane while
for $(k'_0+l'_0)<0$ we have to close it in the lower half plane. The situation
is opposite for the linearization with a minus sign.

\section{Tensor reduction}
\begin{figure}\begin{center}
\epsfig{figure=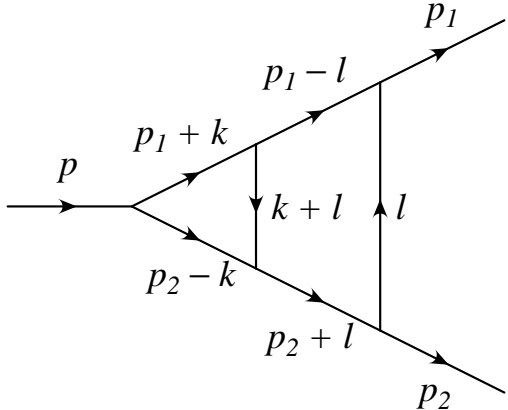, scale=0.5}
\caption{\label{momflow}Momentum flow convention for the planar two-loop
  three-point function}
\end{center}\end{figure}

After having introduced the main tools for the calculation, we can start with
the calculation itself. The starting point is the tensor integral
\begin{equation}
{\cal T}^0_{a_0a_1a_2b_0b_1b_2c}
  =\int\frac{k_0^{a_0}k_1^{a_1}(k_\perp^2)^{a_2}l_0^{b_0}l_1^{b_1}
  (l_\perp^2)^{b_2}(k_\perp l_\perp z)^c}{P_1P_2P_3P_4P_5P_6}d^4k\,d^4l
\end{equation}
where
\begin{eqnarray}\label{propfac}
P_1=(k+p_1)^2-m_1^2+i\eta&&P_4=(l-p_1)^2-m_4^2+i\eta\nonumber\\
P_2=(k-p_2)^2-m_2^2+i\eta&&P_5=(l+p_2)^2-m_5^2+i\eta\nonumber\\
P_3=(k+l)^2-m_3^2+i\eta&&P_6=l^2-m_6^2+i\eta.
\end{eqnarray}
If one is doing Standard Model calculations using
Feynman gauge, the powers are restricted by $0\le a_0,a_1,2a_2\le 3$ and
$0\le b_0,b_1,2b_2,c\le 4$, as well as by $0\le a_0+a_1+2a_2+c\le 3$ and
$1\le b_0+b_1+2b_2+c\le 4$. Nevertheless, the algorithm introduced in this
paper works for general powers. In Fig.~\ref{momflow} we show the momentum
flow for the diagram in order to define the momenta.
For the indices $a_i,b_i,c$ we allow non-negative integer values. In
performing the tensor reduction the integrals are simplified to integrals with
simpler numerator and/or denominator structure.

For one-loop integrals the numerator can always be removed by reduction
procedures \cite{Passarino:1978jh}. For two-loop integrals this need not be
the case. In general there are not enough propagator factors to cancel all
components of the numerator that occur. For the genuine planar two-loop
three-point function all numerator factors related to the second loop momentum
$l$ can be cancelled. This is not the case for the reduced topologies, and it
is not the case for the first loop momentum $k$. The aim of the tensor
reduction in general is to reduce the numerator as far as possible so that the
integrations can be performed similar to what is done for a trivial
numerator~\cite{Czarnecki:1994td}. The reduction procedure will be explained
in the following. The representation of the procedure by diagrams will show
the topologies only. In this spirit plus signs like the one occuring in
Eq.~(\ref{diagmix0}) have to be understood as sums over diagrams with the same
topology but different factors and signs.

\subsection{Cancellation of the mixed contribution}
The first step in the cancellation procedure consist in cancelling powers of
the mixed factor $(k_\perp l_\perp z)$. This factor occurs also in the
propagator factor $P_3$. We can write
\begin{equation}
2k_\perp l_\perp z=N_3-P_3
\end{equation}
where
\begin{equation}
N_3=k_0^2-k_1^2-k_\perp^2+l_0^2-l_1^2-l_\perp^2+2k_0l_0-2k_1l_1-m_3^2+i\eta.
\end{equation}
Used iteratively, we obtain
\begin{equation}
(k_\perp l_\perp z)^c=\pfrac{N_3}2^c-\sum_{i=0}^{c-1}\frac{N_3^i}{2^{i+1}}
(k_\perp l_\perp z)^{c-i-1}P_3.
\end{equation}
This iterative formula is necessary because the propagator factor $P_3$
occurs only once. As applied to the integral, we obtain
\begin{equation}
\int\frac{(k_\perp l_\perp z)^cd^4k\,d^4l}{P_1P_2P_3P_4P_5P_6}
  =\frac1{2^c}\int\frac{N_3^cd^4k\,d^4l}{P_1P_2P_3P_4P_5P_6}
  -\sum_{i=0}^{c-1}\frac1{2^{i+1}}\int
  \frac{N_3^i(k_\perp l_\perp z)^{c-i-1}}{P_1P_2P_4P_5P_6}d^4k\,d^4l
\end{equation}
where we have skipped all the other numerator factors for convenience. The
first part is the same planar integral with a different numerator structure.
However, the second part no longer contains the mixing propagator factor
$P_3$. Instead, the diagram factorizes. In addition, we can conclude that this
vanishes if $c-i-1$ is odd. The reason is that the integration over $z$ is
given by
\begin{equation}
\int_{-1}^{+1}\frac{z^{c-i-1}dz}{\sqrt{1-z^2}}
\end{equation}
which vanishes if the integrand is odd. Diagramatically we can write
\begin{equation}\label{diagmix0}
\raise-27pt\hbox{\epsfig{figure=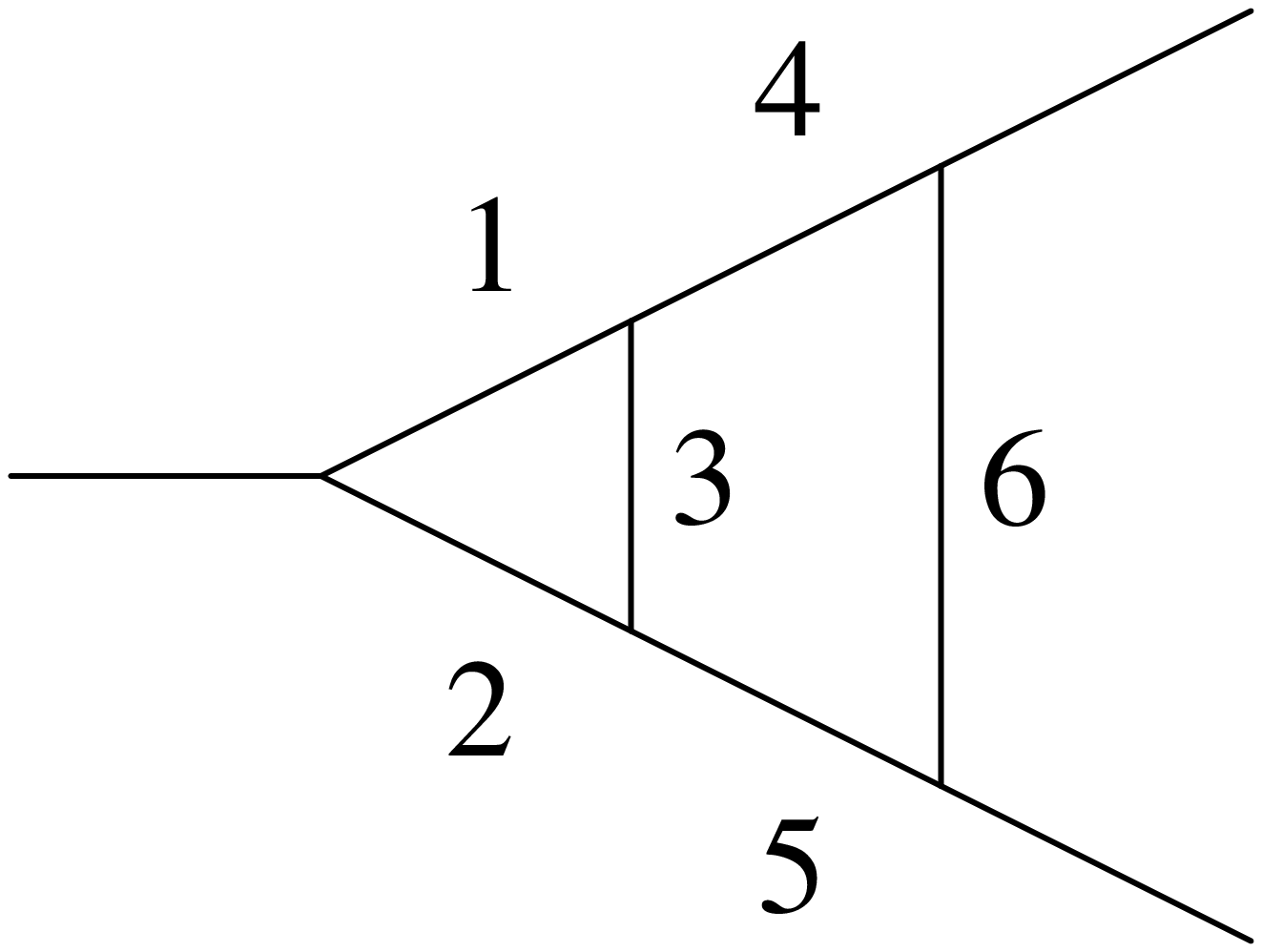, scale=0.2}}
\times(k_\perp l_\perp z)^c\quad\longrightarrow\quad
\raise-27pt\hbox{\epsfig{figure=d123456.eps, scale=0.2}}\quad+\quad
\raise-27pt\hbox{\epsfig{figure=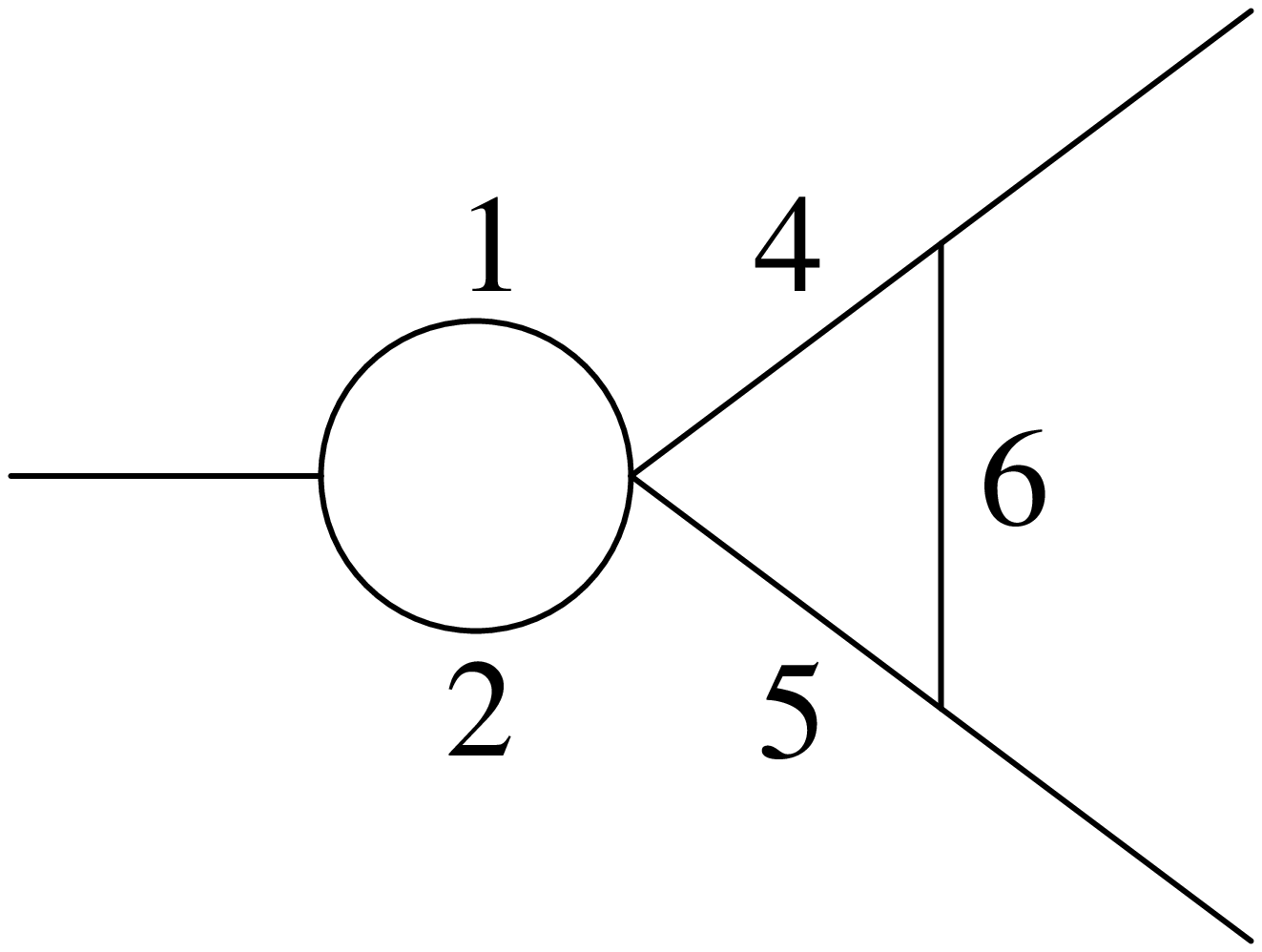, scale=0.2}}
\end{equation}
After having cancelled the mixing part it is irrelevant with which loop
momentum we continue. However, we always start with the orthogonal space
components.

\subsection{Cancellation of $k_\perp^2$}
The factor $k_\perp^2=s$ occurs in the denominator factor $P_1$. We can write
\begin{equation}
k_\perp^2=N_1-P_1,\qquad
N_1=k_0^2-k_1^2+2k_0E_1-2k_1q_z+E_1^2-q_z^2-m_1^2+i\eta.
\end{equation}
The iterative formula
\begin{equation}
(k_\perp^2)^{a_2}=N_1^{a_2}-\sum_{i=0}^{a_2-1}N_1^i(k_\perp^2)^{a_2-i-1}P_1
\end{equation}
can be used for the integral to obtain
\begin{equation}
\int\frac{(k_\perp^2)^{a_2}d^4k\,d^4l}{P_1P_2P_3P_4P_5P_6}
  =\int\frac{N_1^{a_2}d^4k\,d^4l}{P_1P_2P_3P_4P_5P_6}
  -\sum_{i=0}^{a_2-1}\int\frac{N_1^i(k_\perp^2)^{a_2-i-1}}{P_2P_3P_4P_5P_6}
  d^4k\,d^4l.
\end{equation}
Schematically this reduction reads
\begin{equation}
\raise-27pt\hbox{\epsfig{figure=d123456.eps, scale=0.2}}
\times(k_\perp^2)^{a_2}\quad\longrightarrow\quad
\raise-27pt\hbox{\epsfig{figure=d123456.eps, scale=0.2}}\quad+\quad
\raise-27pt\hbox{\epsfig{figure=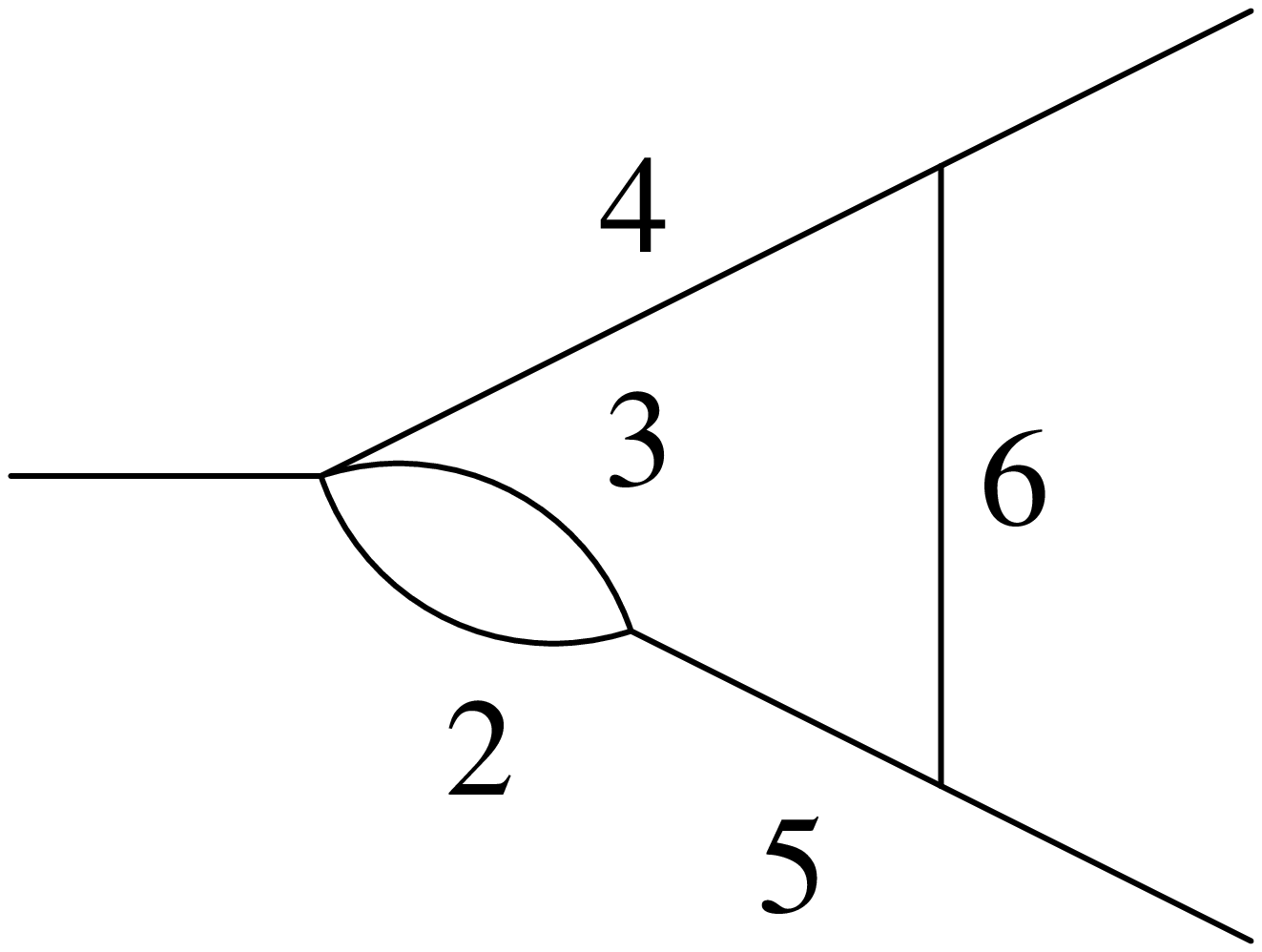, scale=0.2}}
\end{equation}
In the following only the diagrammatic reduction will be shown.

\subsection{Replacement of $k_1$}
In anticipation of the linearization we replace the numerator factors
$k_1$ by $\mp((k_0\mp k_1)-k_0)$, using
\begin{equation}
k_1^{a_1}=(\mp 1)^{a_1}\sum_{i=0}^{a_1}(-k_0)^{a_1-i}
  \left({a_1\atop i}\right)(k_0\mp k_1)^i.
\end{equation}
While the additional factors $k_0^{a_1-i}$ will be cancelled together with
$k_0^{a_0}$ in the next step, the subsequent linearization will replace
$(k_0\mp k_1)$ by $k'_0$.

\subsection{Cancellation of $k_0$}
$P_1$ and $P_2$ are the only propagator factors that contain only $k$ and not
$l$. We can combine these both to obtain
\begin{equation}
2Ek_0=N_2+P_1-P_2,\qquad
  N_2=p_2^2-p_1^2-m_2^2+m_1^2.
\end{equation}
An iterative formula can be constructed as before, the reduction reads
\begin{equation}
\raise-27pt\hbox{\epsfig{figure=d123456.eps, scale=0.2}}
\times k_0^{a_0}\quad\longrightarrow\quad
\raise-27pt\hbox{\epsfig{figure=d123456.eps, scale=0.2}}\quad+\quad
\raise-27pt\hbox{\epsfig{figure=dx23456.eps, scale=0.2}}\quad+\quad
\raise-27pt\hbox{\epsfig{figure=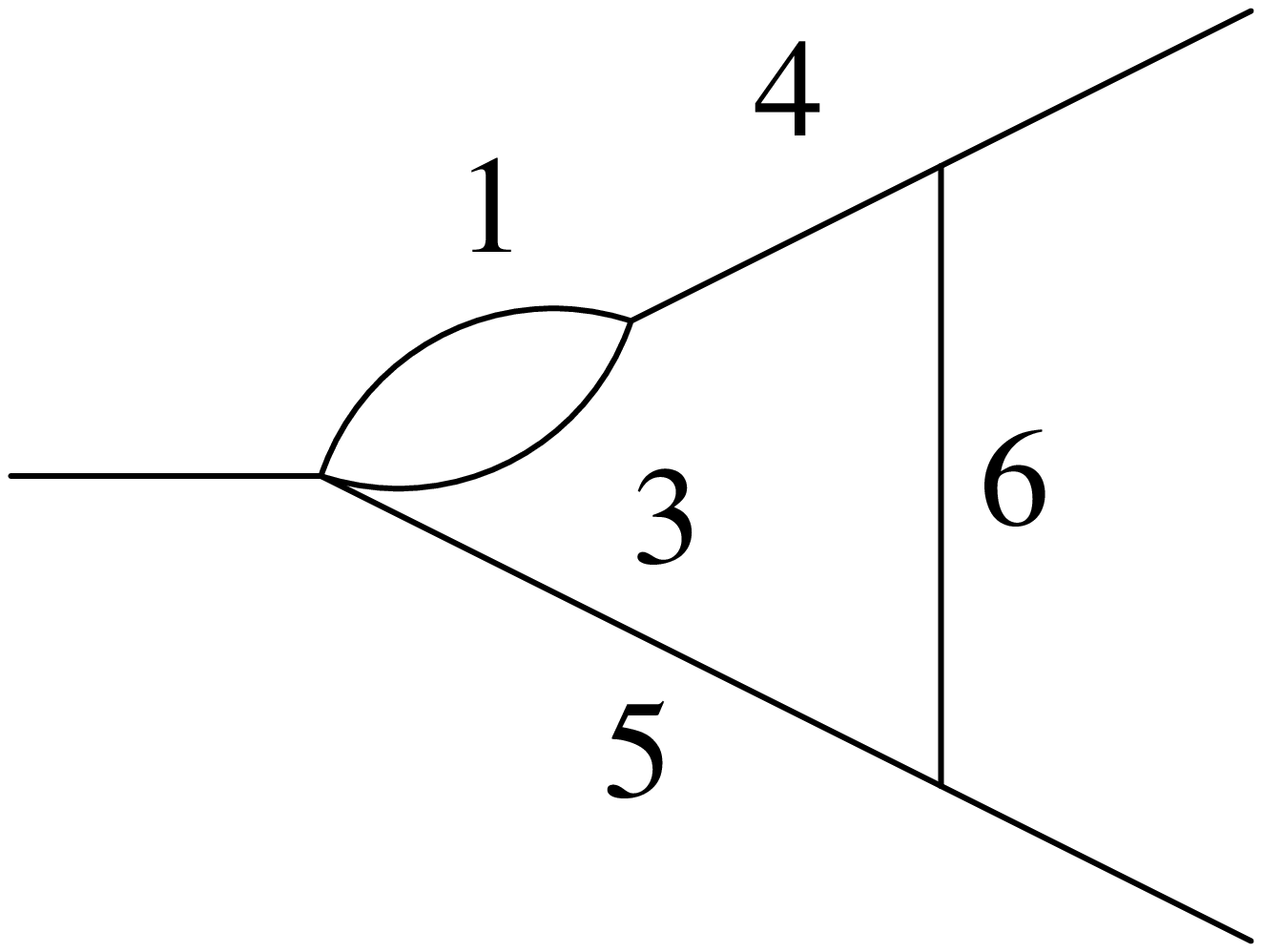, scale=0.2}}
\end{equation}

\subsection{Cancellation of $l_\perp^2$}
In using
\begin{equation}
l_\perp^2=N_6-P_6,\qquad N_6=l_0^2-l_1^2-m_6^2+i\eta
\end{equation}
we can cancel the factors $l_\perp^2$, ending up with the reduction scheme
\begin{equation}
\raise-27pt\hbox{\epsfig{figure=d123456.eps, scale=0.2}}
\times(l_\perp^2)^{b_2}\quad\longrightarrow\quad
\raise-27pt\hbox{\epsfig{figure=d123456.eps, scale=0.2}}\quad+\quad
\raise-27pt\hbox{\epsfig{figure=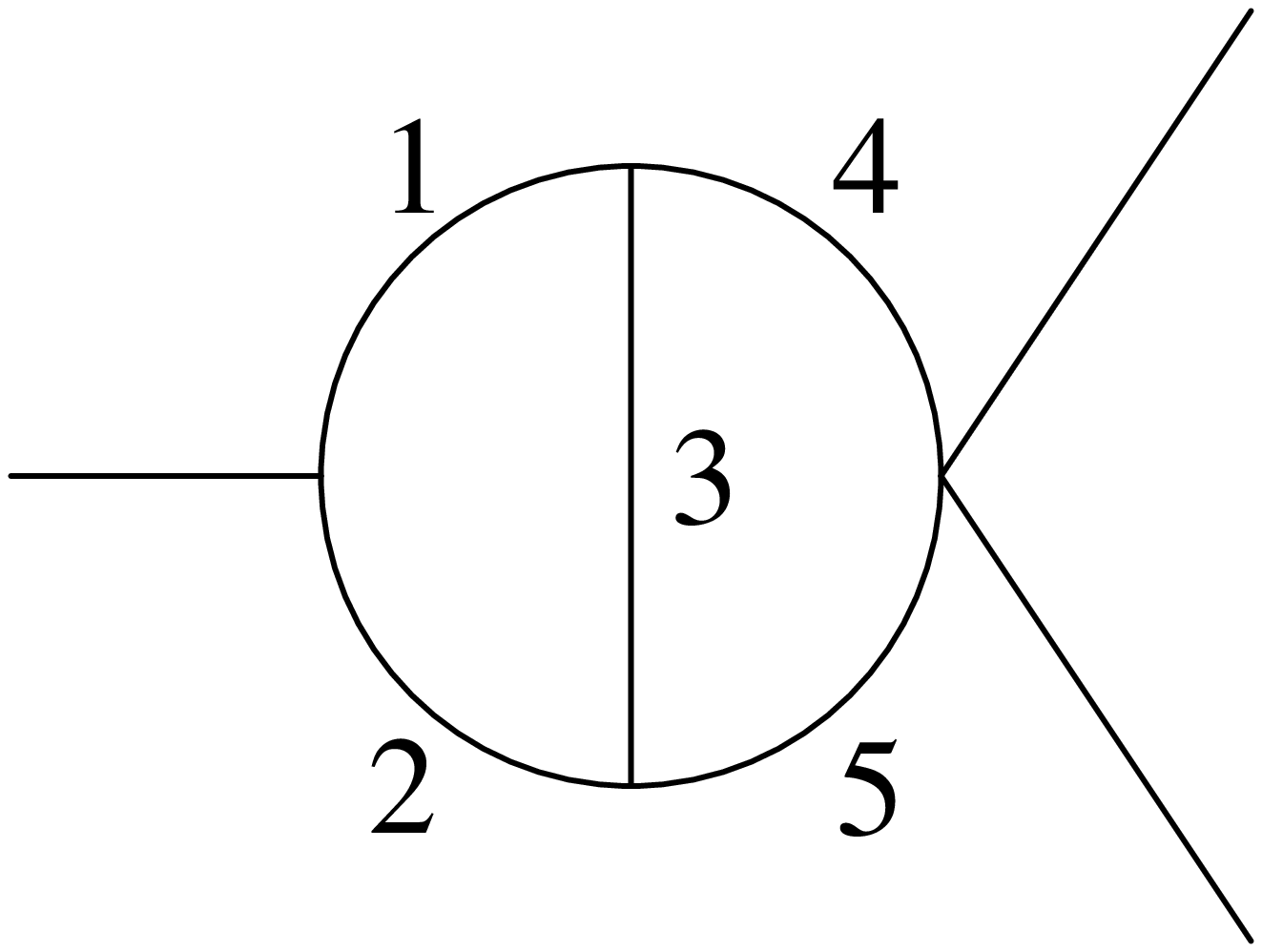, scale=0.2}}
\end{equation}

\subsection{Cancellation of $l_1$}
Different from the situation for the replacement of $k_1$, we have three
propagator factors which do not contain $k$. In this case the factors $l_1$
can be cancelled completely. We use
\begin{equation}
2q_zl_1=N_4+P_4-P_6,\qquad N_4=2l_0E_1-p_1^2+m_4^2-m_6^2
\end{equation}
to reduce powers of $l_1$ according to the scheme
\begin{equation}
\raise-27pt\hbox{\epsfig{figure=d123456.eps, scale=0.2}}
\times l_1^{b_1}\quad\longrightarrow\quad
\raise-27pt\hbox{\epsfig{figure=d123456.eps, scale=0.2}}\quad+\quad
\raise-27pt\hbox{\epsfig{figure=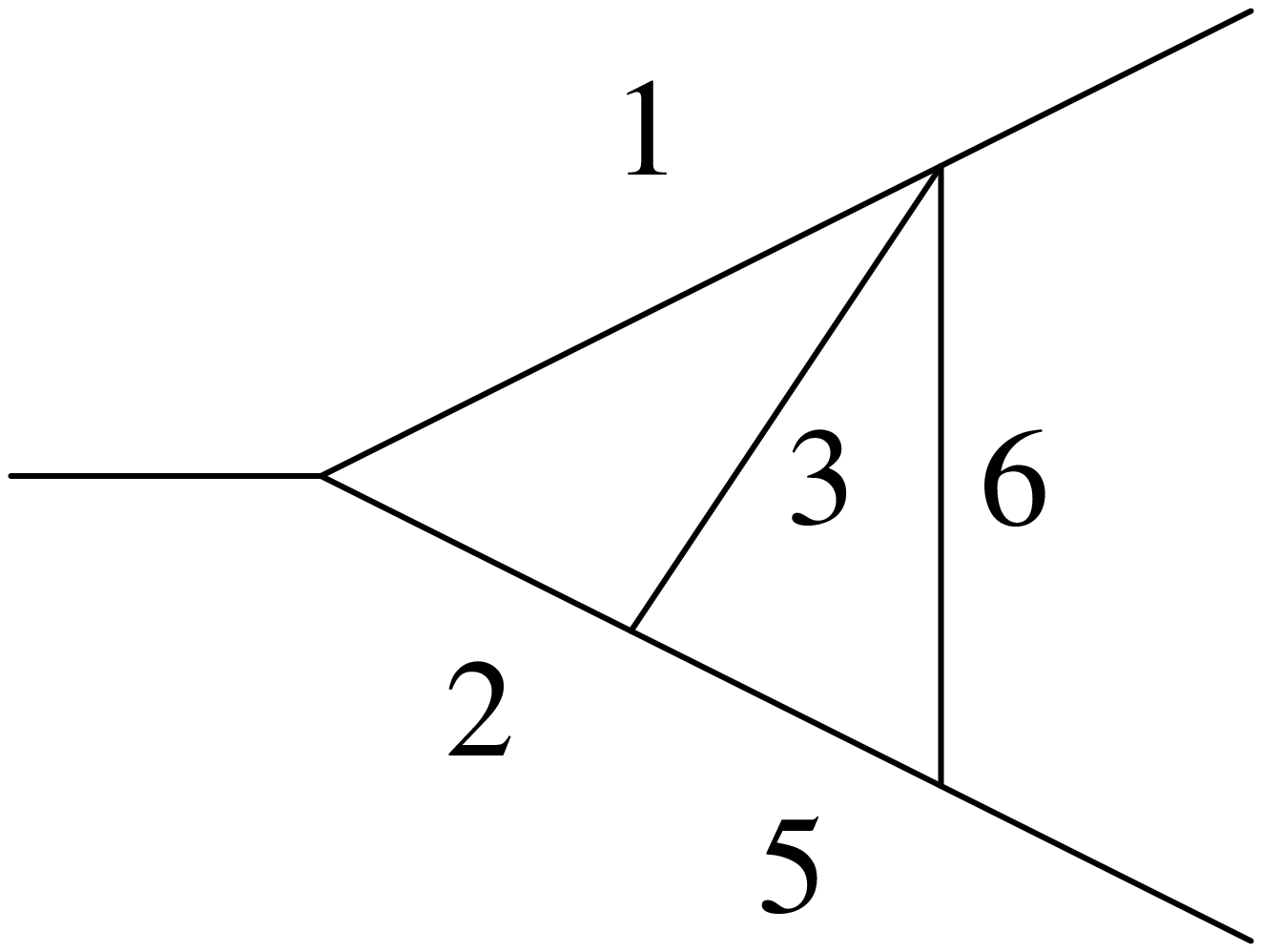, scale=0.2}}\quad+\quad
\raise-27pt\hbox{\epsfig{figure=d12345x.eps, scale=0.2}}
\end{equation}

\subsection{Cancellation of $l_0$}
Finally, we can cancel the factors $l_0$ as well, using
\begin{equation}
2El_0=N_5+P_5-P_4,\qquad N_5=p_1^2-p_2^2-m_4^2+m_5^2.
\end{equation}
The reduction scheme reads
\begin{equation}
\raise-27pt\hbox{\epsfig{figure=d123456.eps, scale=0.2}}
\times l_0^{b_0}\quad\longrightarrow\quad
\raise-27pt\hbox{\epsfig{figure=d123456.eps, scale=0.2}}\quad+\quad
\raise-27pt\hbox{\epsfig{figure=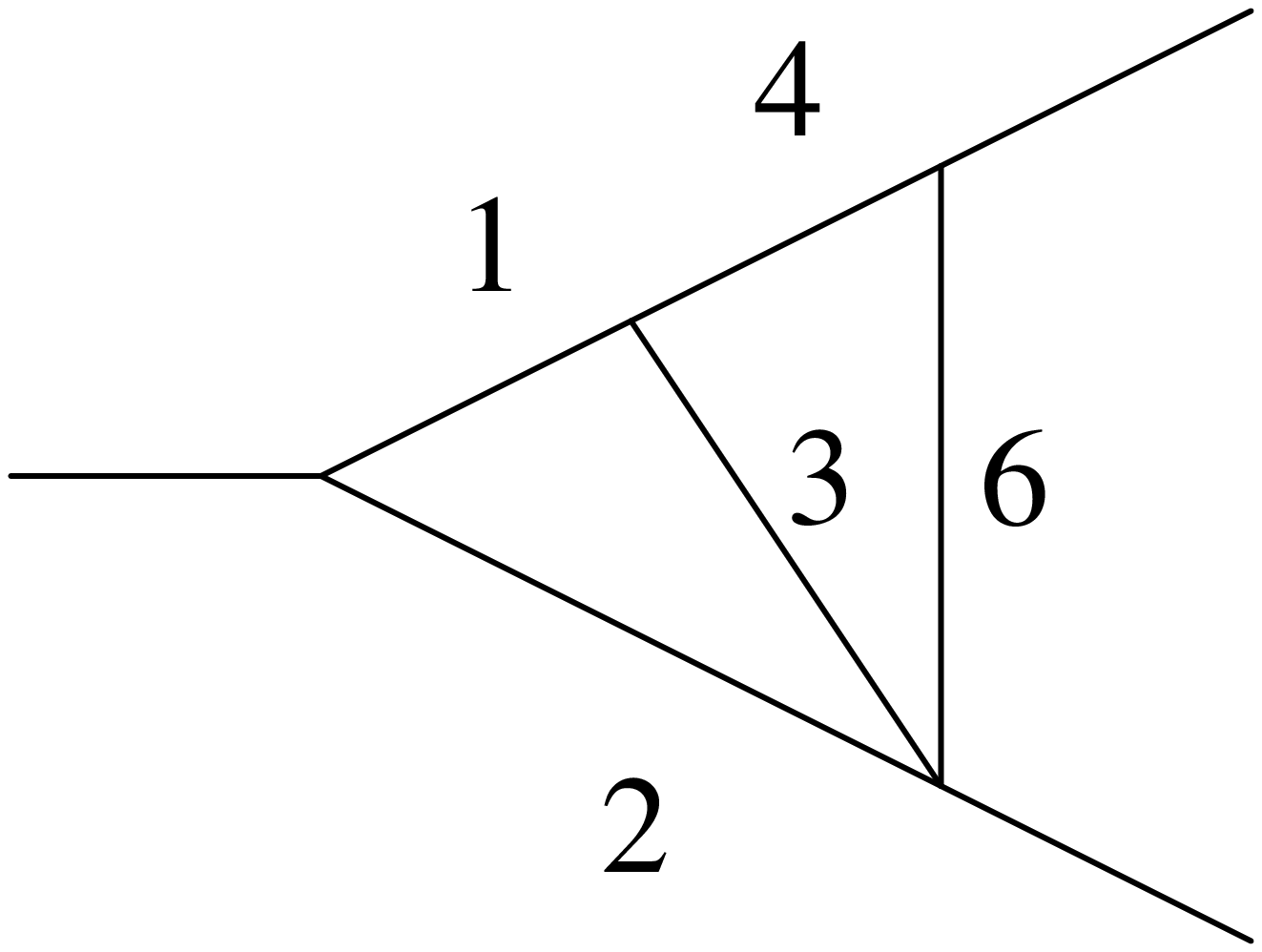, scale=0.2}}\quad+\quad
\raise-27pt\hbox{\epsfig{figure=d123x56.eps, scale=0.2}}
\end{equation}

\subsection{Summary of the reduction procedure\label{sumredpro}}
After performing the reduction procedure, we can collect all the steps into
a single one. All diagrams which are still of the (original) planar topology 
after reduction procedure have a numerator factor $(k_0\mp k_1)$ in different
powers. The maximal power, i.e.\ the maximal value for $\alpha$, is given
by $a_1+2a_2+2c$. The numerator factors of the diagrams with reduced topology
are changed but not easier. Nevertheless, because of the different topology,
reduction procedures which are known in the literature or which will be
explained in the following can be applied. Schematically we write
\begin{eqnarray}
\lefteqn{\raise-27pt\hbox{\epsfig{figure=d123456.eps, scale=0.2}}
\times k_0^{a_0}k_1^{a_1}(k_\perp^2)^{a_2}l_0^{b_0}l_1^{b_1}(l_\perp^2)^{b_2}
  (k_\perp l_\perp z)^c\quad\longrightarrow\quad
\raise-27pt\hbox{\epsfig{figure=d123456.eps, scale=0.2}}
\times(k_0\mp k_1)^\alpha}\nonumber\\&&+\quad
\raise-27pt\hbox{\epsfig{figure=dx23456.eps, scale=0.2}}\quad+\quad
\raise-27pt\hbox{\epsfig{figure=d1x3456.eps, scale=0.2}}\quad+\quad
\raise-27pt\hbox{\epsfig{figure=d12x456.eps, scale=0.2}}\nonumber\\&&+\quad
\raise-27pt\hbox{\epsfig{figure=d123x56.eps, scale=0.2}}\quad+\quad
\raise-27pt\hbox{\epsfig{figure=d1234x6.eps, scale=0.2}}\quad+\quad
\raise-27pt\hbox{\epsfig{figure=d12345x.eps, scale=0.2}}\kern96pt
\end{eqnarray}
Denoting the topologies by ${\cal T}^0,{\cal T}^1,\ldots,{\cal T}^6$ according
to the cancelled propagator factor, we can cite the following references:
\begin{itemize}
\item {\it Two-loop three-point functions with two-point subloop\/}, as
they are given in our case by the topologies ${\cal T}^1$ and ${\cal T}^2$,
can be calculated by using dispersion
relations~\cite{Bauberger:1994by,Post:1997,Post:1996gg}.
\item The {\it factorizing topology ${\cal T}^3$} is calculated as a product
of one-loop topologies. This contributions may be calculated using
\xloops~\cite{Bauer:2001ig} or similar packages like
SANC~\cite{Andonov:2004hi}.
\item The {\it effective two-loop two-point topology ${\cal T}^6$} 
can be calculated using \xloops~\cite{Do:2003}.
\item What is left are the {\it rotated reduced planar topologies\/}
${\cal T}^4$ and ${\cal T}^5$. They can be reduced in a similar manner as the
original planar topologies.\footnote{Because of historical reasons, these
topologies are called rotated reduced planar topologies. In the original
reduced planar topology the line crossing the triangle is going horizontally
from left to right, connecting the left vertex with the middle of the vertical
line on the right.} In the following section we will deal with these
topologies and the necessary modifications of the reduction procedure. 
\end{itemize}

\subsection{Modifications for the rotated reduced planar topologies}
For the rotated reduced planar topologies we assume the same numerator factor
as for the non-reduced one, $k_0^{a_0}k_1^{a_1}(k_\perp^2)^{a_2}l_0^{b_0}
l_1^{b_1}(l_\perp^2)^{b_2}(k_\perp l_\perp z)^c$. The procedure is quite the
same as in the case discussed before. But because one of the denominator
factors $P_4$ and $P_5$ is absent, it is not possible to cancel the factors of
$l_1$ as in the previous case. Instead, we anticipate the linearization as in
the case of $k_1$ (and with the same sign) in writing
\begin{equation}
l_1=\mp\left((l_0\mp l_1)-l_0\right)\quad\Rightarrow\quad
l_1^{b_1}=(\mp 1)^{b_1}\sum_{j=0}^{b_1}(-l_0)^{b_1-j}\left({b_1\atop j}\right)
  (l_0\mp l_1)^j.
\end{equation}
Again, while the additional factors $l_0^{b_1-j}$ will be cancelled together
with $l_0^{b_0}$ in the last step, the subsequent linearization will replace
$(l_0\mp l_1)$ by $l'_0$. However, this last step makes the difference between
the two topologies.

For the topology ${\cal T}^4$ the denominator factor $P_4$ is absent. However,
we can obtain
\begin{equation}
2(E_2\pm q_z)l_0=N'_5-P_6+P_5,\qquad
  N'_5=\pm2(l_0\mp l_1)q_z-p_2^2-m_6^2+m_5^2.
\end{equation}
At this point the different signs for the linearization enter the game. While
we have designed the term $(l_0\mp l_1)$ in the way that it can be combined
with the factors from the cancellation of $l_1$, in order to obtain $l_0$ we
have to divide by $(E_2\pm q_z)$. However, if the second produced particle is
massless, $p_2^2=0$, this factor will become zero. In order to include also
the massless case, we will prefer the linearization $l_1=l'_1+l_0$ for this
topology, i.e.\ the upper sign. The whole reduction procedure for the case of
the topology ${\cal T}^4$ reads
\begin{eqnarray}
\lefteqn{\raise-27pt\hbox{\epsfig{figure=d123x56.eps, scale=0.2}}
\times k_0^{a_0}k_1^{a_1}(k_\perp^2)^{a_2}l_0^{b_0}l_1^{b_1}(l_\perp^2)^{b_2}
  (k_\perp l_\perp z)^c\quad\longrightarrow}\nonumber\\&&\qquad
\raise-27pt\hbox{\epsfig{figure=d123x56.eps, scale=0.2}}
\times(k_0-k_1)^\alpha(l_0-l_1)^\beta\nonumber\\&&+\quad
\raise-27pt\hbox{\epsfig{figure=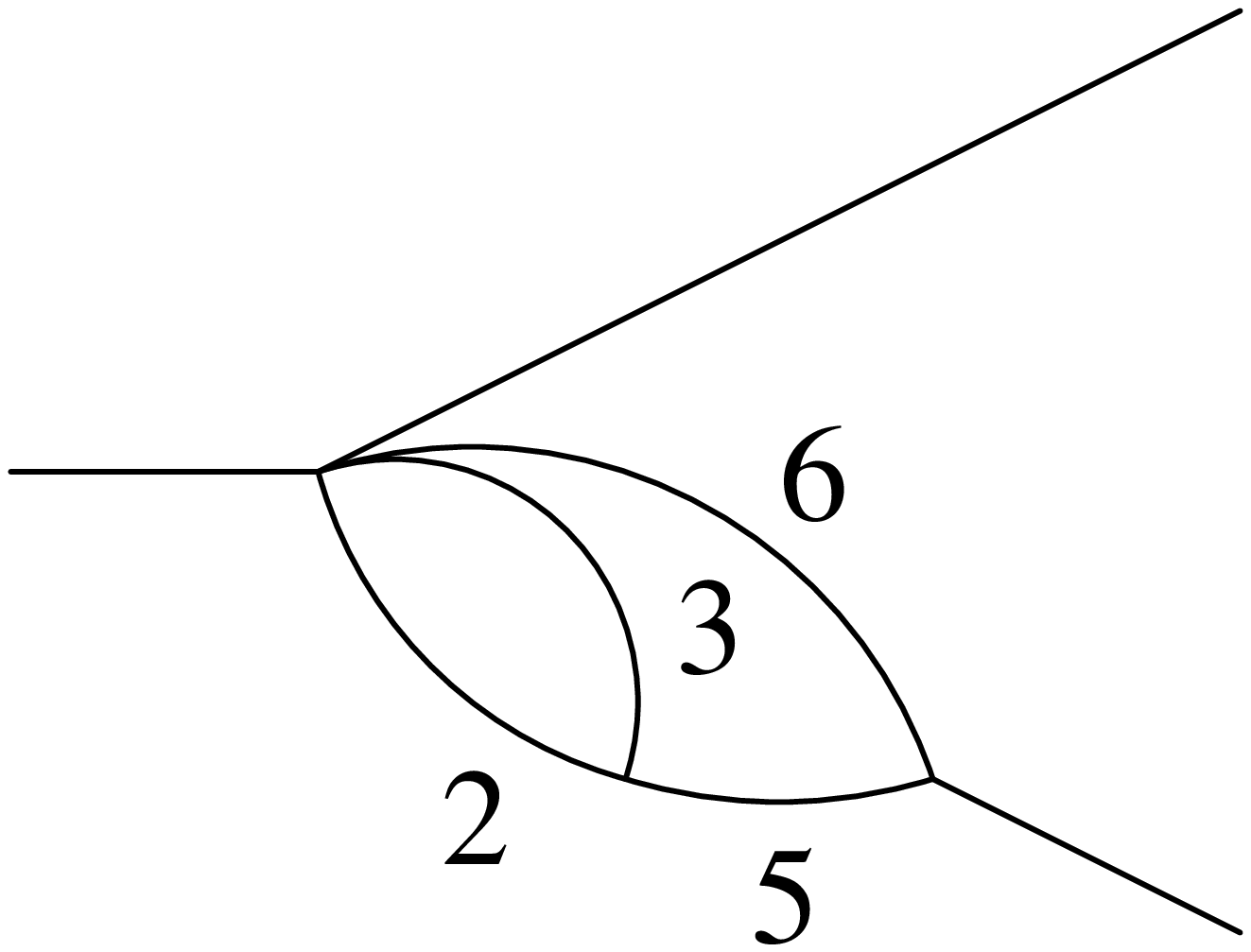, scale=0.2}}\quad+\quad
\raise-27pt\hbox{\epsfig{figure=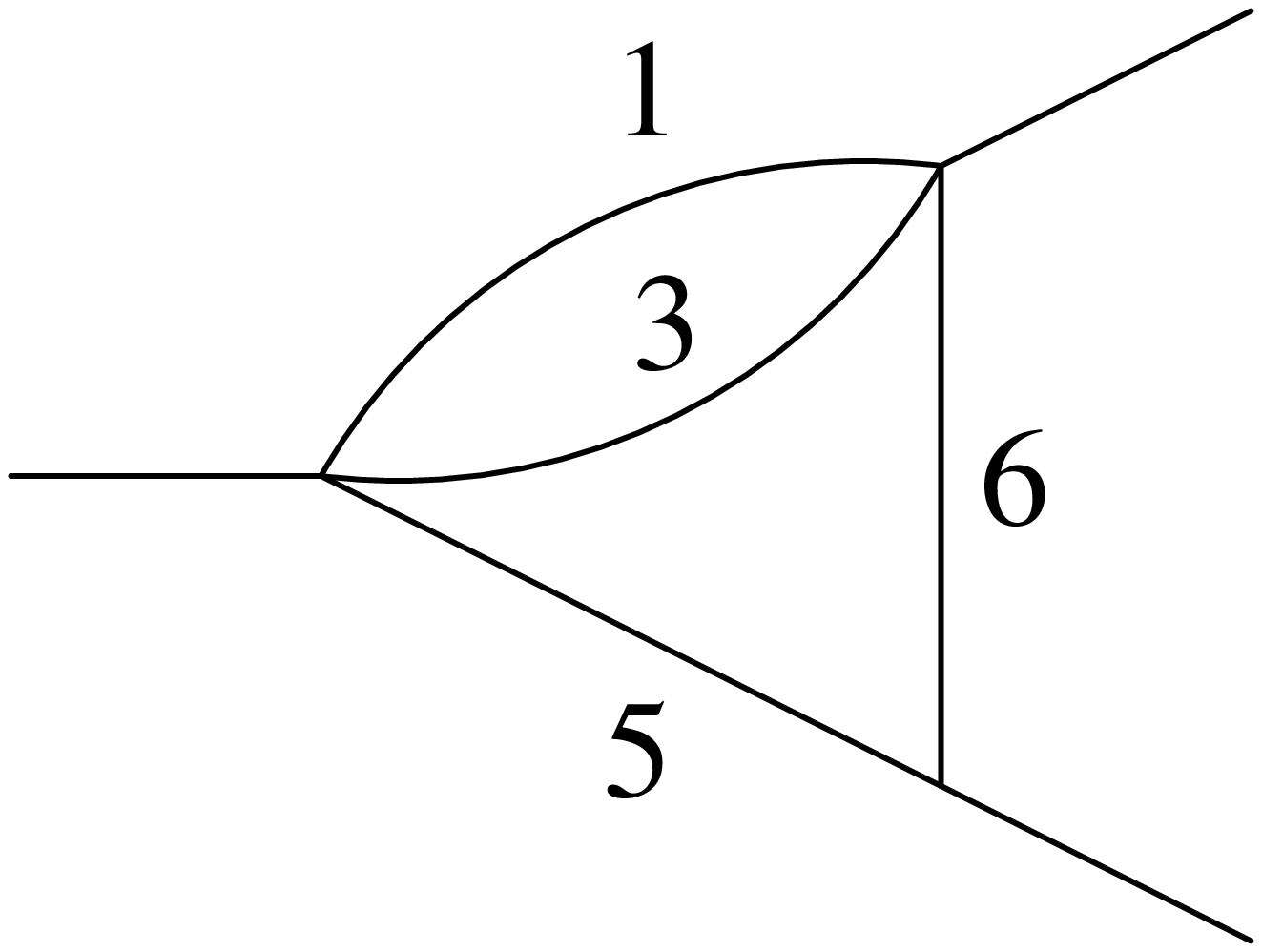, scale=0.2}}\quad+\quad
\raise-27pt\hbox{\epsfig{figure=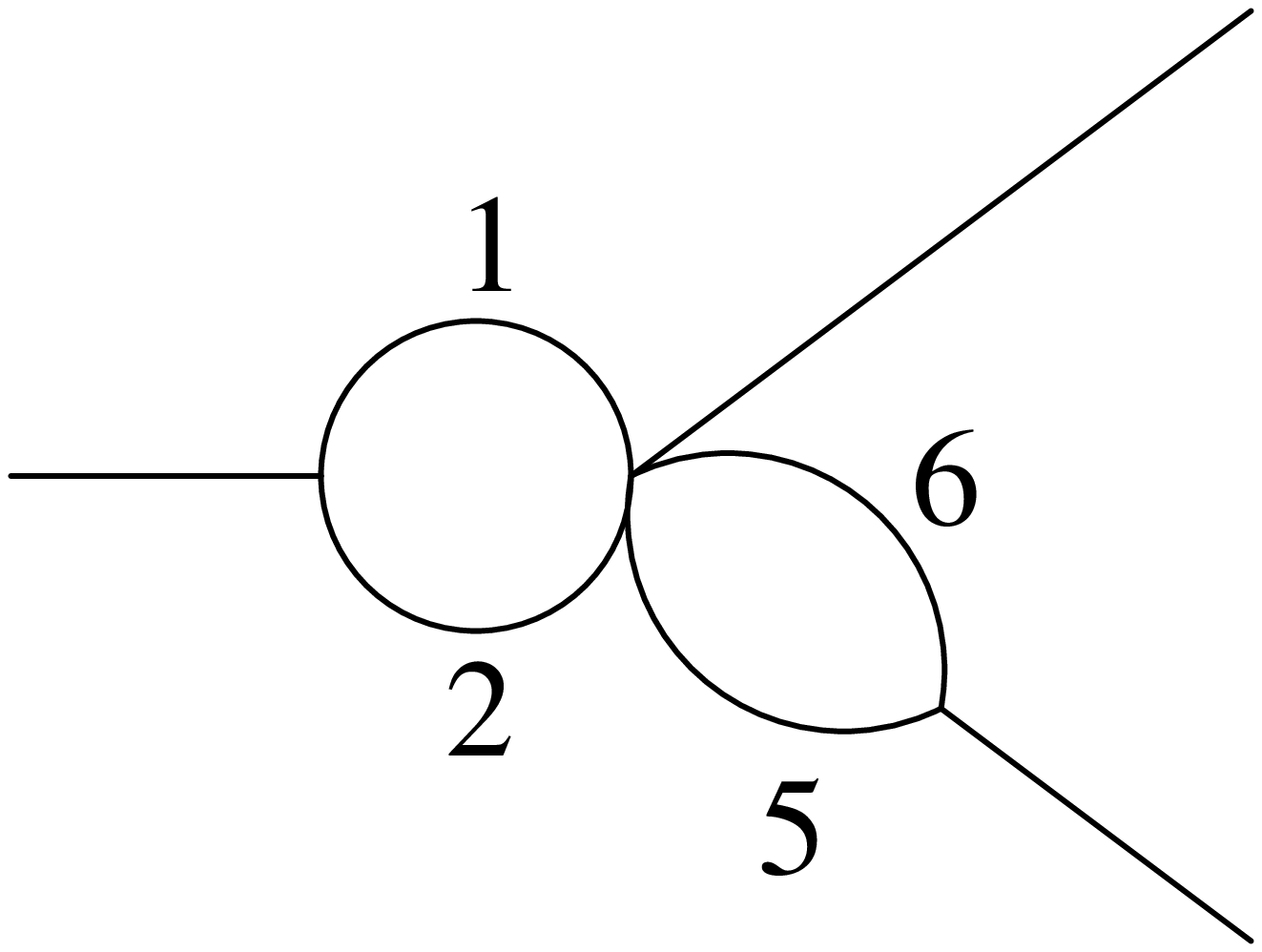, scale=0.2}}\nonumber\\&&+\quad
\raise-27pt\hbox{\epsfig{figure=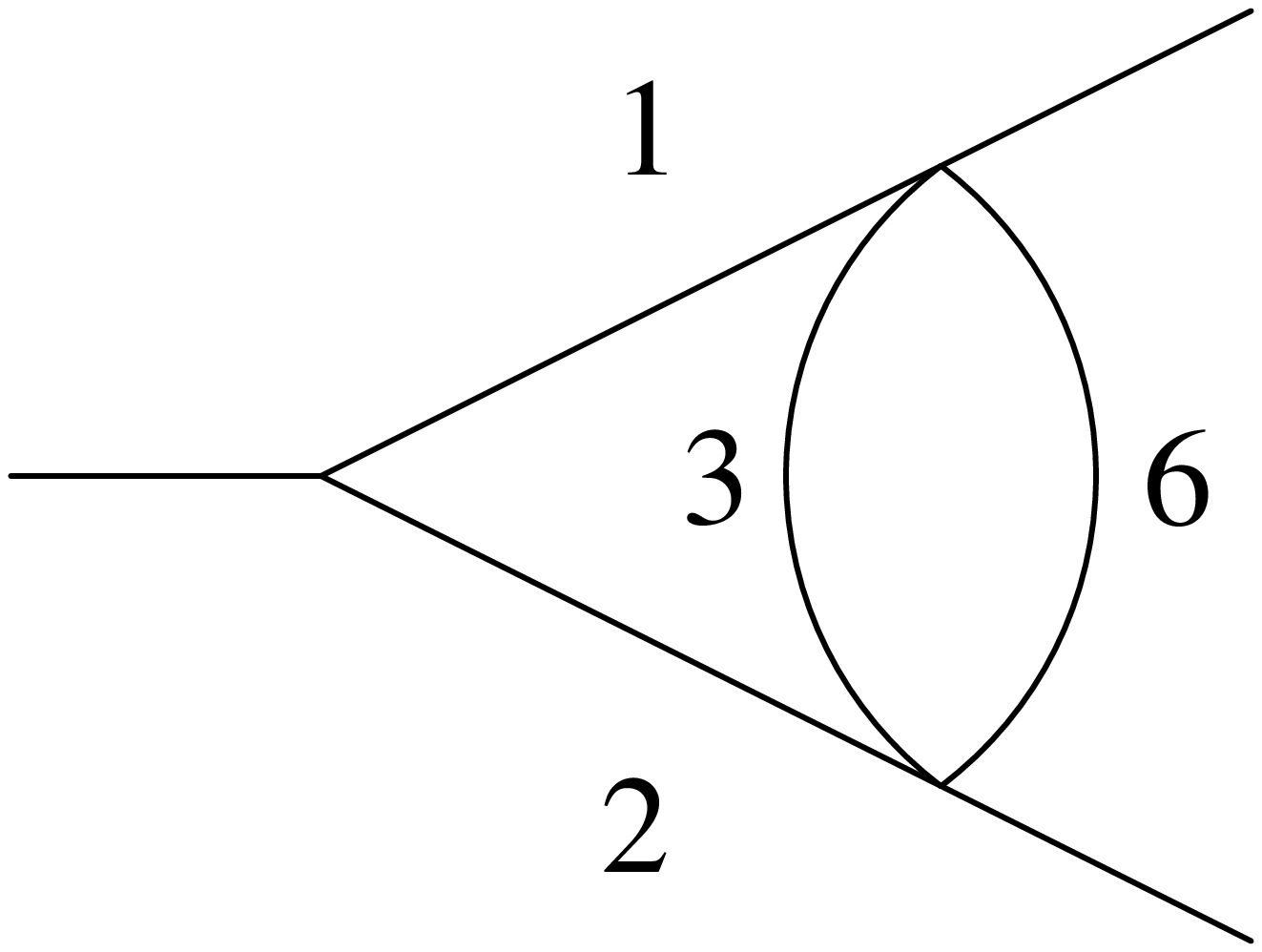, scale=0.2}}\quad+\quad
\raise-27pt\hbox{\epsfig{figure=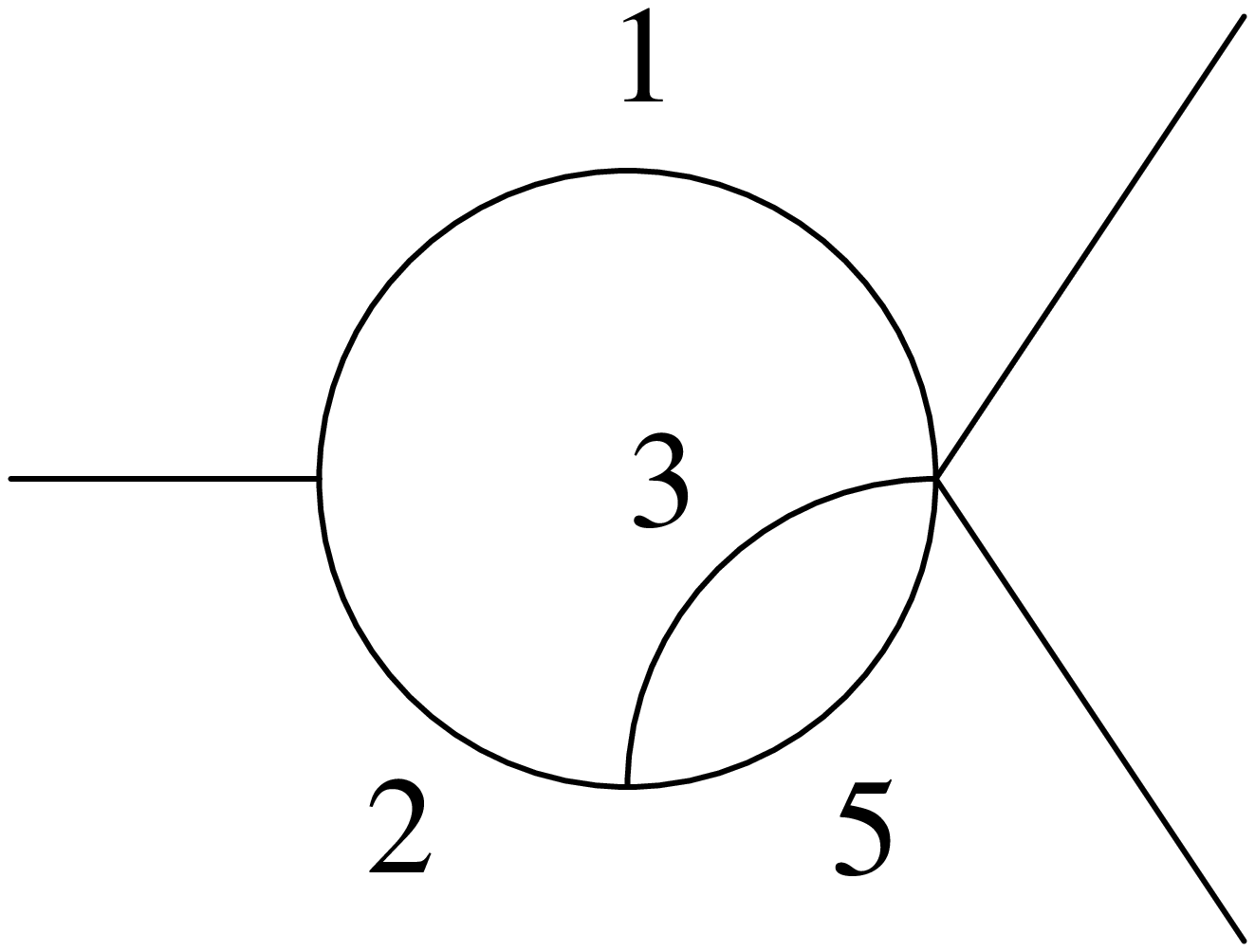, scale=0.2}}\kern96pt
\end{eqnarray}
All the diagrams on the right hand side except for the first one with maximal
powers $a_1+2a_2+2c$ for $\alpha$ and $b_0+b_1+2b_2+2c$ for $\beta$ are again
members of the different simpler topology classes which were mentioned in the
previous paragraph.

Looking at the topology ${\cal T}^5$, the propagator factor $P_5$ is absent.
In this case we can obtain a reduction formula from combining $P_4$ and $P_6$
to obtain
\begin{equation}
2(E_1\mp q_z)l_0=N'_4-P_4+P_6,\qquad
  N'_4=\mp2(l_0\mp l_1)q_z+p_2^2-m_4^2+m_6^2.
\end{equation}
In this case the linearization $l_1=l'_1-l_0$ (lower sign) is more
appropriate. The result of the reduction reads
\begin{eqnarray}
\lefteqn{\raise-27pt\hbox{\epsfig{figure=d1234x6.eps, scale=0.2}}
\times k_0^{a_0}k_1^{a_1}(k_\perp^2)^{a_2}l_0^{b_0}l_1^{b_1}(l_\perp^2)^{b_2}
  (k_\perp l_\perp z)^c\quad\longrightarrow}\nonumber\\&&\qquad
\raise-27pt\hbox{\epsfig{figure=d1234x6.eps, scale=0.2}}
\times(k_0+k_1)^\alpha(l_0+l_1)^\beta\nonumber\\&&+\quad
\raise-27pt\hbox{\epsfig{figure=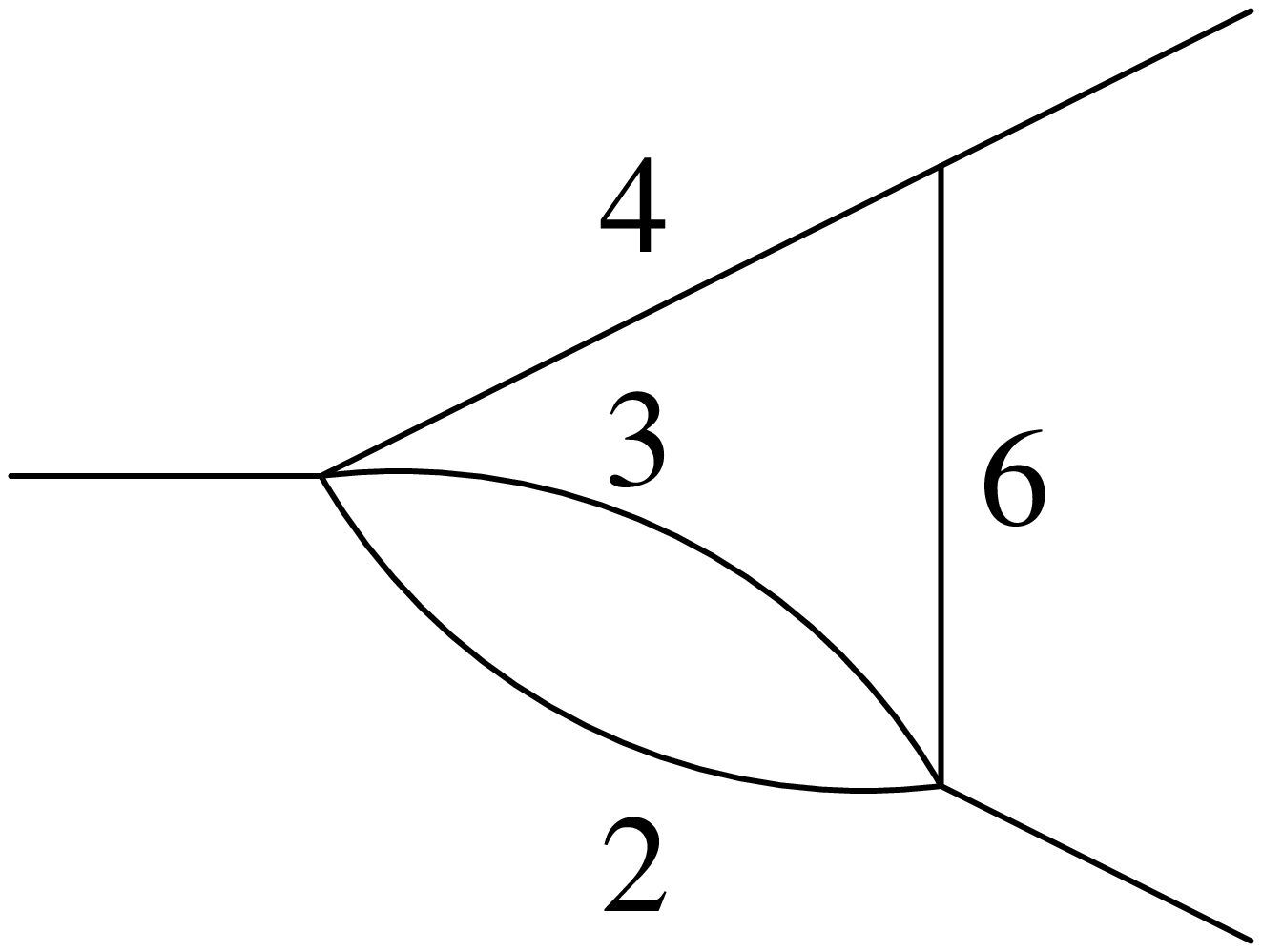, scale=0.2}}\quad+\quad
\raise-27pt\hbox{\epsfig{figure=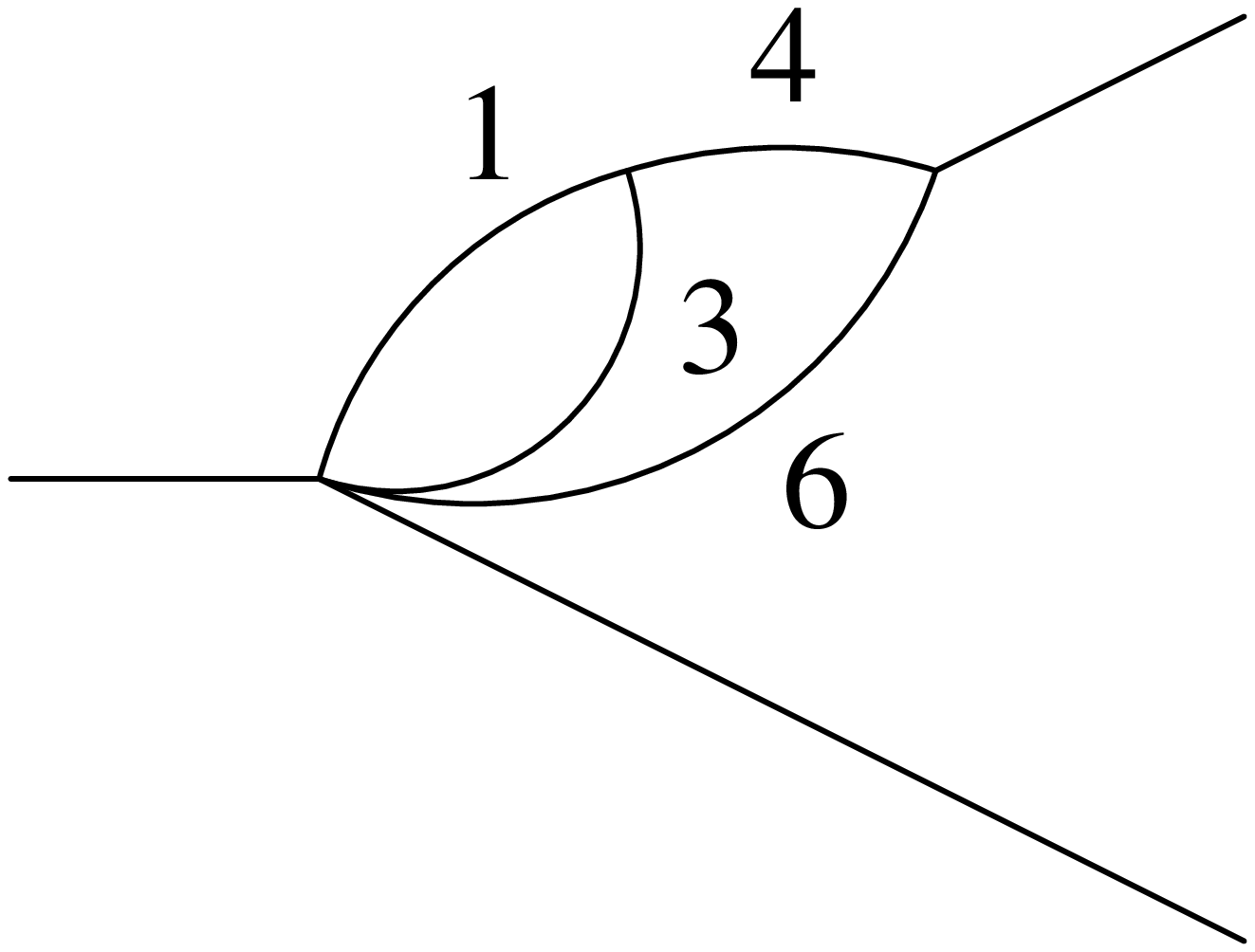, scale=0.2}}\quad+\quad
\raise-27pt\hbox{\epsfig{figure=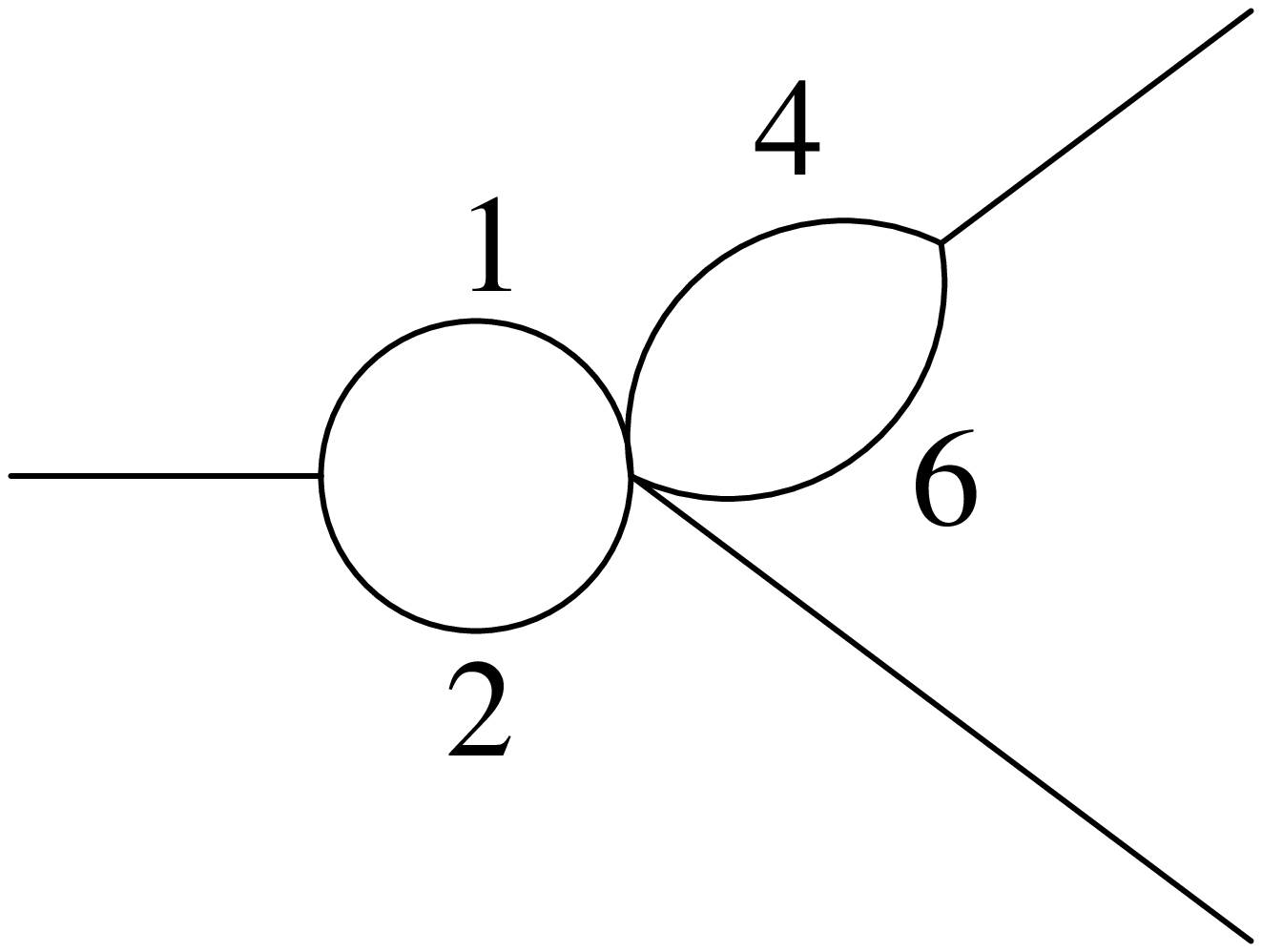, scale=0.2}}\nonumber\\&&+\quad
\raise-27pt\hbox{\epsfig{figure=d123xx6.eps, scale=0.2}}\quad+\quad
\raise-27pt\hbox{\epsfig{figure=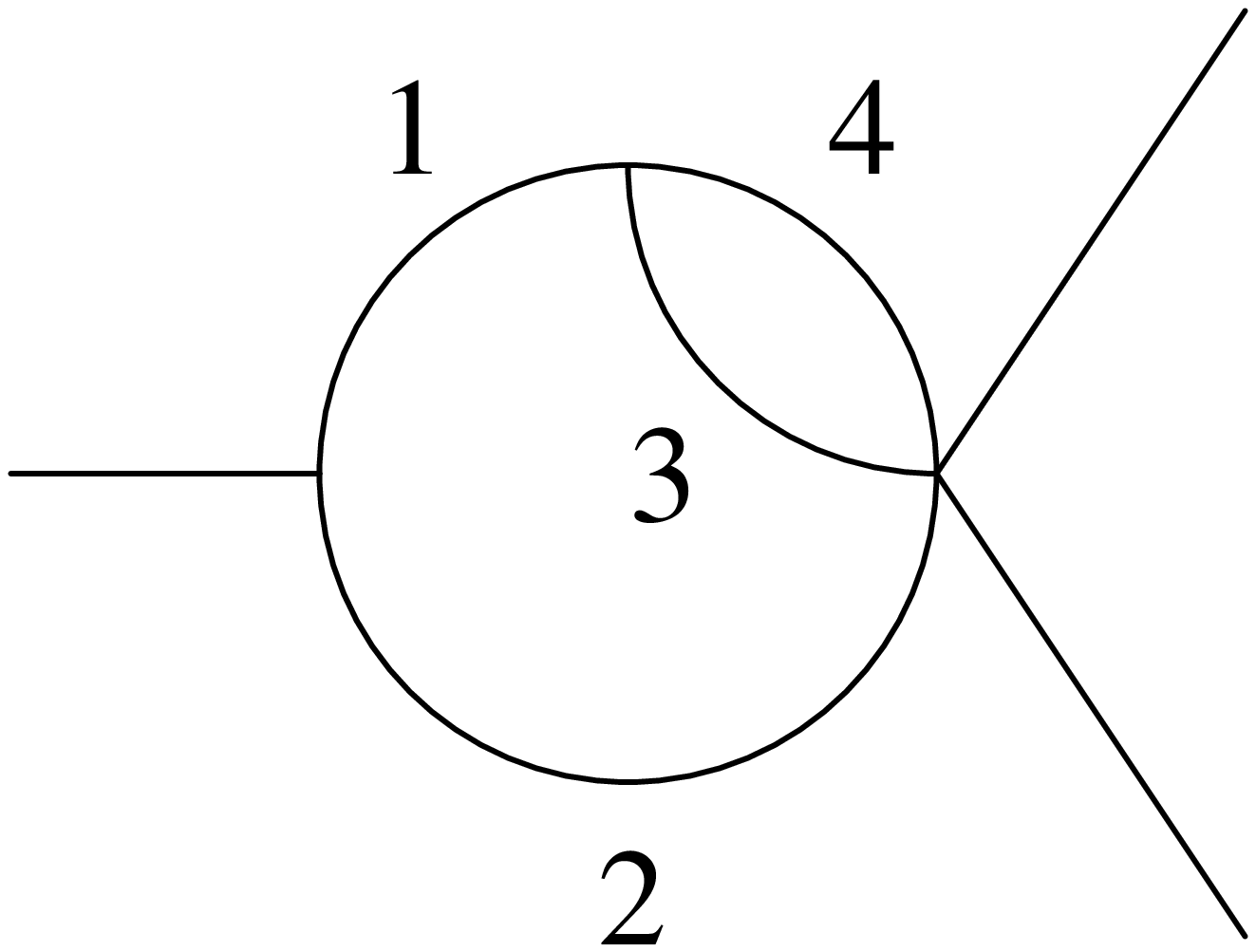, scale=0.2}}\kern96pt
\end{eqnarray}

\section{Integration of the master integrals}
After having reduced the numerator of the integrand to factors
$(k_0\mp k_1)^\alpha$ in case of the planar two-loop topology and
$(k_0\mp k_1)^\alpha(l_0\mp l_1)^\beta$ in case of the rotated reduced planar
two-loop topologies, we can start to integrate this set of master integrals
${\cal T}^0_\alpha,{\cal T}^x_{\alpha,\beta}$. Before we do so, we have to
consider the occurrence of ultraviolet (UV) divergences and their treatment in
terms of appropriate subtractions. As for the treatment of infrared (IR)
divergences we refer to Ref.~\cite{Fleischer:1997bq} for the scalar case. In
the more general case, methods taken from Ref.~\cite{Harlander:1998dq} can be
applied. However, in this paper we consider only IR-finite examples.

\subsection{UV-divergences and subtraction procedure}
The UV-divergences can be subdivided into three classes: divergences with
respect to the two loop momenta and global divergences. It is quite obvious
that if we have $d_k$ propagator factors depending only on $k$, $d_l$
propagator factors depending only $l$, $d_{kl}$ propagator factors depending
both on $k$ and $l$, and a power $k^{n_k}l^{n_l}$ in the numerator, the
corresponding degrees of divergence in $D$ space-time dimensions are given by
\begin{eqnarray}
\omega_k&=&2(d_k+d_{kl})-n_k-D,\nonumber\\
\omega_l&=&2(d_l+d_{kl})-n_l-D,\nonumber\\
\omega&=&2(d_k+d_l+d_{kl})-(n_k+n_l)-2D.
\end{eqnarray}
The two-loop integral is divergent in $D$ dimensions, if at least one of the
degrees $\omega_k$, $\omega_l$, or $\omega$ is zero or negative. If the degree
of divergence is zero, this divergence is called logarithmic divergence.

In order to calculate the integral, we first have to regularize it. We use
dimensional regularization and write $D=4-2\eps$. After that, integrals can
be split off into a divergent and a convergent part. While the convergent part
can be calculated for $\eps=0$, i.e.\ for $D=4$ space-time dimensions, the
integrand of the divergent part is simpler but has the same UV-behaviour.
In an appropriate subtraction procedure, therefore, we subtract and add an
integrand which is simple enough to be integrated analytically (at least the
singular part) but has the same singularities as the integrand itself in order
to cancel the singularities. This subtraction procedure is found and will be
formulated multiplicatively~\cite{Kreimer:1993tx,Frink:2000}. The subtracted
integrand contains the subtraction factors
\begin{equation}\label{subkl}
{\cal K}^{(j_k)}=\prod_{r=1}^{j_k}\left(1-\frac{D_k}{D_{k,r}}\right),\qquad
{\cal L}^{(j_l)}=\prod_{r=1}^{j_l}\left(1-\frac{D_l}{D_{l,r}}\right)
\end{equation}
with
\begin{eqnarray}
D_k=\prod_{i=1}^{d_k}\left((k+p_{k,i})^2-m_{k,i}^2+i\eta\right),&&
D_{k,r}=\prod_{i=1}^{d_k}\left((k+\kappa_rp_{k,i})^2-m_{k,i,r}^2+i\eta\right)
  \nonumber\\
D_l=\prod_{i=1}^{d_l}\left((l+p_{l,i})^2-m_{l,i}^2+i\eta\right),&&
D_{l,r}=\prod_{i=1}^{d_l}\left((l+\lambda_rp_{l,i})^2-m_{l,i,r}^2+i\eta\right)
\end{eqnarray}
where $p_{k,i},p_{l,i}\in\{\pm p_1,\pm p_2\}$,
$m_{k,i}\in\{m_1,m_2\}$, $m_{l,i}\in\{m_4,m_5\}$, and $m_{k,i,r},m_{l,i,r}$
are {\em subtraction masses\/} which have to be introduced
artificially.\footnote{In the following we will simplify the indices according
to the special cases.} The degrees of divergence for this subtracted integrand
increase by $j_k$ and $j_l$, resp.\ and thus may lead to integrals without
singularities for arbitrary values for $\kappa_r$, $\lambda_r$, $m_{k,i,r}$,
and $m_{l,i,r}$. Finally, the limit $\kappa_r,\lambda_r\to 0$ can be
performed, leaving the subtracted integral convergent and enabling the
semi-analytical calculation. Since the subtraction terms always contain a
two-point subloop, the singular part can be calculated analytically. For the
non-singular part one can use $\eps=0$~\cite{Bauberger:1994by,
Post:1997,Post:1996gg}. In the end the sum of subtracted integral and
subtraction terms have to be independent of the subtraction masses. In the
following we will deal with the subtracted integrals only.

\subsection{Subtraction for the planar topology\label{subnonred}}
The reduction procedure explained in the previous section leads to the
integrals
\begin{equation}\label{inttal0}
{\cal T}^0_\alpha=\int\frac{(k_0\mp k_1)^\alpha}{P_1P_2P_3P_4P_5P_6}d^Dk\,d^Dl.
\end{equation}
The degrees of divergence can be calculated in four space-time dimensions,
\begin{equation}
\omega_k=2-\alpha,\qquad
\omega_l=4,\qquad
\omega=4-\alpha.
\end{equation}
There are no subdivergences in $l$, i.e.\ divergences caused by the
integration over the loop momentum $l$. However, for $\alpha\ge 2$ we have to
subtract subdivergences in $k$. We choose $j_k=\alpha-1$ and multiply the
integrand in Eq.~(\ref{inttal0}) by
\begin{equation}\label{subk12}
{\cal K}^{(j_k)}=\prod_{i=1}^{j_k}\left(1-\frac{P_1P_2}{P_{1,i}P_{2,i}}\right)
\end{equation}
where $P_1=(k+p_1)^2-m_1^2+i\eta$ and $P_2=(k-p_2)^2-m_2^2+i\eta$ are given in
Eq.~(\ref{propfac}). The modified propagator factors read
\begin{equation}
P_{1,i}=(k+\kappa_ip_1)^2-m_{1,i}^2+i\eta,\qquad
P_{2,i}=(k-\kappa_ip_2)^2-m_{2,i}^2+i\eta.
\end{equation}
Note that each of the subtraction factors in Eq.~(\ref{subk12}) improves the
degree of divergence at least by $1$ because by power counting the denominator
is given by $k^4$ and terms of lower order in $k$ while the numerator starts
with $k^3$. The masses $m_{1,i},m_{2,i}$ of the subtraction can be chosen
arbitrarily. However, in order to avoid the introduction of spurious
IR-singularities and squared propagator factors in the denominator, the
subtraction masses are usually chosen to be non-zero and different from the
physical ones and from each other~\cite{Frink:2000}. Furthermore, in order not
to introduce new thresholds, they should be larger than the masses of the
decaying particles. On the other hand, they should not be too large because 
of the possible occurence of numerical instabilities.

\subsection{Integration for the planar topology}
After the linearization $k_0=k'_0+k_1$, in the P/O-space representation the
propagator factors are given by
\begin{eqnarray}
P_1&=&2(k'_0+E_1-q_z)k_1+(k'_0+E_1)^2-q_z^2-s-m_1^2+i\eta\nonumber\\
P_2&=&2(k'_0-E_2-q_z)k_1+(k'_0-E_2)^2-q_z^2-s-m_2^2+i\eta\nonumber\\
P_{1,i}&=&2\left(k'_0-\kappa_i(E_1-q_z)\right)k_1+(k'_0+\kappa_iE_1)^2
  -\kappa_i^2q_z^2-s-m_{1,i}^2+i\eta\nonumber\\
P_{2,i}&=&2\left(k'_0-\kappa_i(E_2+q_z)\right)k_1+(k'_0-\kappa_iE_2)^2
  -\kappa_i^2q_z^2-s-m_{2,i}^2+i\eta.
\end{eqnarray}
In case of the (non-reduced) planar diagram where both linearizations are
possible, we select the linearizations $k_0=k'_0+k_1$ and $l_0=l'_0+l_1$ for
convenience. The propagator factors which are linear expressions in $k_1$ can
be written in closed form expressions,
\begin{eqnarray}
P_1=\theta_1k_1+\xi_1-s+i\eta,&&
P_{1,i}=\theta_{1,i}k_1+\xi_{1,i}-s+i\eta\nonumber\\
P_2=\theta_2k_1+\xi_2-s+i\eta,&&
P_{2,i}=\theta_{2,i}k_1+\xi_{2,i}-s+i\eta.
\end{eqnarray}
Because the parameters $\kappa_i$ can take arbitrary values, the denominator
of ${\cal K}^{(j_k)}$ in Eq.~(\ref{subkl}) together with the factors $P_1$ and
$P_2$ from the integrand consists of different linear factors.

Integrating over $k_1$, the integration path $[-\infty,+\infty]$ can be closed
in the upper or lower complex half plane, and Cauchy's theorem can be applied.
The decision for one of these paths depends on the sign of the linearization
and on the sign of $(k'_0+l'_0)$ (cf.\ Eq.~(\ref{k1l1cut}) and its
discussion). The poles are given by the (different) zeros of the propagator
factors present in the calculation, the residues are given by
\begin{eqnarray}
&&\mbox{Res\ }\left[\frac1{P_1P_2}\prod_{i=1}^{j_k}
  \left(1-\frac{P_1P_2}{P_{1,i}P_{2,i}}\right);P_1=0\right]
  =\frac1{P_2}\Bigg|_{P_1=0},\nonumber\\
&&\mbox{Res\ }\left[\frac1{P_1P_2}\prod_{i=1}^{j_k}
  \left(1-\frac{P_1P_2}{P_{1,i}P_{2,i}}\right);P_2=0\right]
  =\frac1{P_1}\Bigg|_{P_2=0},\nonumber\\
&&\mbox{Res\ }\left[\frac1{P_1P_2}\prod_{i=1}^{j_k}
  \left(1-\frac{P_1P_2}{P_{1,i}P_{2,i}}\right);P_{1,j}=0\right]
  =-\frac1{P_{2,j}}\prod_{\scriptstyle i=1\atop i\ne j}^{j_k}
  \left(1-\frac{P_1P_2}{P_{1,i}P_{2,i}}\right)\Bigg|_{P_{1,j}=0},\nonumber\\
&&\mbox{Res\ }\left[\frac1{P_1P_2}\prod_{i=1}^{j_k}
  \left(1-\frac{P_1P_2}{P_{1,i}P_{2,i}}\right);P_{2,j}=0\right]
  =-\frac1{P_{1,j}}\prod_{\scriptstyle i=1\atop i\ne j}^{j_k}
  \left(1-\frac{P_1P_2}{P_{1,i}P_{2,i}}\right)\Bigg|_{P_{2,j}=0}.
\end{eqnarray}
The condition $P_1=0$ results in $k_1=(s-\xi_1-i\eta)/\theta_1=:k_{1(1)}$. 
One obtains
\begin{equation}
\frac1{P_2}\Bigg|_{P_1=0}
  =\frac{\theta_1}{(\theta_2-\theta_1)(s-s_{2,1}-i\eta)}
  =\frac{{\cal N}^k_1}{s-s_{2,1}-i\eta},\qquad
  s_{2,1}:=\frac{\theta_2\xi_1-\theta_1\xi_2}{\theta_2-\theta_1}.
\end{equation}
In the same way $P_2=0$ results in $k_1=(s-\xi_2-i\eta)/\theta_2=:k_{1(2)}$.
One obtains
\begin{equation}
\frac1{P_1}\Bigg|_{P_2=0}
  =\frac{\theta_2}{(\theta_1-\theta_2)(s-s_{1,2}-i\eta)}
  =\frac{{\cal N}^k_2}{s-s_{1,2}-i\eta},\qquad
  s_{1,2}:=\frac{\theta_1\xi_2-\theta_2\xi_1}{\theta_1-\theta_2}.
\end{equation}
Note that $s_{1,2}=s_{2,1}$. For the other two types of residues, however, a
simplification appears after inserting the poles. Because of the fact that
for the semi-analytical calculation of the convergent part of the integral we
use the limit $\kappa_i\to 0$ for all $i$, we obtain
\begin{equation}
\theta_{1,i},\theta_{2,i}\to 2k_0=:\theta_0\quad\mbox{for\ }i=1,2,\ldots,j_k.
\end{equation}
In this limit, and on the pole $P_{1,j}=0$ given by
$k_1=(s-\xi_{1,j}-i\eta)/\theta_0=:k_{1(1,j)}$, the different propagator
factors read
\begin{eqnarray}
P_{1,i}\Big|_{P_{1,j}=0}&=&\xi_{1,i}-\xi_{1,j},\nonumber\\
P_{2,i}\Big|_{P_{1,j}=0}&=&\xi_{2,i}-\xi_{1,j},\nonumber\\
P_1\Big|_{P_{1,j}=0}&=&\frac1{\theta_0}\left((\theta_1-\theta_0)(s-i\eta)
  +(\xi_1\theta_0-\theta_1\xi_{1,j})\right),\nonumber\\
P_2\Big|_{P_{1,j}=0}&=&\frac1{\theta_0}\left((\theta_2-\theta_0)(s-i\eta)
  +(\xi_2\theta_0-\theta_2\xi_{1,j})\right).
\end{eqnarray}
Obviously, the dependence on $s$ appears only in the numerator. We can write
\begin{equation}\label{n1jadef}
-\frac1{P_{2,j}}\prod_{\scriptstyle i=1\atop i\ne j}^{j_k}
  \left(1-\frac{P_1P_2}{P_{1,i}P_{2,i}}\right)\Bigg|_{P_{1,j}=0}
  =\sum_{a=0}^{2j_k-2}{\cal N}^k_{1,j,a}s^a,
\end{equation}
and similarly
\begin{equation}\label{n2jadef}
-\frac1{P_{1,j}}\prod_{\scriptstyle i=1\atop i\ne j}^{j_k}
  \left(1-\frac{P_1P_2}{P_{1,i}P_{2,i}}\right)\Bigg|_{P_{2,j}=0}
  =\sum_{a=0}^{2j_k-2}{\cal N}^k_{2,j,a}s^a.
\end{equation}
The different poles are given by
\begin{eqnarray}
k_{1(1)}=\frac{s-\xi_1-i\eta}{\theta_1}&&
k_{1(2)}=\frac{s-\xi_2-i\eta}{\theta_2}\nonumber\\
k_{1(1,i)}=\frac{s-\xi_{1,i}-i\eta}{\theta_0}&&
k_{2(2,i)}=\frac{s-\xi_{2,i}-i\eta}{\theta_0}\qquad
  \mbox{for\ }i=1,2,\ldots,j_k.
\end{eqnarray}
For $\theta_i<0$ ($i=0,1,2$) the corresponding pole can be found in the upper
complex half plane. For the linearization $k_0=k'_0+k_1$ and $k'_0+l'_0>0$,
the path has to be closed in the upper half plane. In this case we obtain a
non-vanishing residue. For $k'_0+l'_0<0$, however, the residue occurs only in
case of $\theta_i>0$ and has the opposite sign. Using
\begin{equation}
{\cal B}^k_n=2\pi i\left[\theta(k'_0+l'_0)\theta(-\theta_n)
  -\theta(-(k'_0+l'_0))\theta(\theta_n)\right],\qquad n=0,1,2
\end{equation}
where $\theta(x)$ is the step function, we can integrate over $k_1$ to obtain
\begin{eqnarray}\label{residuek}
\lefteqn{\int_{-\infty}^{+\infty}\frac{dk_1}{P_1P_2}
  \left(\prod_{i=1}^{j_k}\left(1-\frac{P_1P_2}{P_{1,i}P_{2,i}}\right)\right)
  f(k_1,l_1)}\nonumber\\
  &=&\sum_{n=1}^2\left(\frac{{\cal B}^k_n{\cal N}^k_n}{s-s_{1,2}-i\eta}
  f(k_{1(n)},l_1)
  +\sum_{i=1}^{j_k}\sum_{a=0}^{2j_k-2}{\cal B}^k_0{\cal N}^k_{n,i,a}
  s^af(k_{1(n,i)},l_1)\right)\qquad
\end{eqnarray}
where $f(k_1,l_1)$ is given e.g.\ by the inverse of
$\sqrts{(A(k_1,l_1)+i\eta)^2-4st}$ (cf.\ Eq.~(\ref{sqrts})).

For the integration over $l_1$ we have to consider the propagator factors
$P_4$, $P_5$, and $P_6$. Because the degree of divergence $\omega_l$ is
positive, no subtractions are needed. After the linearization $l_0=l'_0+l_1$
we can write
\begin{eqnarray}
P_4&=&2(l'_0-E_1+q_z)l_1+(l'_0-E_1)^2-q_z^2-t-m_4^2+i\eta,\nonumber\\
P_5&=&2(l'_0+E_2+q_z)l_1+(l'_0+E_2)^2-q_z^2-t-m_5^2+i\eta,\nonumber\\
P_6&=&2l'_0l_1+l_0^{\prime2}-t-m_6^2+i\eta
\end{eqnarray}
which again can be cast into the closed form expression 
\begin{equation}
P_i=\phi_il_1+\psi_i-t+i\eta,\qquad i=4,5,6.
\end{equation}
The procedure which was explained for the integration over $k_1$ works
accordingly. The poles of
\begin{equation}
\frac1{P_4P_5P_6}=\left(\prod_{j=4}^6(\phi_jl_1+\psi_j-t+i\eta)\right)^{-1}
\end{equation}
are found at
\begin{equation}
l_{1(m)}=\frac{t-\psi_m-i\eta}{\phi_m}\qquad\mbox{for\ }m=4,5,6.
\end{equation}
From Eq.~(\ref{k1l1cut}) we read off that the cut is located in the same
complex half plane as in case of the integration over $k_1$. Therefore, the
occurrence and the sign of the residue is determined by
\begin{equation}
{\cal B}^l_m=2\pi i\left[\theta(k'_0+l'_0)\theta(-\phi_m)
  -\theta(-(k'_0+l'_0))\theta(\phi_m)\right],\qquad m=4,5,6.
\end{equation}
Using Cauchy's theorem, we obtain
\begin{eqnarray}\label{residuel0}
\lefteqn{\int_{-\infty}^{+\infty}\frac{dl_1}{P_4P_5P_6}f(k_1,l_1)}
  \nonumber\\
  &=&\sum_{m=4}^6\left(\prod_{\scriptstyle j=4\atop j\ne m}^6
  (\phi_jl_{1(m)}+\psi_j-t+i\eta)\right)^{-1}{\cal B}^l_m
  f(k_1,l_{1(m)})\\
  &=&\sum_{m=4}^6\left(\prod_{\scriptstyle j=4\atop j\ne m}^6
  \Big(t(\phi_j-\phi_m)+(\psi_j\phi_m-\psi_m\phi_j)
  -i\eta(\phi_j-\phi_m)\Big)\right)^{-1}\phi_m^2{\cal B}^l_m
  f(k_1,l_{1(m)}).\nonumber
\end{eqnarray}
Of course, the same is valid if $k_1$ is replaced by $k_{1(n)}$ or
$k_{1(n,i)}$. Usually, $\phi_j$ and $\phi_m$ are different for $j\ne m$. In
this case we can write
\begin{equation}
t(\phi_j-\phi_m)-(\phi_j\psi_m-\phi_m\psi_j)-i\eta(\phi_j-\phi_m)
  =(\phi_j-\phi_m)(t-t_{j,m}-i\eta)
\end{equation}
with
\begin{equation}
t_{j,m}=\frac{\phi_j\psi_m-\phi_m\psi_j}{\phi_j-\phi_m}.
\end{equation}
We perform a partial fraction decomposition to obtain
\begin{equation}
\int_{-\infty}^{+\infty}\frac{dl_1}{P_4P_5P_6}f(k_1,l_1)
  =\sum_{m=4}^5\sum_{j=m+1}^6
  \frac{{\cal B}^l_m{\cal N}^l_{m,j}}{t-t_{m,j}-i\eta}f(k_1,l_{1(m)})
\end{equation}
where ${\cal N}^l_{m,j}$ are the corresponding coefficients including
$\phi_m^2$. A special situation occurs if $p_1^2=0$. Because of $q_z=E_1$, in
this case we have $\phi_4=\phi_6$ and the corresponding denominator factor
contributes in the form
$(\psi_4-\psi_6)\phi_6$ only. However, if $p_2^2=0$, no special case has to be
taken.\footnote{For the other linearization $l_0=l'_0-l_1$ the situation is the
opposite.}

The integrations over $k_1$ and $l_1$ and the integration over $z$ can be
combined in the closed form expression for the convergent part
${\cal V}^0_\alpha$ of the master integral ${\cal T}^0_\alpha$ in case of
$\kappa_i\to 0$,
\begin{eqnarray}\label{k1l1sint0}
\lefteqn{{\cal V}^0_\alpha
  \ =\ \int\frac{(k_0-k_1)^\alpha}{P_1P_2P_3P_4P_5P_6}
  {\cal K}^{(j_k)}d^4k\,d^4l\ =\ \pi^2\sum_{n=1}^2
  \sum_{m=4}^5\sum_{j=m+1}^6\int_0^\infty k_0^{\prime\alpha}dk'_0\,dl'_0
  \int_0^\infty\frac{{\cal N}^l_{m,j}dt}{t-t_{m,j}-i\eta}}
  \nonumber\\
  &&\qquad\times\int_0^\infty\left(\frac{{\cal B}^{kl}_{n,m}{\cal N}^k_n
  ds}{(s-s_{1,2}-i\eta)\ \sqrts{(A_{n,m}+i\eta)^2-4st}}
  +\sum_{i=1}^{j_k}\sum_{a=0}^{2j_k-2}\frac{{\cal B}^{kl}_{0,m}
  {\cal N}^k_{n,i,a}s^ads}{\sqrts{(A_{n,i,m}+i\eta)^2-4st}}\right)\nonumber\\
\end{eqnarray}
where
\begin{equation}\label{calBkl0}
{\cal B}^{kl}_{n,m}={\cal B}^k_n{\cal B}^l_m
  =-4\pi^2\left[\theta(k'_0+l'_0)\theta(-\theta_n)\theta(-\phi_m)
  +\theta(-(k'_0+l'_0))\theta(\theta_n)\theta(\phi_m)\right]
\end{equation}
and $A_{n,m},A_{n,i,m}$ are the coefficients $A$ in Eq.~(\ref{defAB}) with
$k_1$ replaced by $k_{1(n)},k_{1(n,i)}$ and $l_1$ replaced by $l_{1(m)}$,
i.e.\ $A_{n,m}=A(k_{1(n)},l_{1(m)})$ and $A_{n,i,m}=A(k_{1(n,i)},l_{1(m)})$.

\subsection{Subtraction for the rotated reduced planar topologies}
For the rotated reduced planar topologies we are left with the integrals
\begin{equation}
{\cal T}^{9-x}_{\alpha\beta}=\int\frac{(k_0\mp k_1)^\alpha(l_0\mp l_1)^\beta}
  {P_1P_2P_3P_xP_6}d^Dk\,d^Dl
\end{equation}
where in case of the topology ${\cal T}^4$ ($x=5$) the upper sign, in case of
the topology ${\cal T}^5$ ($x=4$) the lower sign is valid. The degrees of
divergence can be calculated to be
\begin{equation}
\omega_k=2-\alpha,\qquad
\omega_l=2-\beta,\qquad
\omega=2-\alpha-\beta.
\end{equation}
For $\beta\ge 2$ the integration over $l$ becomes divergent and a subtraction
has to be performed in addition to the subtraction in $k$. The subtraction
factor is given by
\begin{equation}
{\cal L}^{(j_l)}=\prod_{j=1}^{j_l}\left(1-\frac{P_x}{P_{x,j}}\right),\qquad
  x=4,5
\end{equation}
where in addition to $P_4=(l-p_1)^2-m_4^2+i\eta$ and
$P_5=(l+p_2)^2-m_5^2+i\eta$ we have
\begin{equation}
P_{4,j}=(l-\lambda_jp_1)^2-m_{4,j}^2+i\eta,\qquad
P_{5,j}=(l+\lambda_jp_2)^2-m_{5,j}^2+i\eta.
\end{equation}
The same limitations as they were mentioned at the end of Sec.~\ref{subnonred}
for the subtraction masses $m_{1,i}$ and $m_{2,i}$ apply for the masses
$m_{4,j}$ and $m_{5,j}$ as well.

\subsection{Integration for the rotated reduced planar topologies}
The integration over $k_1$ can be done in the same way as for the original
(non-reduced) planar topology. However, special care has to be taken
concerning the different linearization in case of the topology ${\cal T}^5$.
For the topology ${\cal T}^5$ we have $\theta_0=-2k'_0$ and the step function
term ${\cal B}^k_n$ has to be replaced by
\begin{equation}
{\cal B}^k_n=2\pi i\left[\theta(-(k'_0+l'_0))\theta(-\theta_n)
  -\theta(k'_0+l'_0)\theta(\theta_n)\right].
\end{equation}

For the integration over $l_1$ we have to perform a subtraction. After the
linearization $l_0=l'_0\pm k_1$ the propagator factors in the
P/O-space representation are given by
\begin{eqnarray}
P_5&=&2(l'_0+(E_2+q_z))l_1+(l'_0+E_2)^2-q_z^2-t-m_5^2+i\eta\nonumber\\
P_{5,j}&=&2(l'_0+\lambda_j(E_2+q_z))l_1+(l'_0+\lambda_jE_2)^2
  -\lambda_j^2q_z^2-t-m_5^2+i\eta\nonumber\\
P_6&=&2l'_0l_1+l_0^{\prime2}-t-m_6^2+i\eta\qquad
  \mbox{for the topology\ }{\cal T}_4,\\
P_4&=&-2(l'_0-(E_1+q_z))l_1+(l'_0-E_1)^2-q_z^2-t-m_4^2+i\eta\nonumber\\
P_{4,j}&=&-2(l'_0-\lambda_j(E_2+q_z))l_1+(l'_0-\lambda_jE_2)^2
  -\lambda_j^2q_z^2-t-m_4^2+i\eta\nonumber\\
P_6&=&-2l'_0l_1+l_0^{\prime2}-t-m_6^2+i\eta\qquad
  \mbox{for the topology\ }{\cal T}_5.
\end{eqnarray}
These propagator factors can be written as
\begin{eqnarray}
P_x=\phi_xl_1+\psi_x-t+i\eta,&&
P_{x,j}=\phi_{x,j}l_1+\psi_{x,j}-t+i\eta\qquad x=4,5\nonumber\\
P_6=\phi_6l_1+\psi_6-t+i\eta.&&
\end{eqnarray}
In the limit $\lambda_i\to 0$, which will be considered in the following, we
see that
\begin{equation}
\phi_{x,j}\to\phi_6=\pm 2l'_0.
\end{equation}
In case of the linearization $l_0=l'_0+l_1$ the integration path
$[-\infty,+\infty]$ over $l_1$ can be closed in the upper complex half plane
for $(k'_0+l'_0)>0$ and in the lower complex half plane for $(k'_0+l'_0)<0$.
For the other linearization $k_0=k'_0-k_1$ the situation is just the opposite.
Because the poles are given by
\begin{equation}
l_{1(x)}=\frac{t-\psi_x-i\eta}{\phi_x},\qquad
l_{1(x,j)}=\frac{t-\psi_{x,j}-i\eta}{\phi_6},\qquad
l_{1(6)}=\frac{t-\psi_6-i\eta}{\phi_6},
\end{equation}
the integration ranges in $t$ are constrained by
\begin{equation}
{\cal B}^l_m=\cases{2\pi i\left[\theta(-(k'_0+l'_0))\theta(-\phi_m)
  -\theta(k'_0+l'_0)\theta(\phi_m)\right]&for $x=4$ (topology ${\cal T}^5$)\cr
  2\pi i\left[\theta(k'_0+l'_0)\theta(-\phi_m)
  -\theta(-(k'_0+l'_0))\theta(\phi_m)\right]&for $x=5$
  (topology ${\cal T}^4$).\cr}
\end{equation}
In order to use Cauchy's theorem, we calculate the residues
\begin{eqnarray}
&&\mbox{Res\ }\left[\frac1{P_xP_6}\prod_{i=1}^{j_l}\left(1-\frac{P_x}{P_{x,i}}
  \right);P_x=0\right]=\frac1{P_6}\Bigg|_{P_x=0},\nonumber\\[7pt]
&&\mbox{Res\ }\left[\frac1{P_xP_6}\prod_{i=1}^{j_l}\left(1-\frac{P_x}{P_{x,i}}
  \right);P_{x,j}=0\right]=\frac1{P_6}\prod_{\scriptstyle i=1\atop i\ne j}
  \left(1-\frac{P_x}{P_{x,i}}\right)\Bigg|_{P_{x,j}=0},\nonumber\\
&&\mbox{Res\ }\left[\frac1{P_xP_6}\prod_{i=1}^{j_l}\left(1-\frac{P_x}{P_{x,i}}
  \right);P_6=0\right]=\frac1{P_x}\prod_{i=1}^{j_l}\left(1-\frac{P_x}{P_{x,i}}
  \right)\Bigg|_{P_6=0}.
\end{eqnarray}
For the first residue we obtain
\begin{equation}
\frac1{P_6}\bigg|_{P_x=0}=\frac{\phi_x}{(\phi_6-\phi_x)(t-t_{6,x}-i\eta)}
  =:\frac{{\cal N}^l_x}{t-t_{6,x}-i\eta},\qquad
  t_{6,x}=\frac{\phi_6\psi_x-\phi_x\psi_6}{\phi_6-\phi_x}.
\end{equation}
For the second one we first calculate
\begin{eqnarray}
P_6\Big|_{P_{x,j}=0}&=&\psi_6-\psi_{x,j},\nonumber\\
P_x\Big|_{P_{x,j}=0}&=&\frac1{\phi_6}\left((\phi_x-\phi_6)(t-i\eta)
  -(\phi_x\psi_{x,j}-\phi_6\psi_x)\right),\nonumber\\
P_{x,i}\Big|_{p_{x,j}=0}&=&\psi_{x,i}-\psi_{x,j}.
\end{eqnarray}
Therefore, the denominator of this residuum is independent of $t$ while the
numerator is a power series up to the power $t^{j_l-1}$. We define the
coefficients ${\cal N}^l_{x,j,a}$ by
\begin{equation}
\frac1{P_6}\prod_{\scriptstyle i=1\atop i\ne j}
  \left(1-\frac{P_x}{P_{x,i}}\right)\Bigg|_{P_{x,j}=0}
  =:\sum_{a=0}^{j_l-1}{\cal N}^l_{x,j,a}t^a.
\end{equation}
The last residue appears to be a combination of a pole in $t$ and a power
series. However, we can separate these two parts by adding and subtracting an
appropriate term. This term is given by
\begin{equation}
\frac1{P_x}\Bigg|_{P_6=0}=\frac{\phi_6}{(\phi_x-\phi_6)(t-t_{x,6}-i\eta)}
  =:\frac{{\cal N}^l_6}{t-t_{x,6}-i\eta},\qquad
  t_{x,6}=\frac{\phi_x\psi_6-\phi_6\psi_x}{\phi_x-\phi_6}.
\end{equation}
If we subtract this term from the result for the last residue, we obtain
\begin{equation}
\frac1{P_x}\prod_{i=1}^{j_l}\left(1-\frac{P_x}{P_{x,i}}\right)-\frac1{P_x}
  =\frac1{P_x}\left\{1-\sum_{i=1}^{j_l}\frac{P_x}{P_{x,i}}+O(P_x^2)\right\}
  -\frac1{P_x}=-\sum_{i=1}^{j_l}\frac1{P_{x,i}}+O(P_x).
\end{equation}
The difference no longer has a factor $P_x$ in the denominator. Instead,
we obtain again a power series up to the power $t^{j_l-1}$. We define
the coefficients ${\cal N}^l_{6,j,a}$ by
\begin{equation}
\frac1{P_x}\prod_{i=1}^{j_l}\left(1-\frac{P_x}{P_{x,i}}\right)\bigg|_{P_6=0}
  -\frac1{P_x}\bigg|_{P_6=0}=:\sum_{a=0}^{j_l-1}{\cal N}^l_{6,j,a}t^a.
\end{equation}
Having calculated the residues, we can perform the integration over $l_1$ to
obtain
\begin{eqnarray}
\lefteqn{\int_{-\infty}^{+\infty}\frac{dl_1}{P_xP_6}\prod_{i=1}^{j_l}
  \left(1-\frac{P_x}{P_{x,i}}\right)f(k_1,l_1)}\nonumber\\
  &=&\frac{{\cal B}^l_x{\cal N}^l_x}{t-t_{6,x}-i\eta}f(k_1,l_{1(x)})
  +\frac{{\cal B}^l_6{\cal N}^l_6}{t-t_{x,6}-i\eta}f(k_1,l_{1(6)})
  \nonumber\\&&
  +\sum_{i=1}^{j_l}\sum_{a=0}^{j_l-1}\left({\cal B}^l_6{\cal N}^l_{x,i,a}t^a
  +{\cal B}^l_6{\cal N}^l_{6,i,a}t^a\right)f(k_1,l_{1(6,i)})\nonumber\\
  &=&\sum_{m=x,6}\left(\frac{{\cal B}^l_m{\cal N}^l_m}{t-t_{x,6}-i\eta}
  f(k_1,l_{1(m)})+\sum_{i=1}^{j_l}\sum_{a=0}^{j_l-1}{\cal B}^l_6
  {\cal N}^l_{m,i,a}t^af(k_1,l_{1(m,i)})\right).
\end{eqnarray}

The integrations over $k_1$ and $l_1$ and the integration over $z$ can be
combined in the closed form expression for the convergent part
${\cal V}^{9-x}_\alpha$ of the master integral ${\cal T}^{9-x}_\alpha$ in case
of $\kappa_i,\lambda_i\to 0$. One has
\begin{eqnarray}\label{k1l1sintx}
\lefteqn{{\cal V}^{9-x}_\alpha\ =\ \int\frac{(k_0\mp k_1)^\alpha
  (l_0\mp l_1)^\beta}{P_1P_2P_3P_xP_6}
  {\cal K}^{(j_k)}{\cal K}^{(j_l)}d^4k\,d^4l
  \ =\ \pi^2\int_0^\infty k_0^{\prime\alpha}dk'_0l_0^{\prime\beta}dl'_0
  \int_0^\infty ds\,dt}\nonumber\\&&\times\ \sum_{n=1}^2\sum_{m=x,6}
  \Bigg\{\frac{{\cal B}^{kl}_{n,m}{\cal N}^k_n{\cal N}^l_m}{(s-s_{1,2}-i\eta)
  (t-t_{x,6}-i\eta)\ \sqrts{(A_{n,m}+i\eta)^2-4st}}\nonumber\\&&\qquad\qquad
  +\sum_{i=1}^{j_k}\sum_{a=0}^{2j_k-2}\frac{{\cal B}^{kl}_{0,m}
  {\cal N}^k_{n,i,a}s^a{\cal N}^l_m}{(t-t_{x,6}-i\eta)\
  \sqrts{(A_{n,i,m}+i\eta)^2-4st}}\nonumber\\&&\qquad\qquad
  +\sum_{j=1}^{j_l}\sum_{b=0}^{j_l-1}\frac{{\cal B}^{kl}_{n,6}
  {\cal N}^k_n{\cal N}^l_{m,j,b}t^b}{(s-s_{1,2}-i\eta)\
  \sqrts{(A_{n,m,j}+i\eta)^2-4st}}\nonumber\\&&\qquad\qquad
  +\sum_{i=1}^{j_k}\sum_{j=1}^{j_l}\sum_{a=0}^{2j_k-2}\sum_{b=0}^{j_l-1}
  \frac{{\cal B}^{kl}_{0,6}{\cal N}^k_{n,i,a}s^a{\cal N}^l_{m,j,b}
  t^b}{\sqrts{(A_{n,i,m,j}+i\eta)^2-4st}}\Bigg\}.\kern96pt
\end{eqnarray}
For the topology ${\cal T}^4$ ($x=5$), ${\cal B}^{kl}_{n,m}$ is given by
Eq.~(\ref{calBkl0}), i.e.\ it is the same as for the non-reduced planar
topology. For the topology ${\cal T}^5$ ($x=4$), however, we obtain
\begin{equation}\label{calBkl5}
{\cal B}^{kl}_{n,m}={\cal B}^k_n{\cal B}^l_m
  =-4\pi^2\left[\theta(-(k'_0+l'_0))\theta(-\theta_n)\theta(-\phi_m)
  +\theta(k'_0+l'_0)\theta(\theta_n)\theta(\phi_m)\right].
\end{equation}
The coefficients in the square roots read
$A_{n,m}=A(k_{1(n)},l_{1(m)})$, $A_{n,i,m}=A(k_{1(n,i)},l_{1(m)})$,
$A_{n,m,j}=A(k_{1(n)},l_{1(m,j)})$, and
$A_{n,i,m,j}=A(k_{1(n,i)},l_{(m,j)})$.

\subsection{Integral basis for the orthogonal space quadrature}
The integrals over $s$ and $t$ resulting for both the non-reduced and the
rotated reduced planar topology are called basic integrals. They are of the
types
\begin{eqnarray}\label{basicint}
{\cal F}(r_s,r_t,r_0,s_0,t_0)&=&\int_0^\infty
  \frac{ds\,dt}{(s-s_0-i\eta)(t-t_0-i\eta)\
  \sqrts{(r_0-r_ss-r_tt)^2-4st}},\nonumber\\
{\cal F}^s_\alpha(r_s,r_t,r_0,t_0)&=&\int_0^\infty
  \frac{s^\alpha ds\,dt}{(t-t_0-i\eta)\
  \sqrts{(r_0-r_ss-r_tt)^2-4st}},\nonumber\\
{\cal F}^t_\beta(r_s,r_t,r_0,s_0)&=&\int_0^\infty
  \frac{t^\beta ds\,dt}{(s-s_0-i\eta)\
  \sqrts{(r_0-r_ss-r_tt)^2-4st}},\nonumber\\
{\cal F}^{st}_{\alpha,\beta}(r_s,r_t,r_0)&=&\int_0^\infty
  \frac{s^\alpha t^\beta ds\,dt}{\sqrts{(r_0-r_ss-r_tt)^2-4st}}.
\end{eqnarray}
All these integrals are divergent except for the first one. However, after the
subtraction they occur in sums in which the divergences vanish. While the
integral ${\cal F}(r_s,r_t,r_0,s_0,t_0)$ is calculated in
Refs.~\cite{Frink:1996ya,Frink:1996,Kilian:1996}, the other integrals are new.
The results can be found in Appendix~\ref{appa} in terms of the parameters
$r_s$, $r_t$, $r_0$, $s_0$, and $t_0$. In this paragraph we are dealing only
with the dependence on the parameters $r_s$, $r_t$, and $r_0$ and on the poles
found in the residue integration. In case of the integral
${\cal F}(r_s,r_t,r_0,s_0,t_0)$ we obtain
\begin{eqnarray}\label{r0stdef}
r_0&=&(k'_0+l'_0)^2\mp2\left(\frac{\xi_n+i\eta}{\theta_n}
  +\frac{\psi_m+i\eta}{\phi_m}\right)(k'_0+l'_0)-m_3^2-i\eta\ =:\ r_{n,m},
  \nonumber\\
r_s&=&1\mp2\frac{k'_0+l'_0}{\theta_n},\qquad
r_t\ =\ 1\mp2\frac{k'_0+l'_0}{\phi_m}
\end{eqnarray}
where the signs correspond to the two linearizations $l_0=l'_0\pm l_1$. Note
that $r_s,r_t>0$ and $r_sr_t>1$ in regions where the numerical integration has
to be done. For ${\cal F}^s_\alpha$ we have to replace $\theta_n$ by
$\theta_0$ and $\xi_n$ by $\xi_{n,i}$. For ${\cal F}^t_\beta$ we have to
replace $\phi_m$ by $\phi_6$ and $\psi_m$ by $\psi_{m,j}$. For
${\cal F}^{st}_{\alpha,\beta}$, finally, both replacements have to be
performed. Note that up to the choice for the parameters $s_0$, $t_0$, $r_s$
and $r_t$, ${\cal F}^s_\alpha$ and ${\cal F}^t_\beta$ are the same integrals.

The imaginary parts in the rational factors in the numerator are used to
separate real and imaginary parts according to the Sokhotsky--Plemely
relations
\begin{equation}
\lim_{\eta\to+0}\int_a^b\frac{f(x)dx}{x-x_0\pm i\eta}
  ={\rm P}\int_a^b\frac{f(x)dx}{x-x_0}\mp i\pi\int_a^b\delta(x-x_0)f(x)dx
\end{equation}
where ``P'' indicates the principal value integral.

\subsection{Analytic behaviour of the modified square root}
We have to consider different cases in order to analyze the analytic
behaviour of the square root occurring in the
integrand~\cite{Czarnecki:1994td,Frink:1996ya,Frink:1996}. 
The equation
\begin{equation}
R(s,t)=(r_0-r_ss-r_tt)^2-4st=0
\end{equation}
parametrizes an ellipse which separates the positive and negative values of
the radicand. Outside of the ellipse the radicand is positive, while for
points inside the ellipse it takes negative values. The ellipse touches the
axes at $s=r_0/r_s$ and $t=r_0/r_t$. For $r_0<0$, therefore, the ellipse is
located in the third quadrant. But because the integration is performed
in the first quadrant ($s\in[0,\infty]$, $t\in[0,\infty]$), the square root
will give real values and no imaginary part is produced. For $r_0>0$, however,
the ellipse moves into the integration region and imaginary parts enter the
calculations. 

\begin{figure}\begin{center}
\epsfig{figure=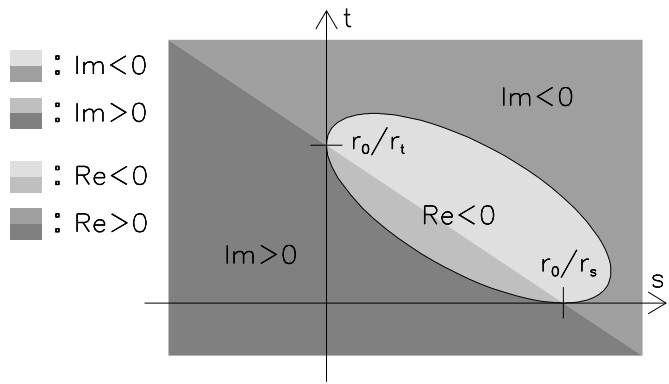, scale=0.6}
\caption{\label{ellipse}Regions in the $(s,t)$-plane with different real and
imaginary parts for the radicand $R(s,t)=(r_0-r_ss-r_tt)^2-4st$}
\end{center}\end{figure}

In order to evaluate the square root uniquely, we also consider the imaginary
part of the radicand, caused by the term $i\eta$. It turns out that the sign
of the imaginary part is determined by the sign of $r_0-r_ss-r_tt$, the zeros
of which are given by the straight line connecting the points $s=r_0/r_s$ and
$t=r_0/r_t$ on the axes where the ellipse touches the axes. Above this line
(as seen from the origin for $r_0>0$), the imaginary part is negative, below
this line it is positive. The situation is shown in Fig.~\ref{ellipse}.

We distinguish the following cases:
\begin{itemize}
\item outside the ellipse and below the line
($\real R(s,t)>0$, $\imag R(s,t)>0$) we have
\[\sqrts{(r_0-r_ss-r_tt)^2-4st}=+\sqrt{(r_0-r_ss-r_tt)^2-4st}\]
\item outside the ellipse and above the line
($\real R(s,t)>0$, $\imag R(s,t)<0$) we have
\[\sqrts{(r_0-r_ss-r_tt)^2-4st}=-\sqrt{(r_0-r_ss-r_tt)^2-4st}\]
\item inside the ellipse ($\real R(s,t)<0$) we have
\[\sqrts{(r_0-r_ss-r_tt)^2-4st}=i\sqrt{4st-(r_0-r_ss-r_tt)^2}\]
\end{itemize}

\section{Landau singularities and thresholds}
Multiloop integrals with many propagator factors are influenced by their
singularity structure. The pole structure can be analyzed by using Feynman
parametrization~\cite{Landau:1959fi,Bohm:2001}. Starting from
\begin{equation}
F(p_j,m_k)=\int\left(\prod_{l=1}^Ld^4k_l\right)f(p_j,k_l,m_k)
  \prod_{n=1}^N\frac{i}{A_n},\qquad A_n=q_n^2-m_n^2+i\eta
\end{equation}
with $E$ outer momenta $p_j$, $L$ loop momenta $k_l$, $N$ inner masses $m_n$,
and $q_n$ a linear combination of $p_j$ and $k_l$, we can apply the Feynman
parametrization
\begin{equation}
\frac1{A_1A_2\cdots A_N}=(N-1)!\int_0^1\frac{\delta(\lambda_1+\lambda_2+\ldots
  +\lambda_N-1)d\lambda_1d\lambda_2\cdots d\lambda_N}{(\lambda_1A_1
  +\lambda_2A_2+\ldots+\lambda_NA_N)^N}.
\end{equation}
By shifting the inner momenta $k_l$ to $k'_l$ by finite amounts (which are
determined by the outer momenta), the denominator can be rewritten as
\begin{equation}
{\cal D}=\sum_{n=1}^N\lambda_nA_n=\phi+K(k'_1,k'_2,\ldots,k'_L)
\end{equation}
where $\phi$ is independent of inner momenta and $K$ is a quadratic form in
the shifted momenta $k'_l$. It can then be shown~\cite{Landau:1959fi,Bohm:2001}
that the denominator vanishes, i.e.\ the integral has so-called Landau
singularities, if
\begin{eqnarray}
\lambda_n(q_n^2-m_n^2)&=&0\qquad\mbox{for all $n=1,\ldots N$
  (first Landau equation)}\\\mbox{and}\quad
\sum_{n=1}^N\lambda_nq_n&=&0\qquad\mbox{for each loop
  (second Landau equation)}.
\end{eqnarray}

\begin{figure}[ht]\begin{center}
\epsfig{figure=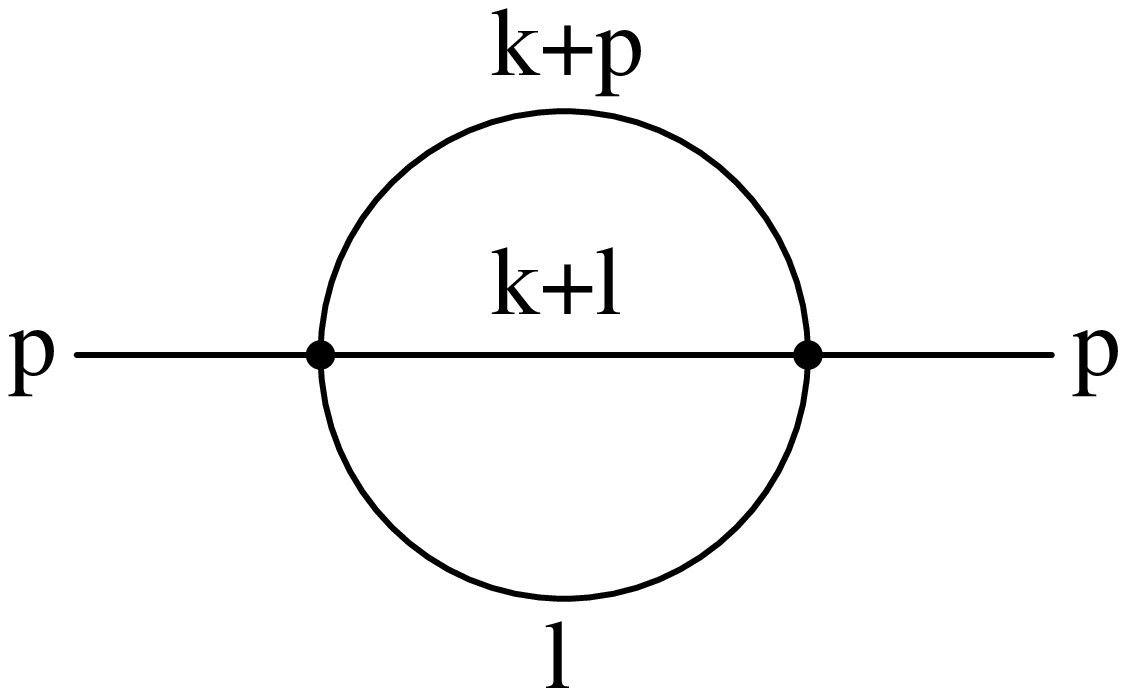, scale=0.3}\qquad\qquad
\raise2pt\hbox{\epsfig{figure=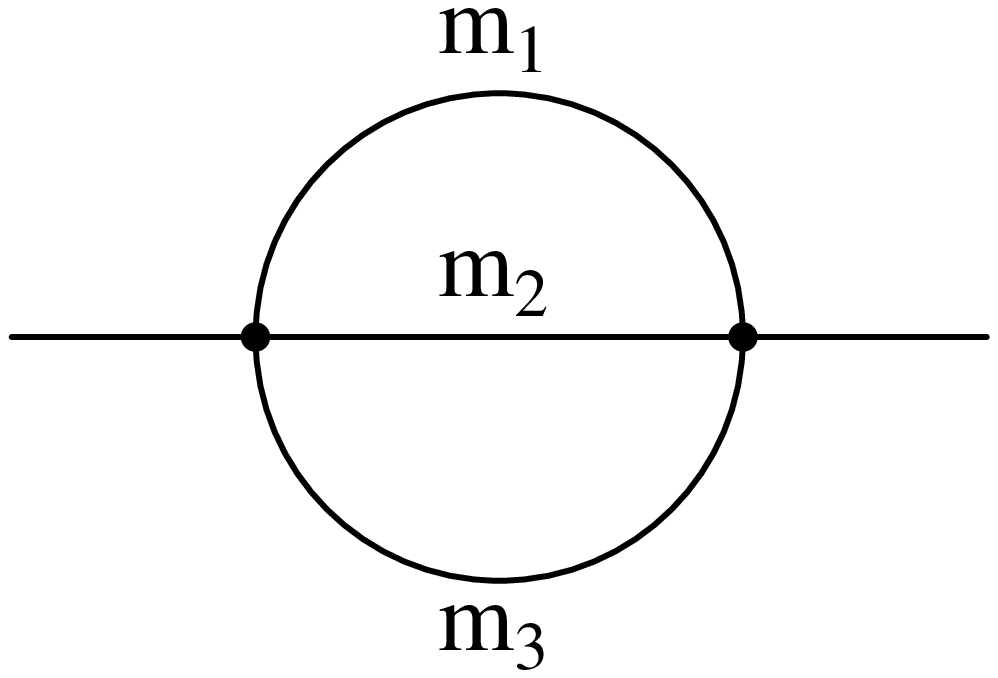, scale=0.3}}
\caption{\label{sunset}Sunset diagram with momenta (left hand
side) and masses (right hand side)}
\end{center}\end{figure}

\subsection{Two- and three-particle thresholds}
The Landau equations can be used to analyze a given Feynman diagram. For
two-point functions, singularities occur if the squared outer momentum crosses
thresholds or pseudothresholds. For a genuine sunset diagram as shown in
Fig.~\ref{sunset}, a two-loop two-point function with three lines of mass
values $m_1$, $m_2$ and $m_3$ connecting the incoming with the outgoing leg,
these thresholds and pseudothresholds are given by
\begin{equation}
p^2=(m_1\pm m_2\pm m_3)^2
\end{equation}
where the three-particle threshold is located at $p^2=(m_1+m_2+m_3)^2$.
In general, the genuine threshold is the one with the highest value. If the
squared momentum crosses this threshold, an imaginary part appears. According
to the Cutkosky rules~\cite{Cutkosky:1960sp}, this imaginary part corresponds
to the situation where the particles move onto their mass shell. In terms of
Feynman diagrams, inner lines are cut into outer legs for physical particles.
This can be visualized in Feynman diagrams by drawing a cutting line.

In the diagrams we are working with we expect two- and three-particle
thresholds. While the two-particle thresholds are related to the values of
$s_0=s_{n,m}$ and $t_0=t_{n,m}$, all the three-particle thresholds are related
to the values of $r_0=r_{n,m}$,
\begin{eqnarray}
s_{1,2}>0\quad\Leftrightarrow\quad p^2>(m_1+m_2)^2,&&
r_{1,5}>0\quad\Leftrightarrow\quad p^2>(m_1+m_3+m_5)^2,\nonumber\\
t_{4,5}>0\quad\Leftrightarrow\quad p^2>(m_4+m_5)^2,&&
r_{2,4}>0\quad\Leftrightarrow\quad p^2>(m_2+m_3+m_4)^2,\nonumber\\
t_{4,6}>0\quad\Leftrightarrow\quad p^2>(m_4+m_6)^2,&&
r_{1,6}>0\quad\Leftrightarrow\quad p_1^2>(m_1+m_3+m_6)^2,\nonumber\\
t_{5,6}>0\quad\Leftrightarrow\quad p^2>(m_5+m_6)^2,&&
r_{2,6}>0\quad\Leftrightarrow\quad p_2^2>(m_2+m_3+m_6)^2.\qquad
\end{eqnarray}
Anticipating examples to follow, for the subtraction terms some of the
two-particle thresholds vanish while new three-particle thresholds appear. For
the diagram shown on the right hand side of Fig.~\ref{planard0}, additional
three-particle thresholds appear for $p_1^2>(m_{n,j}+m_3+m_4)^2$
($r_{n,j,4}>0$) and $p_2^2>(m_{n,j}+m_3+m_5)^2$ ($r_{n,j,5}>0$) while
two-particle thresholds including the masses $m_1$ and $m_2$ no longer occur.

\subsection{Landau singularities for the scalar topology}
\begin{figure}\begin{center}
\epsfig{figure=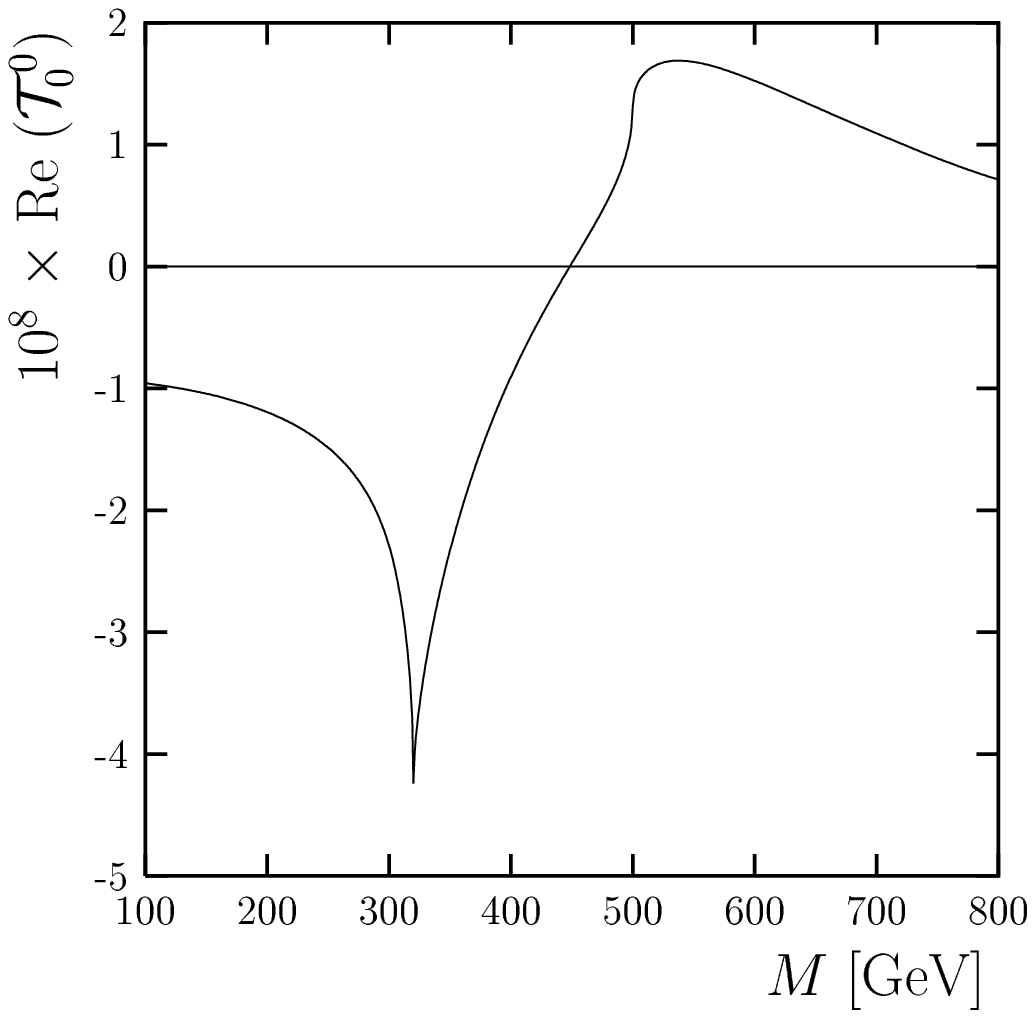, scale=0.8}\vspace{7pt}
\strut\kern-17pt\epsfig{figure=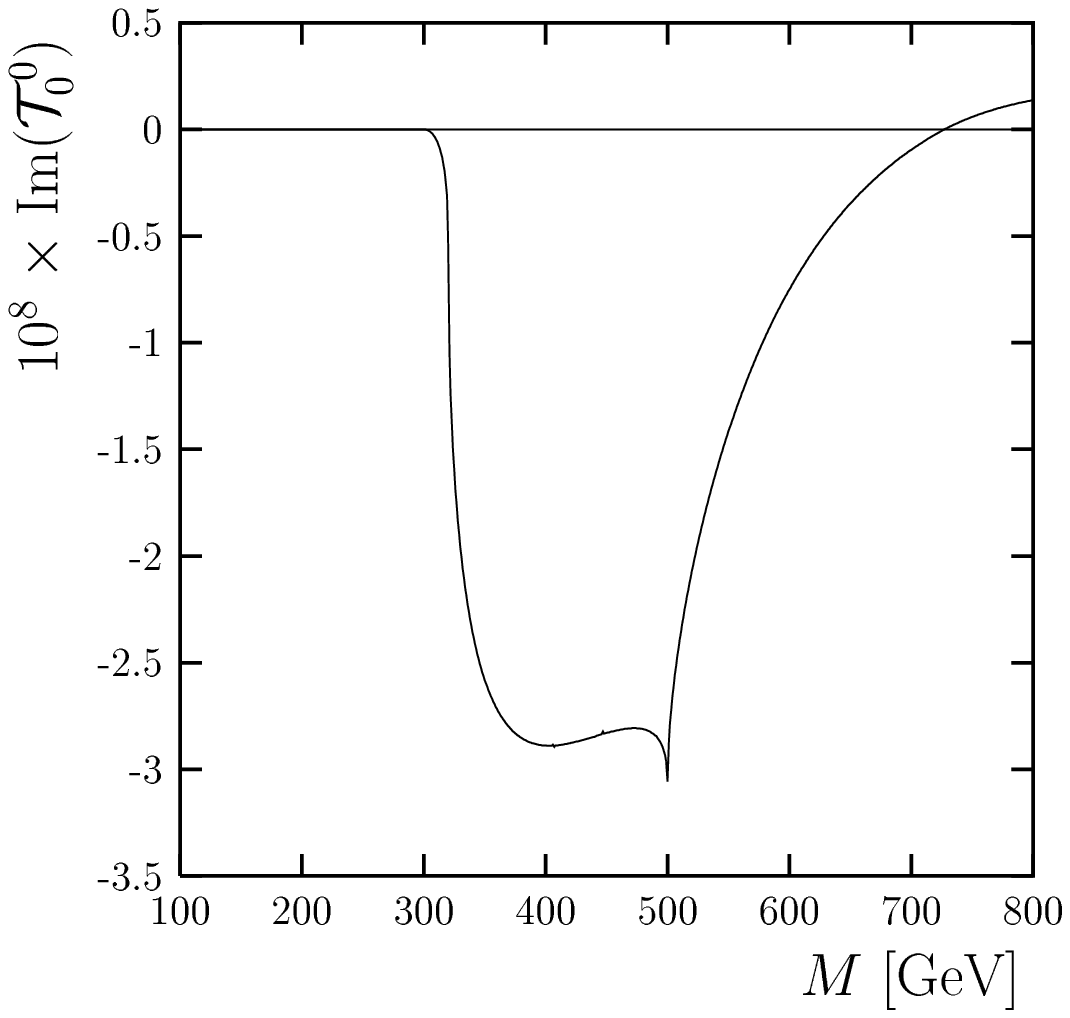, scale=0.8}
\caption{\label{landaus}Real part (top) and and imaginary part (bottom) for
the scalar planar two-loop three-point diagram ${\cal T}^0_0$ of
Eq.~(\ref{inttal0}) with decay mass $M$ between $100\GeV$ and $800\GeV$. For
the values of masses and momenta we use the standard set given in
Eq.~(\ref{standmass}).}
\end{center}\end{figure}

As an example we consider the original planar topology in Eq.~(\ref{inttal0})
with $\alpha=0$. Because we are only interested in a qualitative analysis, we
use the following fictitious values for masses and outgoing momenta,
\begin{equation}\label{standmass}
\matrix{m_1=420\GeV&m_3=100\GeV&m_4=120\GeV&\sqrt{p_1^2}=60\GeV\cr
  m_2=80\GeV&m_5=200\GeV&m_6=300\GeV&\sqrt{p_2^2}=20\GeV.\cr}
\end{equation}
For the decaying particle we vary the value of $M=\sqrt{p^2}$ in a range
between $M=100\GeV$ and $M=800\GeV$ and plot the real and imaginary part. The
result is shown in Fig.~\ref{landaus}.

A three-particle threshold is expected for $M=m_2+m_3+m_4=300\GeV$. At this
point we see that the imaginary part starts to differ from zero. The
two-particle threshold at $M=m_4+m_5=320\GeV$ is characterized by a sharp peak
in the real part, accompanied by an instant decrease of the imaginary part,
leading to a vertical tangent to the curve at this point. For the second
two-particle threshold in this energy range at $M=m_1+m_2=500\GeV$, the
situation is the opposite. The imaginary part shows a sharp peak while the
real part increases with vertical slope.

\subsection{The topology of the subtraction terms}
\begin{figure}\begin{center}
\epsfig{figure=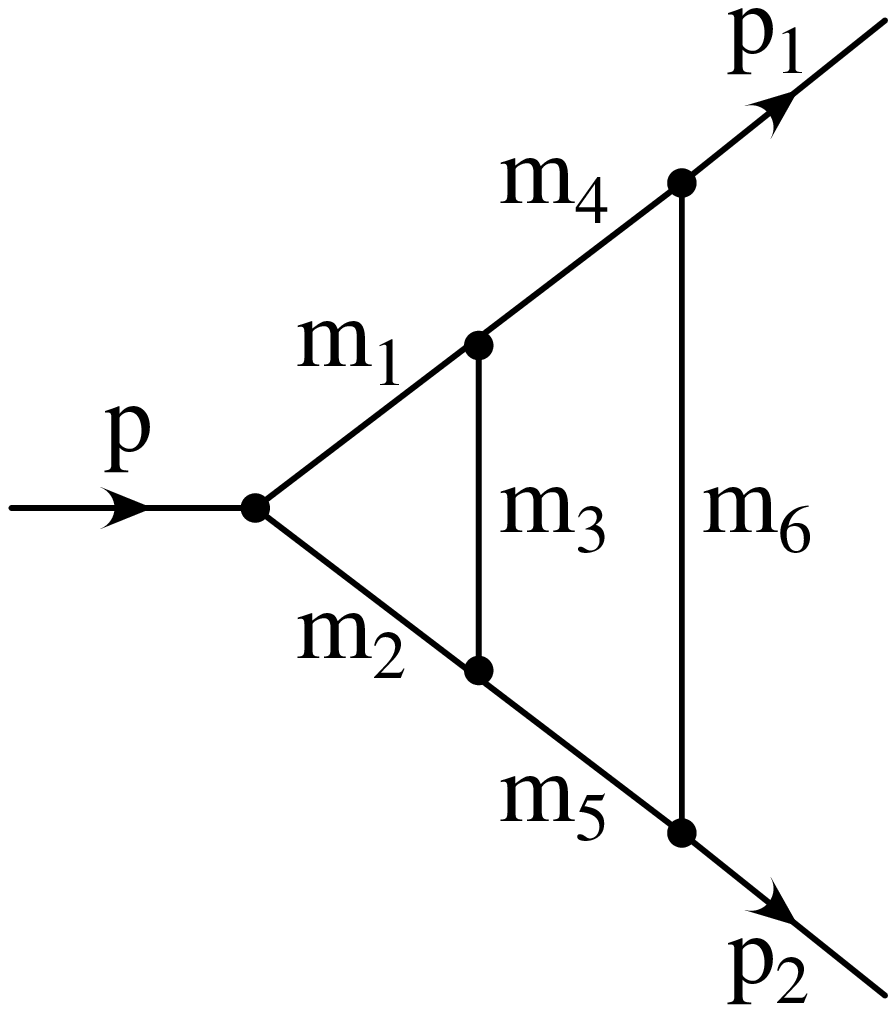, scale=0.4}\quad
\raise56pt\hbox{$\displaystyle-\sum_{n=1}^2\sum_{i=1}^{j_k}$}
\epsfig{figure=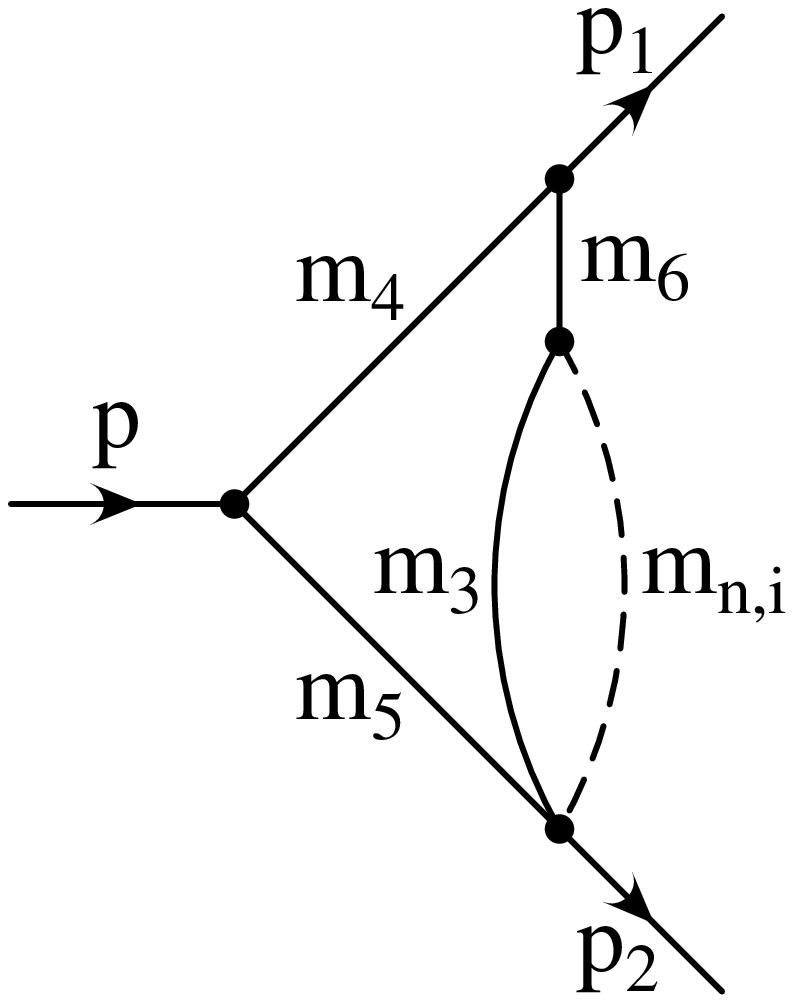, scale=0.4}
\caption{\label{planard0}Planar topology ${\cal T}^0$ (left hand side) and the
subtraction terms (right hand side). For convenience, relative factors are not
mentioned in the diagrammatic representation. The numerator factor for the
master integral and the subtraction terms read $(k_0-k_1)^\alpha$.}
\end{center}\end{figure}

If we consider tensor integrals we have to take subtraction terms into
account. As it was shown in the previous section, for the planar two-loop
three-point master diagram with power $(k_0-k_1)^2$ we need a single
subtraction. Each subtraction replaces the propagator factors by the
subtracted ones. In the present case one has
\begin{equation}\label{V02int}
{\cal V}^0_2=\int\frac{(k_0-k_1)^2d^4kd^4l}{P_1P_2P_3P_4P_5P_6}
  \left(1-\frac{P_1P_2}{P_{1,1}P_{2,1}}\right)
  =\int\frac{(k_0-k_1)^2d^4kd^4l}{P_1P_2P_3P_4P_5P_6}
  -\int\frac{(k_0-k_1)^2d^4kd^4l}{P_{1,1}P_{2,1}P_3P_4P_5P_6}
\end{equation}
(for higher order subtractions the scaled propagator factors appear in higher
powers). If we perform the limit $\kappa\to 0$ for
$P_{1,1}=(k+\kappa p_1)^2-m_{1,1}^2-i\eta$ and
$P_{2,1}=(k-\kappa p_2)^2-m_{2,1}^2-i\eta$, the propagator factors loose their
dependence on the outer momenta $p_i$. In this case the momentum scheme given
for the original diagram is no longer valid. We first have to perform a
partial fraction decomposition for the subtraction term,
\begin{equation}
\int\frac{(k_0-k_1)^2d^Dk\,d^Dl}{P_{1,1}P_{2,1}P_3P_4P_5P_6}
  =\frac1{m_{2,1}^2-m_{1,1}^2}
\left[\int\frac{(k_0-k_1)^2d^Dk\,d^Dl}{P_{1,1}P_3P_4P_5P_6}
  -\int\frac{(k_0-k_1)^2d^Dk\,d^Dl}{P_{2,1}P_3P_4P_5P_6}\right].
\end{equation}
After that, the lines can be rearranged according to their momenta. The
planar diagram is shown in Fig.~\ref{planard0} together with the subtraction.
For convenience, relative factors are not mentioned in the diagrammatic
representation.

\begin{figure}\begin{center}
\epsfig{figure=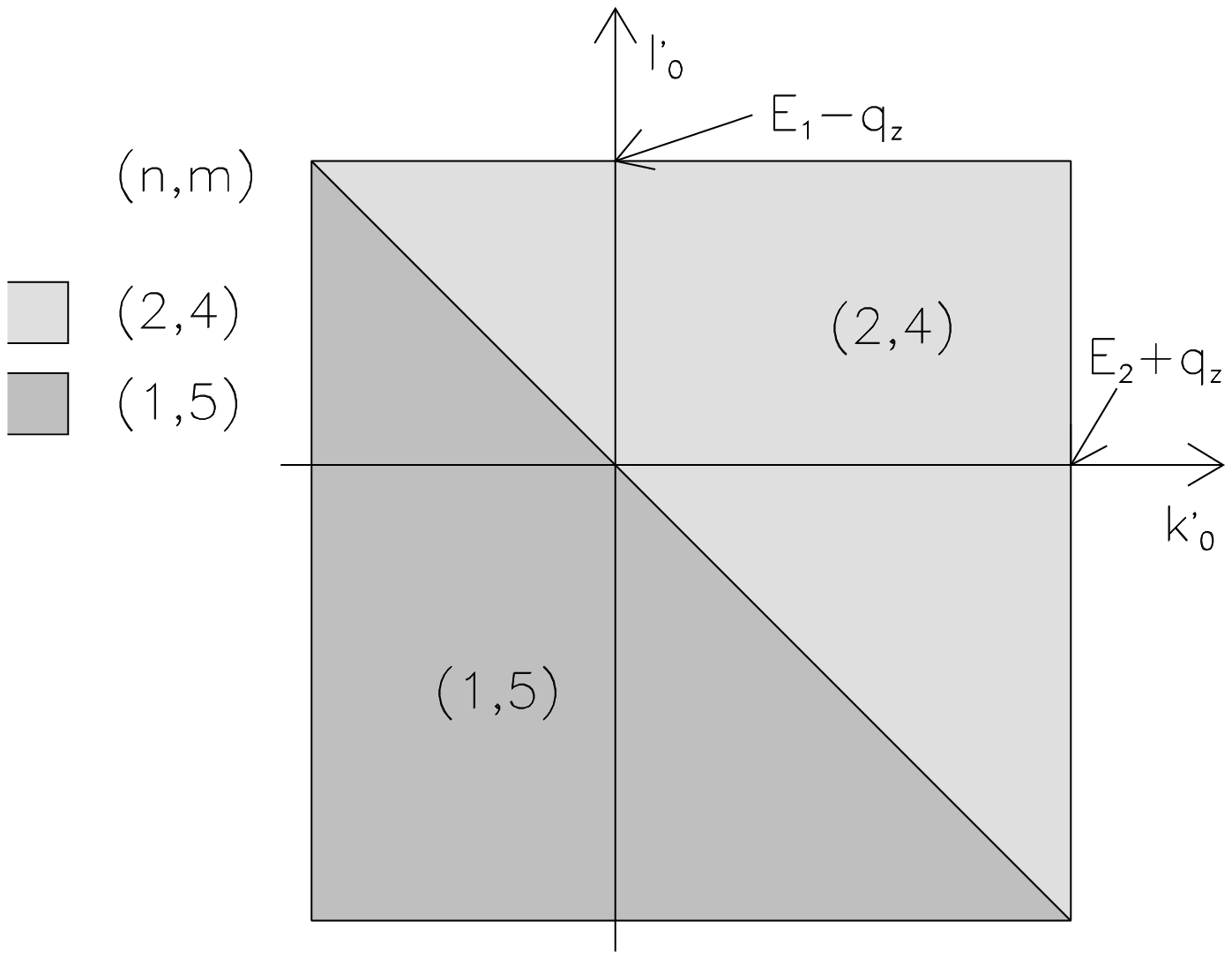, scale=0.8}\vspace{3pt}
\epsfig{figure=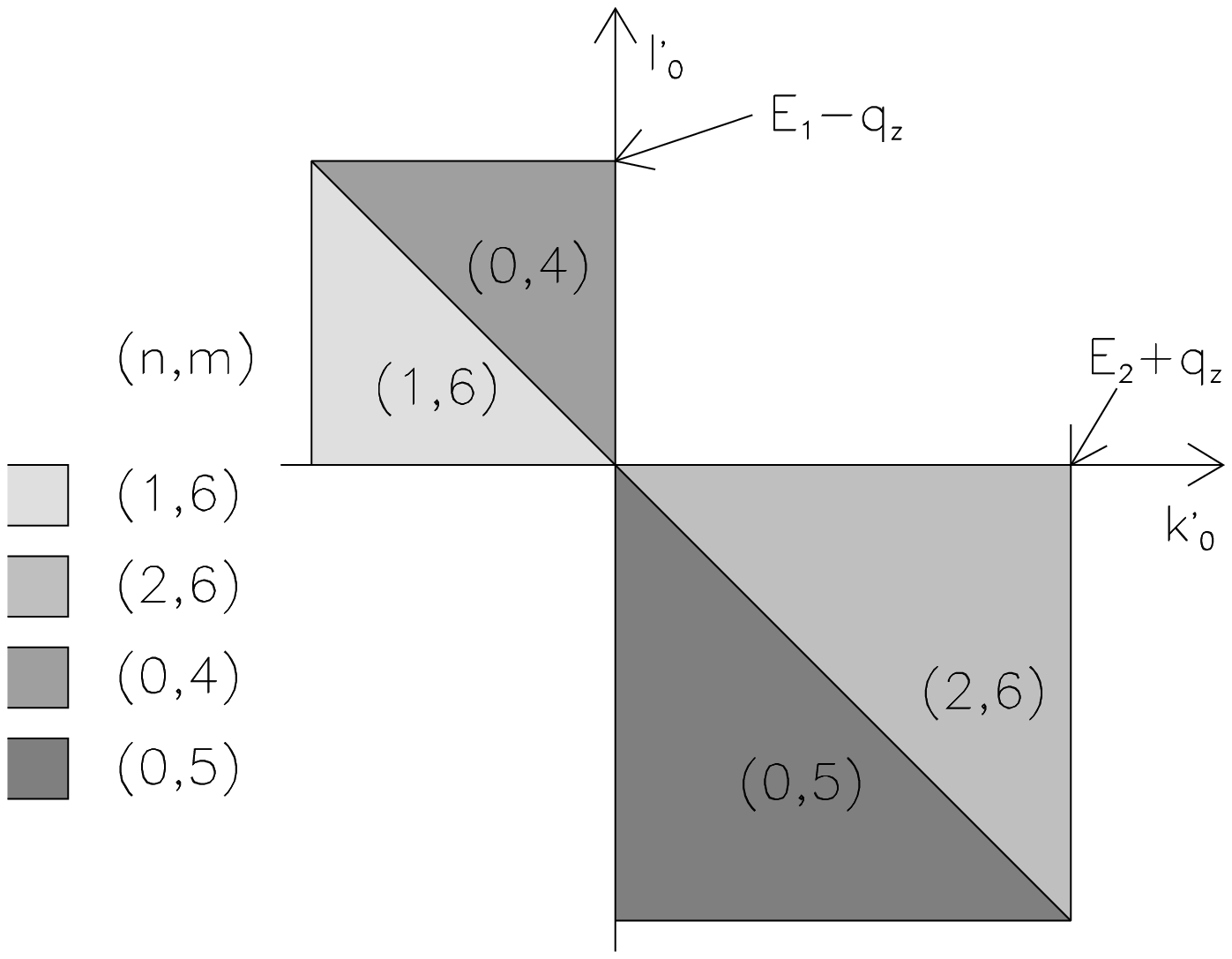, scale=0.8}
\caption{\label{triangles}The integration regions for the last (numerical)
two-dimensional integration are triangles which are bound by the line
$k'_0+l'_0=0$, a vertical and a horizontal line, according to the condition
that ${\cal B}^{kl}_{n,m}$ in Eqs.~(\ref{calBkl0}) or~(\ref{calBkl5}) does not
vanish. The integration regions for different values of index couples $(n,m)$
are shown for the linearizations $k_0=k'_0+k_1$ and $l_0=l'_0+l_1$. In case of
the opposite linearizations, the signs in front of the $q_z$ change.}
\end{center}\end{figure}

\section{Numerical integration}
After having performed all other integrations analytically, only the two
integrations over $k'_0$ and $l'_0$ are left. The integration region is given
by the non-vanishing of the factors ${\cal B}^{kl}_{n,m}$ which read
\begin{equation}
{\cal B}^{kl}_{n,m}
  =-4\pi^2\left[\theta(\pm(k'_0+l'_0))\theta(-\theta_n)\theta(-\phi_m)
  +\theta(\pm(k'_0+l'_0))\theta(\theta_n)\theta(\phi_m)\right]
\end{equation}
depending on the linearizations $k_0=k'_0\pm k_1$ and $l_0=l'_0\pm l_1$. If
we take into account that
\begin{eqnarray}
\theta_1=\pm2(k'_0+E_1\mp q_z),&&\theta_2=\pm2(k'_0-E_2\mp q_z),\qquad
  \theta_0=\pm2k'_0\nonumber\\
\phi_4=\pm2(l'_0-E_1\pm q_z),&&\phi_5=\pm2(l'_0+E_2\pm q_z),\qquad
  \phi_6=\pm2l'_0,
\end{eqnarray}
the integration regions can be seen to be triangles which are bound by the
off-diagonal $k'_0+l'_0=0$ and the different unequalities coming from
$\theta_n$ and $\phi_m$. In case of the index couples
$(n,m)=(1,4),(2,5),(0,6)$ the integration region vanishes identically. The
non-vanishing integration regions are shown in Fig.~\ref{triangles} in case of
the linearizations $k_0=k'_0+k_1$ and $l_0=l'_0+l_1$.
Two remarks are in order at this point.
\begin{itemize}
\item Special care has to be taken for the case $(n,m)=(0,6)$. In this case,
the integration region vanishes only in the limit $\kappa_i,\lambda_i\to 0$.
However, we have thoroughly checked that in this limit the integrand does not
diverge, i.e.\ the integral does not give a non-vanishing contribution even
with a vanishing integration region~\cite{Knodel:2005}.
\item In case that one of the masses vanishes, integration regions might
vanish. In Fig.~\ref{triangles} we see that if $E_1=q_z$, i.e.\ if $p_1^2=0$,
the triangles $(n,m)=(1,6),(0,4)$ vanish. In case of the opposite
linearization and with $p_2^2=0$, the same holds for $(n,m)=(2,6),(0,5)$
instead. However, the integrand does not diverge in these cases.
\end{itemize}
The integrations are performed using numerical routines like the Monte Carlo
integration routine VEGAS~\cite{Lepage:1980dq,Ohl:1998jn} and the program
Divonne of the library CUBA~\cite{Hahn:2004fe}. While numerical routines
sometimes overestimate the precision of the
result~\cite{Kleiss:2005du,vanHameren:2001ng}, the use of two different
routines enables us to have a measure for the accuracy of the result. In this
section we will present a few examples in order to demonstrate the reliability
of our method.

\subsection{The $Z^0$ decay via top quark loop}
\begin{figure}\begin{center}
\epsfig{figure=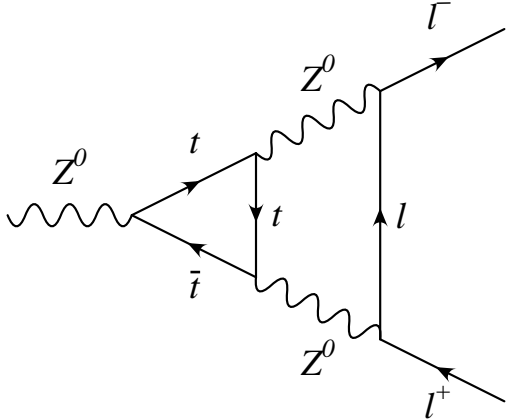, scale=0.5}
\caption{\label{V02dia}Feynman diagram corresponding to ${\cal V}^0_2$
in Eq.~(\ref{V02int})}
\end{center}\end{figure}

As a first example we consider the basic integral in Eq.~(\ref{V02int})
occuring in a process where the $Z^0$ boson couples to a triangle with top
quark current (cf.\ Fig.~\ref{V02dia}).
We have started the discussion already in Eq.~(\ref{V02int}) and will continue
at this point. The masses are chosen to be
\begin{equation}
m_1=m_2=m_3=m_t,\qquad m_4=m_5=m_Z,\quad\mbox{\rm and}\quad m_6=m_l
\end{equation}
where we use the values $m_t=178\GeV$,
$m_Z=91.1876\GeV$~\cite{Eidelman:2004wy}, and $m_l=0$. The subtraction masses
which should have different values are taken as $m_{1,1}=150\GeV$ and
$m_{2,1}=160\GeV$. Because the squared mass of the $Z^0$ boson is below all
possible thresholds, the imaginary part of this integral vanishes. Even though
physical masses are chosen in this paragraph, we stress that instead of
calculating the whole process we are dealing with the occuring planar basic
diagram only.

\begin{figure}\begin{center}
\epsfig{figure=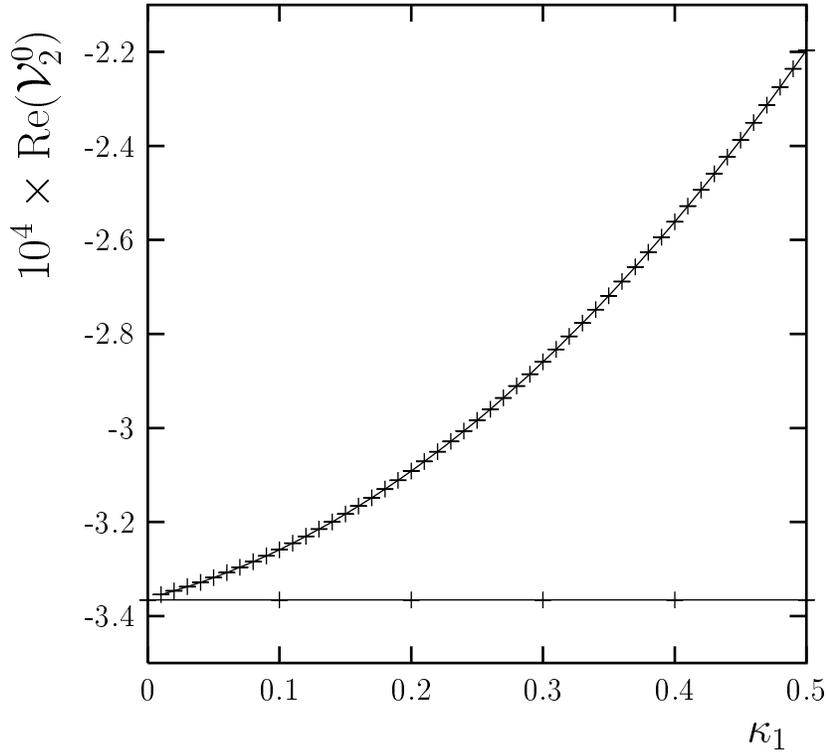, scale=1.0}
\caption{\label{V02rgra}Real part of the integral ${\cal V}^0_2$
in Eq.~(\ref{V02int}) as a function of the parameter $\kappa_1$. For the
masses and momenta we used values related to the decay of the $Z^0$ boson into
a top-quark loop, as explained in the text (cf.\ Fig.~\ref{V02dia}). Shown are
the results for the numerical calculation (curve) and the result for the
semi-analytical calculation with $\kappa_1=0$ (horizontal straight line). In
both cases we have used VEGAS.}
\end{center}\end{figure}

We calculate the integral for $50$ points between $\kappa_1=0.01$ and
$\kappa_1=0.5$ for the (dimensionless) parameter $\kappa_1$. As a next step
we approximate the points by a low degree polynomial in order to extrapolate
to $\kappa_1=0$. The result of this extrapolation can then be compared with
the result for the integral where we used $\kappa_1=0$ from the very beginning.
For the integration with VEGAS, 260\,000 internal points were used, whereas
for Divonne we use a standard precision of $10^{-3}$. The results of this fit
are shown in Tables~\ref{V02veg} and~\ref{V02div}.
\begin{table}[tb]\begin{center}
\renewcommand{\arraystretch}{0.8}
\begin{tabular}{|r|lll|}\hline
degree&real part&$\sigma_{\rm Re}$&$\kern-7pt\chi^2_{\rm Re}/(n-1)$\\\hline
$0$&$-2.941141\cdot 10^{-4}$&$1.4\cdot 10^{-9}$&$1.2\cdot 10^7$\\
$1$&$-3.484637\cdot 10^{-4}$&$2.6\cdot 10^{-9}$&$3.8\cdot 10^5$\\
$2$&$-3.361064\cdot 10^{-4}$&$3.9\cdot 10^{-9}$&$1.4\cdot 10^1$\\
$3$&$-3.361950\cdot 10^{-4}$&$5.2\cdot 10^{-9}$&$1.4\cdot 10^0$\\
$4$&$-3.361836\cdot 10^{-4}$&$6.8\cdot 10^{-9}$&$1.3\cdot 10^0$\\
$5$&$-3.361884\cdot 10^{-4}$&$8.6\cdot 10^{-9}$&$1.3\cdot 10^0$\\
$6$&$-3.361693\cdot 10^{-4}$&$1.1\cdot 10^{-8}$&$1.1\cdot 10^0$\\\hline
 x &$-3.365555\cdot 10^{-4}$&$7.2\cdot 10^{-9}$&\\\hline
\end{tabular}
\caption{\label{V02veg}Extrapolation of the contributions
  for ${\cal V}^0_2$ using VEGAS}
\vspace{12pt}
\begin{tabular}{|r|lll|}\hline
degree&real part&$\sigma_{\rm Re}$&$\chi^2_{\rm Re}/(n-1)$\\\hline
$0$&$-2.81249\cdot 10^{-4}$&$3.9\cdot 10^{-8}$&$1.7\cdot 10^4$\\
$1$&$-3.52842\cdot 10^{-4}$&$8.7\cdot 10^{-8}$&$5.0\cdot 10^2$\\
$2$&$-3.36111\cdot 10^{-4}$&$1.3\cdot 10^{-7}$&$1.0\cdot 10^{-1}$\\
$3$&$-3.36265\cdot 10^{-4}$&$1.9\cdot 10^{-7}$&$7.4\cdot 10^{-2}$\\
$4$&$-3.36301\cdot 10^{-4}$&$2.5\cdot 10^{-7}$&$7.3\cdot 10^{-2}$\\
$5$&$-3.36294\cdot 10^{-4}$&$3.2\cdot 10^{-7}$&$7.3\cdot 10^{-2}$\\
$6$&$-3.36209\cdot 10^{-4}$&$4.1\cdot 10^{-7}$&$7.1\cdot 10^{-2}$\\\hline
 x &$-3.36530\cdot 10^{-4}$&$3.3\cdot 10^{-7}$&\\\hline
\end{tabular}
\caption{\label{V02div}Extrapolation of the contributions
  for ${\cal V}^0_2$ using Divonne}
\end{center}\end{table}
If $\kappa_1=0$ is used from the very beginning and all integrations are done
analytically except for the last two ones, we obtain the results shown in the
last lines of Tables~\ref{V02veg} and~\ref{V02div} marked by a ``x''.
Obviously, these values are already reliably approximated by the second order
fit. In case of VEGAS, the quantity $\chi^2/(n-1)$ for $n$ sampling points
as a measure for the reliability of the polynomial fit of the numerical
integrations approaches the optimal value $1.0$ quite fast. The deviation of
the order of $1$ in case of Divonne is not problematic because we use this
second method only to check the VEGAS calculation. The result is shown
graphically in Fig.~\ref{V02rgra}.
In this figure we show the results of the numerical calculation as a function
of $\kappa_1$ together with the result for the semi-analytical calculation for
$\kappa_1=0$. The error bars for the latter are plotted at different points in
order to allow for a better comparison. Also optically the results for
numerical and semi-analytical calculation match very well. This gives us
confidence in the correct implementation of the algorithm.

\begin{figure}\begin{center}
\epsfig{figure=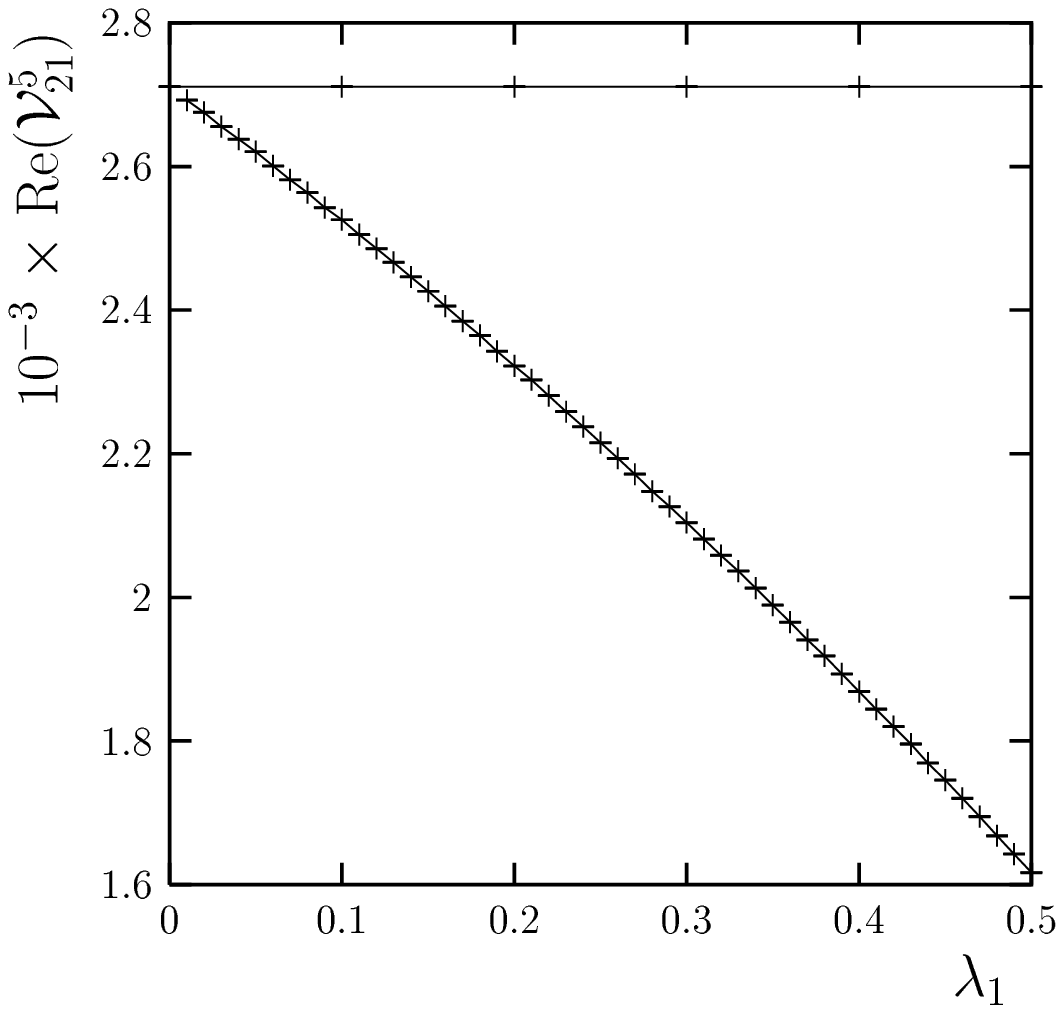, scale=0.8}
\epsfig{figure=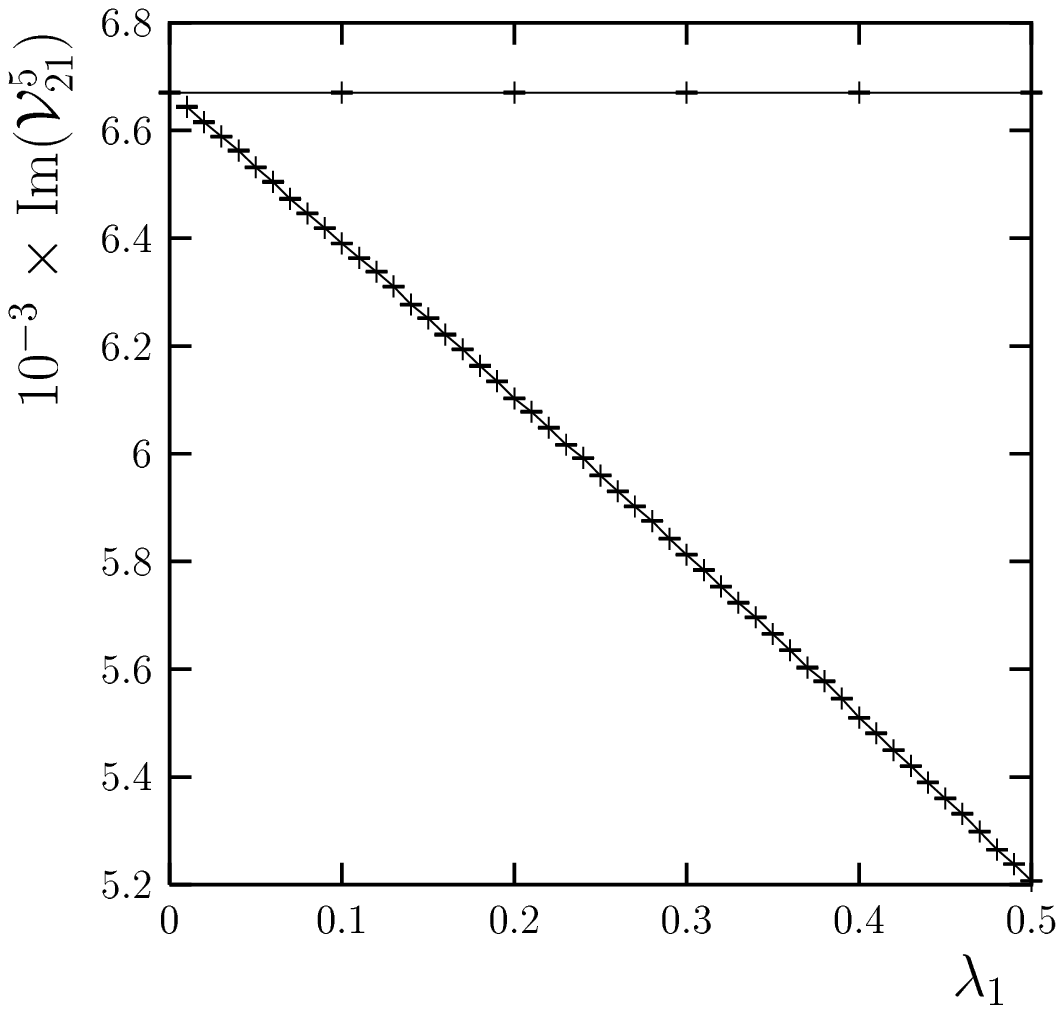, scale=0.8}
\caption{\label{V521gra}Real part (top) and imaginary part (bottom) of the
integral ${\cal V}^5_{21}$ in Eq.~(\ref{V521int}) as a function of the
parameters $\lambda_1$ and $\kappa_1=0.8\cdot\lambda_1$. Shown are the results
for the numerical calculation (curve) and for the semi-analytical calculation
with $\kappa_1=\lambda_1=0$ (horizontal straight line), both obtained by using
VEGAS.}
\end{center}\end{figure}

\begin{figure}\begin{center}
\epsfig{figure=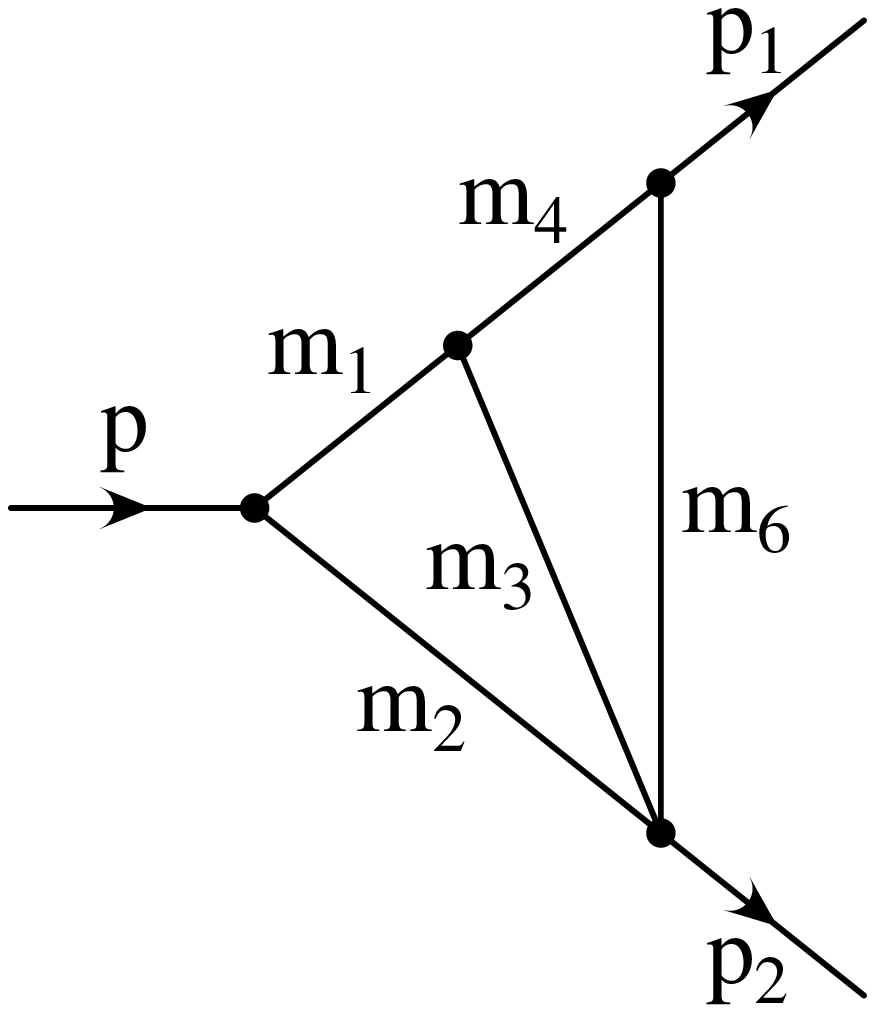, scale=0.4}
\raise56pt\hbox{$-$\ }
\raise2pt\hbox{\epsfig{figure=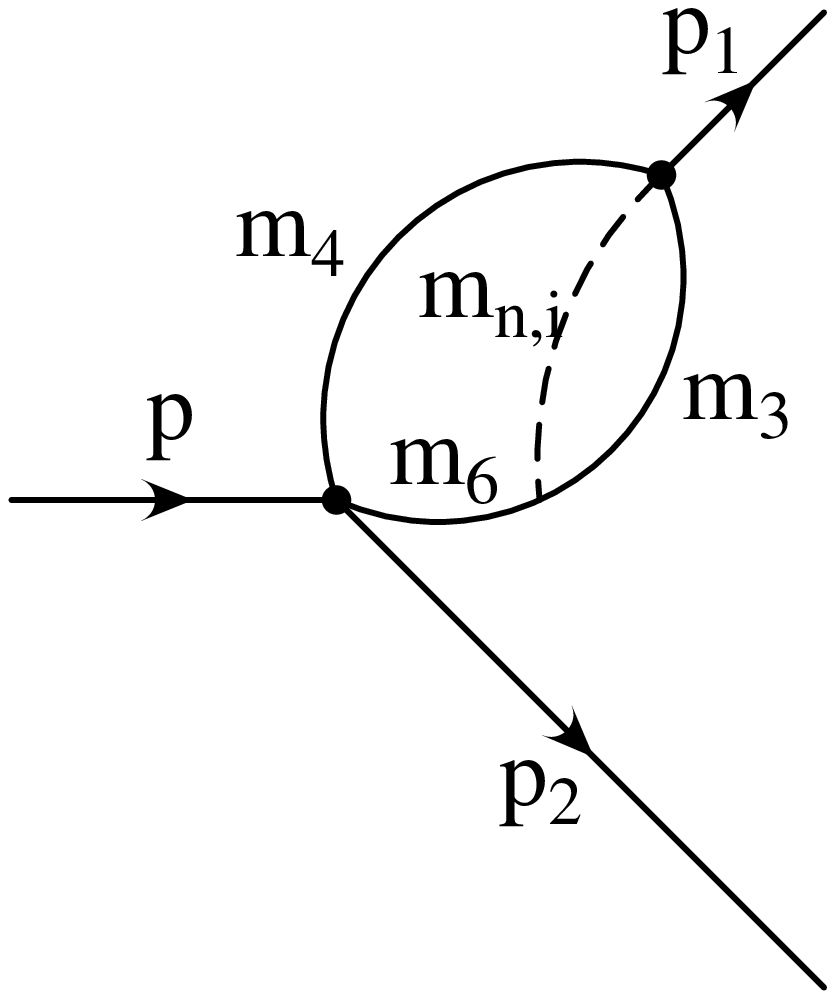, scale=0.4}}
\raise56pt\hbox{$-$\ }
\raise-2pt\hbox{\epsfig{figure=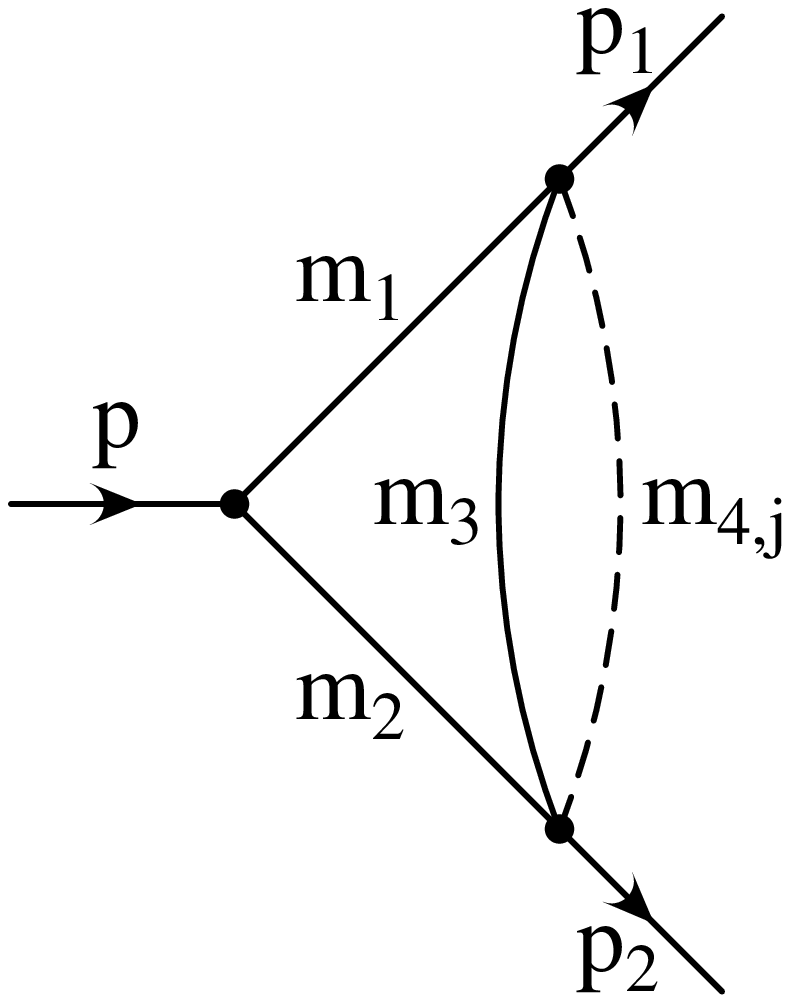, scale=0.4}}
\raise56pt\hbox{$+$\ }
\raise16pt\hbox{\epsfig{figure=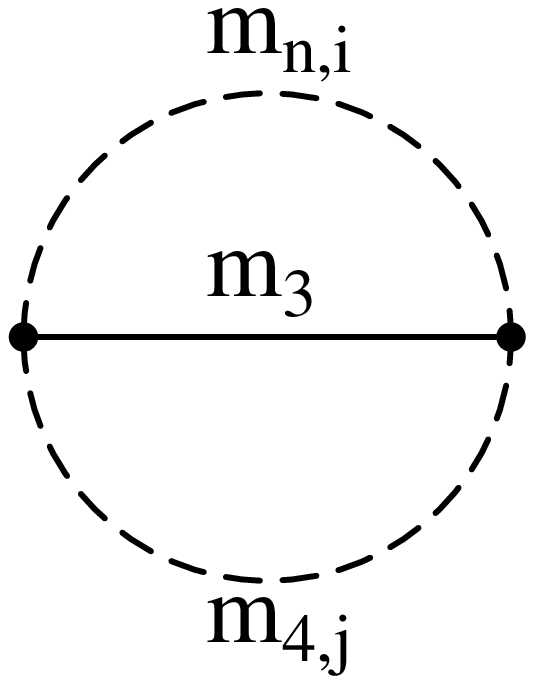, scale=0.4}}
\caption{\label{planard5}Rotated reduced planar topology ${\cal T}^5$ (first
diagram) and the subtraction terms. 
For convenience, relative factors are not mentioned in the diagrammatic
representation. Summations over $n=1,2$, $i=1,\ldots,j_k$ and $j=1,\ldots,j_l$
are assumed (cf.\ Fig.~\ref{planard0}). The numerator factor for the master
integral and the subtraction terms reads $(k_0+k_1)^\alpha(l_0+l_1)^\beta$.}
\end{center}\end{figure}

\subsection{Rotated reduced planar topology} 
Next we consider the numerical results for a specific example of the
rotated reduced planar topology ${\cal T}^5$. The physical starting point is
a decay of the $Z^0$ boson via a lepton loop, coupling to the outer lepton
and anti-lepton legs via $W$ bosons. The integral
\begin{equation}\label{V521int}
{\cal V}^5_{21}=\int \frac{(k_0+k_1)^2(l_0+l_1)}{P_1P_2P_3P_4P_6}
  \left(1-\frac{P_1P_2}{P_{1,1}P_{2,1}}\right)
  \left(1-\frac{P_4}{P_{4,1}}\right)d^4k\,d^4l
\end{equation}
contains two subtraction terms, for the $k$-loop as well as for the $l$-loop.
In this case we consider the limit $\kappa_1=0.8\cdot\lambda_1\to 0$. The
diagram with rotated reduced planar topology is shown in Fig.~\ref{planard5}
together with the three different types of subtraction terms. The masses take
the values $m_1=m_2=m_3=m_l=0$, $m_4=m_W=80.425\GeV$~\cite{Eidelman:2004wy},
and $m_6=m_l=0$. The momenta are given by $\sqrt{p_1^2}=\sqrt{p_2^2}=m_l=0$ and
$\sqrt{p^2}=m_Z$. For the three subtraction masses we take $m_{1,1}=200\GeV$,
$m_{2,1}=300\GeV$, and $m_{4,1}=110\GeV$. In Table~\ref{V521veg} we show the
results for the numerical integration with VEGAS for the real and the
imaginary part. Only the values at polynomial degrees~$2$ and~$3$ are shown.
They are in good agreement with the result at polynomial degree~$2$ obtained
by using Divonne (Table~\ref{V521div}).
\begin{table}[bt]\begin{center}
\renewcommand{\arraystretch}{0.8}
\begin{tabular}{|r|lll|lll|}\hline
degree&real part&$\sigma_{\rm Re}$&$\kern-12pt\chi^2_{\rm Re}/(n-1)$
&imag.\ part&$\sigma_{\rm Im}$&$\kern-12pt\chi^2_{\rm Im}/(n-1)$\\\hline
$2$&$2.71015\cdot 10^3$&$3.0\cdot 10^{-1}$&$8.5\cdot 10^{-1}$
   &$6.67079\cdot 10^3$&$7.7\cdot 10^{-1}$&$1.1\cdot 10^0$\\
$3$&$2.71047\cdot 10^3$&$4.2\cdot 10^{-1}$&$8.3\cdot 10^{-1}$
   &$6.6712\cdot 10^3$&$1.1\cdot 10^0$&$1.1\cdot 10^0$\\\hline
 x &$2.71125\cdot 10^3$&$7.0\cdot 10^{-1}$&
   &$6.6703\cdot 10^3$&$1.7\cdot 10^0$&\\\hline
\end{tabular}
\caption{\label{V521veg}Extrapolation of the contributions
  for ${\cal V}^5_{21}$ using VEGAS}
\vspace{12pt}
\begin{tabular}{|r|lll|lll|}\hline
degree&real part&$\sigma_{\rm Re}$&$\chi^2_{\rm Re}/(n-1)$
&imag.\ part&$\sigma_{\rm Im}$&$\chi^2_{\rm Im}/(n-1)$\\\hline
$2$&$2.7094\cdot 10^3$&$1.0\cdot 10^0$&$4.4\cdot 10^{-1}$
   &$6.6691\cdot 10^3$&$2.7\cdot 10^0$&$3.7\cdot 10^{-1}$\\\hline
 x &$2.7042\cdot 10^3$&$2.6\cdot 10^0$&
   &$6.6736\cdot 10^3$&$6.6\cdot 10^0$&\\\hline
\end{tabular}
\caption{\label{V521div}Extrapolation of the contributions
  for ${\cal V}^5_{21}$ using Divonne}
\end{center}\end{table}
The results for the semi-analytical calculation are again given in the last
lines of the two tables. Looking at the values obtained by using VEGAS in
Fig.~\ref{V521gra}, one sees good agreement.

\begin{figure}\begin{center}
\epsfig{figure=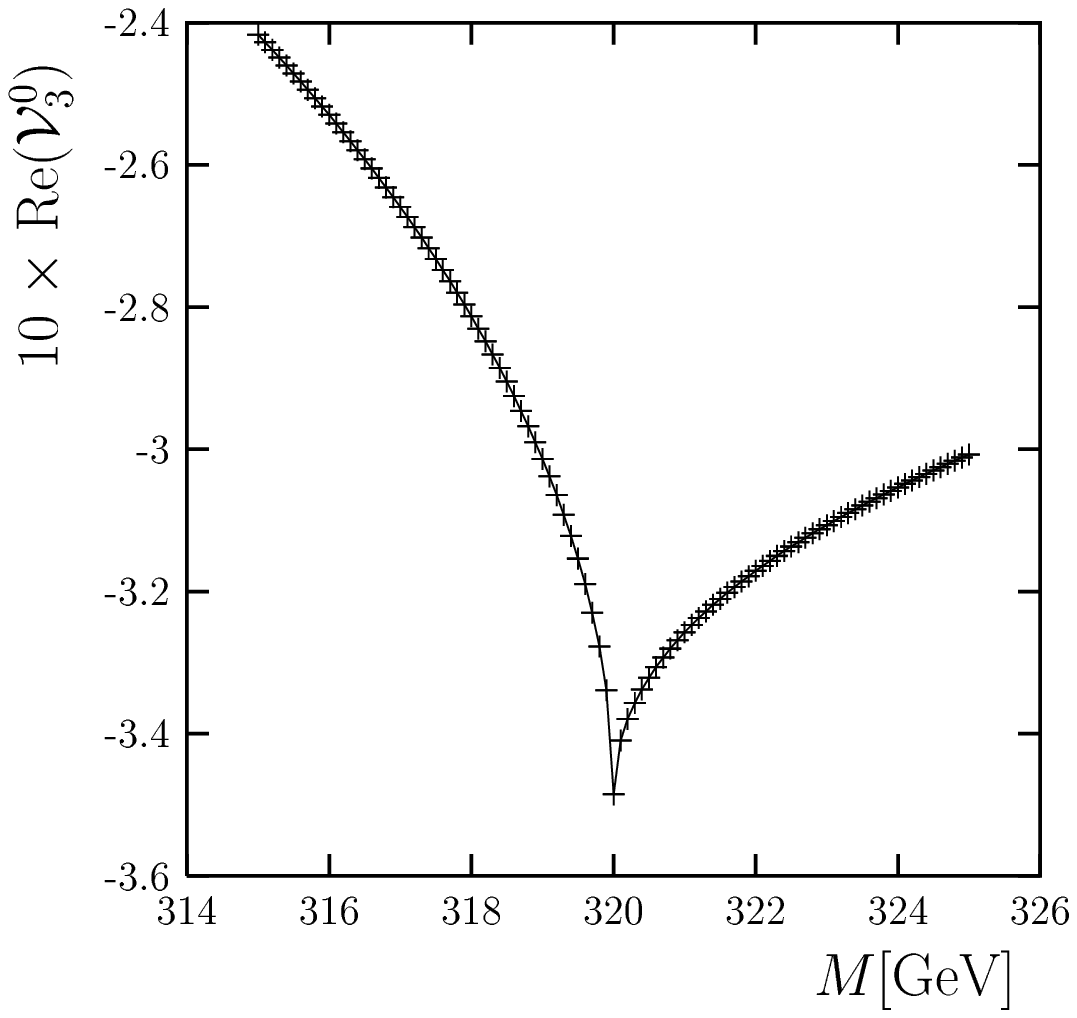, scale=0.8}
\epsfig{figure=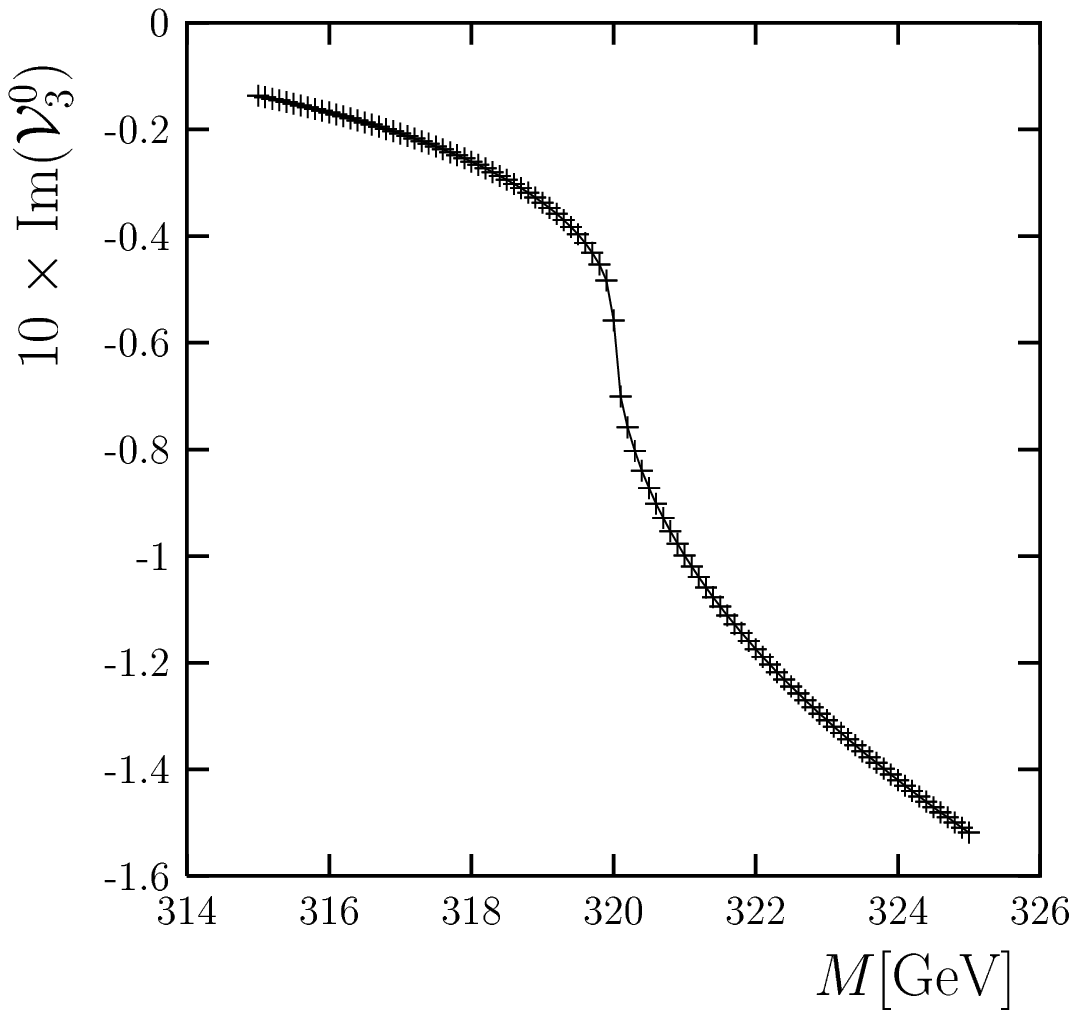, scale=0.8}
\caption{\label{V03run}Real part (top) and imaginary part (bottom) of the
integral ${\cal V}^0_3$ in Eq.~(\ref{V03int}) as a function of the decay mass
$M=\sqrt{p^2}$ close to $M=320\GeV$. The values for the masses and outgoing
momenta are taken from Eq.~(\ref{standmass}).}
\end{center}\end{figure}

\subsection{Results for varying decay mass}
For the integral with two subtractions for the $k$-loop,
\begin{equation}\label{V03int}
{\cal V}^0_3=\int\frac{(k_0-k_1)^3d^4kd^4l}{P_1P_2P_3P_4P_5P_6}
  \prod_{i=1}^2\left(1-\frac{P_1P_2}{P_{1,i}P_{2,i}}\right),
\end{equation}
we vary the decay mass $M=\sqrt{p^2}$ in order to identify different
thresholds. The values for physical masses and outgoing momenta are again
taken from Eq.~(\ref{standmass}), for the subtraction masses we use
$m_{1,1}=100\GeV$, $m_{2,1}=200\GeV$, $m_{1,2}=350\GeV$, and $m_{2,2}=450\GeV$.
In varying $M$ we can observe the behaviour of the real and the imaginary part.
The behaviour is shown in Fig.~\ref{V03run} for decay mass values close to
$M=320\GeV$. As in case of the simple example shown in Fig.~\ref{landaus} we
notice a sharp peak for the real part and a vertical slope for the imaginary
part at the point $M=320\GeV$ which corresponds to the two-particle threshold
$M^2=p^2=(m_4+m_5)^2$. In a second step we are looking more closely at the
threshold region. Taking the value $M=325\GeV$, we analyze the real and
imaginary parts for $\kappa_2=0.6\cdot\kappa_1\to 0$. In Table~\ref{V03veg} we
show results of the polynomial fit for the results obtained by VEGAS. The last
line of Table~\ref{V03veg} and Table~\ref{V03div} are the semi-analytic
results obtained by using VEGAS and Divonne, respectively.
\begin{table}[tb]\begin{center}
\renewcommand{\arraystretch}{0.8}
\begin{tabular}{|r|lll|lll|}\hline
degree&real part&$\sigma_{\rm Re}$&$\kern-20pt\chi^2_{\rm Re}/(n-1)$
&imag.\ part&$\sigma_{\rm Im}$&$\kern-20pt\chi^2_{\rm Im}/(n-1)$\\\hline
$3$&$-3.00814\cdot 10^{-1}$&$1.2\cdot 10^{-5}$&$\kern-4pt 4.2\cdot 10^0$
   &$-1.519135\cdot 10^{-1}$&$5.2\cdot 10^{-6}$&$\kern-4pt 3.6\cdot 10^0$\\
$4$&$-3.00692\cdot 10^{-1}$&$1.5\cdot 10^{-5}$&$\kern-4pt 8.9\cdot 10^{-1}$
   &$-1.518660\cdot 10^{-1}$&$6.8\cdot 10^{-6}$&$\kern-4pt 1.2\cdot 10^0$\\
\hline
 x &$-3.00689\cdot 10^{-1}$&$1.7\cdot 10^{-5}$&
   &$-1.51865\cdot 10^{-1}$&$1.4\cdot 10^{-5}$&\\\hline
\end{tabular}
\caption{\label{V03veg}Extrapolation of the contributions
  for ${\cal V}^0_3$ using VEGAS}
\vspace{12pt}
\begin{tabular}{|r|lll|lll|}\hline
degree&real part&$\sigma_{\rm Re}$&$\kern-10pt\chi^2_{\rm Re}/(n-1)$
&imag.\ part&$\sigma_{\rm Im}$&$\kern-10pt\chi^2_{\rm Im}/(n-1)$\\\hline
 x &$-3.0083\cdot 10^{-1}$&$2.9\cdot 10^{-4}$&
   &$-1.5196\cdot 10^{-1}$&$1.4\cdot 10^{-4}$&\\\hline
\end{tabular}
\caption{\label{V03div}Extrapolation of the contributions
  for ${\cal V}^0_3$ using Divonne}
\end{center}\end{table}
\begin{figure}\begin{center}
\epsfig{figure=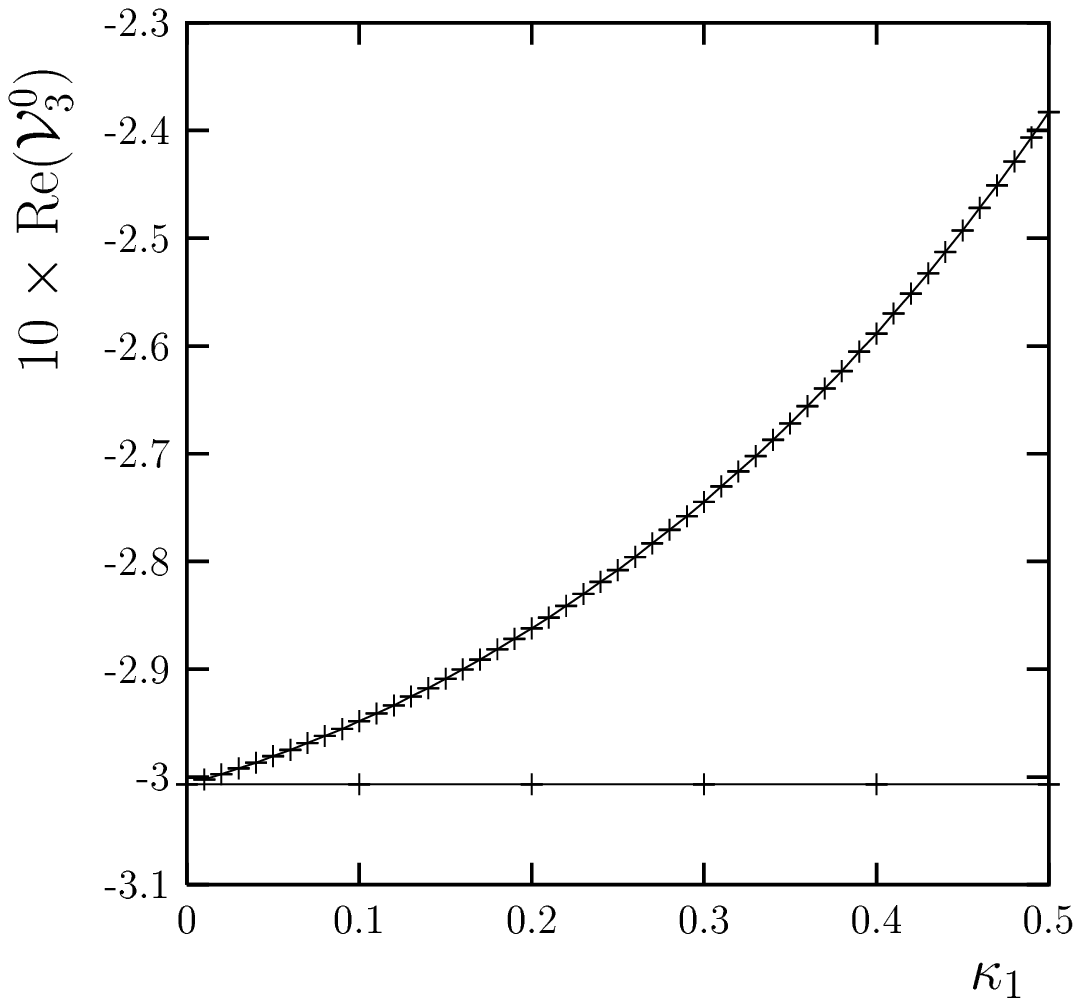, scale=0.8}
\epsfig{figure=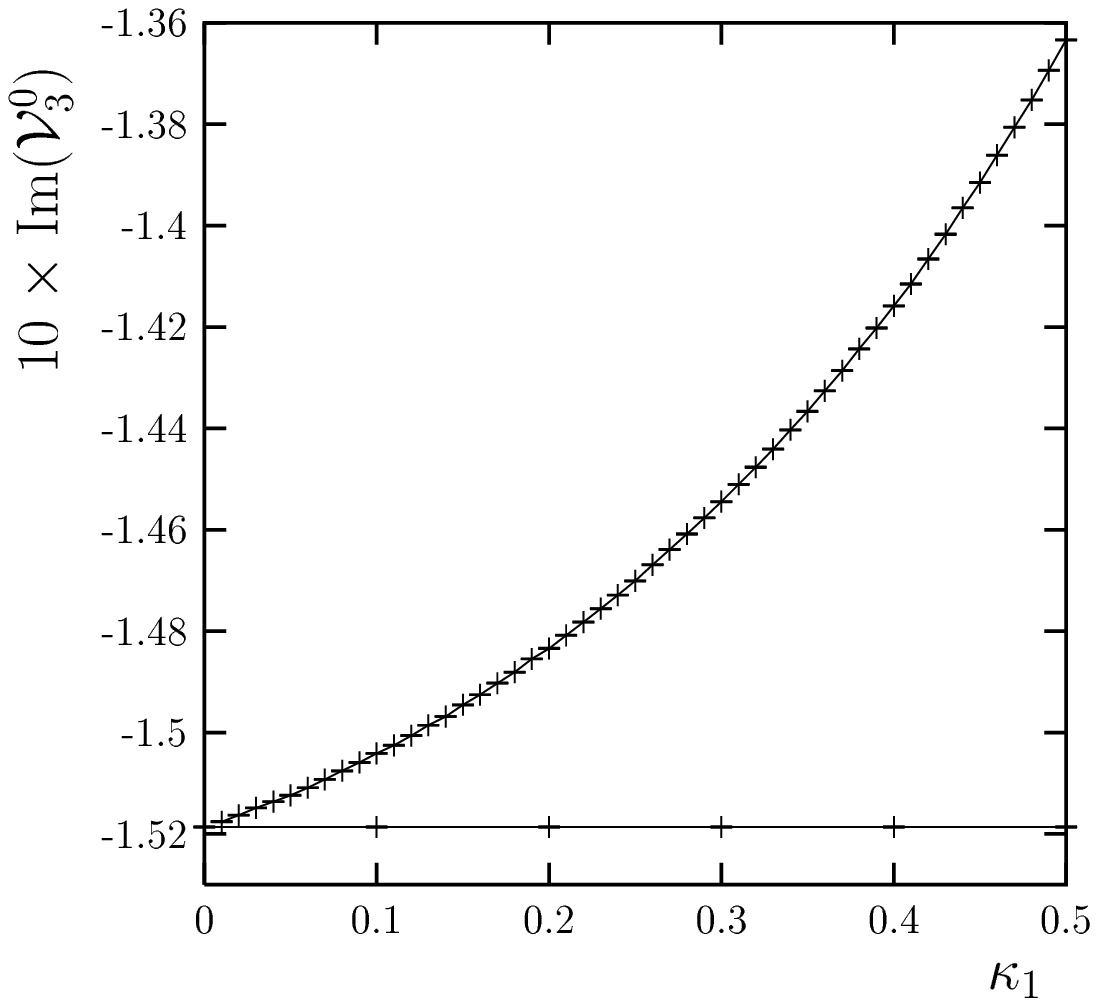, scale=0.8}
\caption{\label{V03gra}Real part (top) and imaginary part (bottom) of the
integral ${\cal V}^0_3$ in Eq.~(\ref{V03int}) as a function of the parameters
$\kappa_1$ and $\kappa_2=0.6\cdot\kappa_1$ for the decay mass value
$M=\sqrt{p^2}=325\GeV$. Shown are the results for the numerical calculation
(curve) and for the semi-analytical calculation with $\kappa_2=\kappa_1=0$
(horizontal straight line), both obtained by using VEGAS.}
\end{center}\end{figure}
The results are shown in Fig.~\ref{V03gra}. Again both methods lead to the
same result in the limit $\kappa_2=0.6\cdot\kappa_1\to 0$.

\subsection{Combined integration}
After having performed these tests, the semi-analytical evaluation of the
partial results from the different master integrals appears to be feasible.
This might be done by performing the numerical integrations separately for
each master integral and afterwards combining the results. However, it turns
out that the more stable method is to first perform the analytical integrations
separately and then to do the numerical integration for the sum of all basic
integrands relevant for a given process. This is especially true for squared
momenta on the threshold or close to it. As an example we use
\begin{equation}\label{topex}
{\cal T}_{\rm ex}=\int\frac{(kl)(p_1k)(p_2k+p_2l)}{P_1P_2P_3P_4P_5P_6}
  d^Dk\,d^Dl.
\end{equation}
After the tensor reduction explained in Sec.~3, we are left with the original
planar topology and the two rotated reduced planar topologies, both with
different powers of $(k_0\mp k_1)$ and $(l_0\mp l_1)$, and diagrams of simpler
topologies. The master integrals for the non-reduced and rotated reduced planar
topologies can be calculated by applying the subtractions explained in Sec.~4.
For the convergent part the limit $\kappa_i,\lambda_i\to 0$ can be performed.
All integrations up to the last two can be done analytically. The combination
of the different UV-finite contributions with planar and rotated reduced
planar topology at this point is called ${\cal V}_{\rm ex}$.

As explained before, there are two possibilities to proceed. We can combine
all analytical results to a single expression and integrate this expression
numerically (combined integration), or we can perform the numerical
integration of the analytical expression for each of the master integrals
separately and add up the results afterwards (separated integration). We will
perform both calculations in the following to compare them.

\begin{figure}\begin{center}
\epsfig{figure=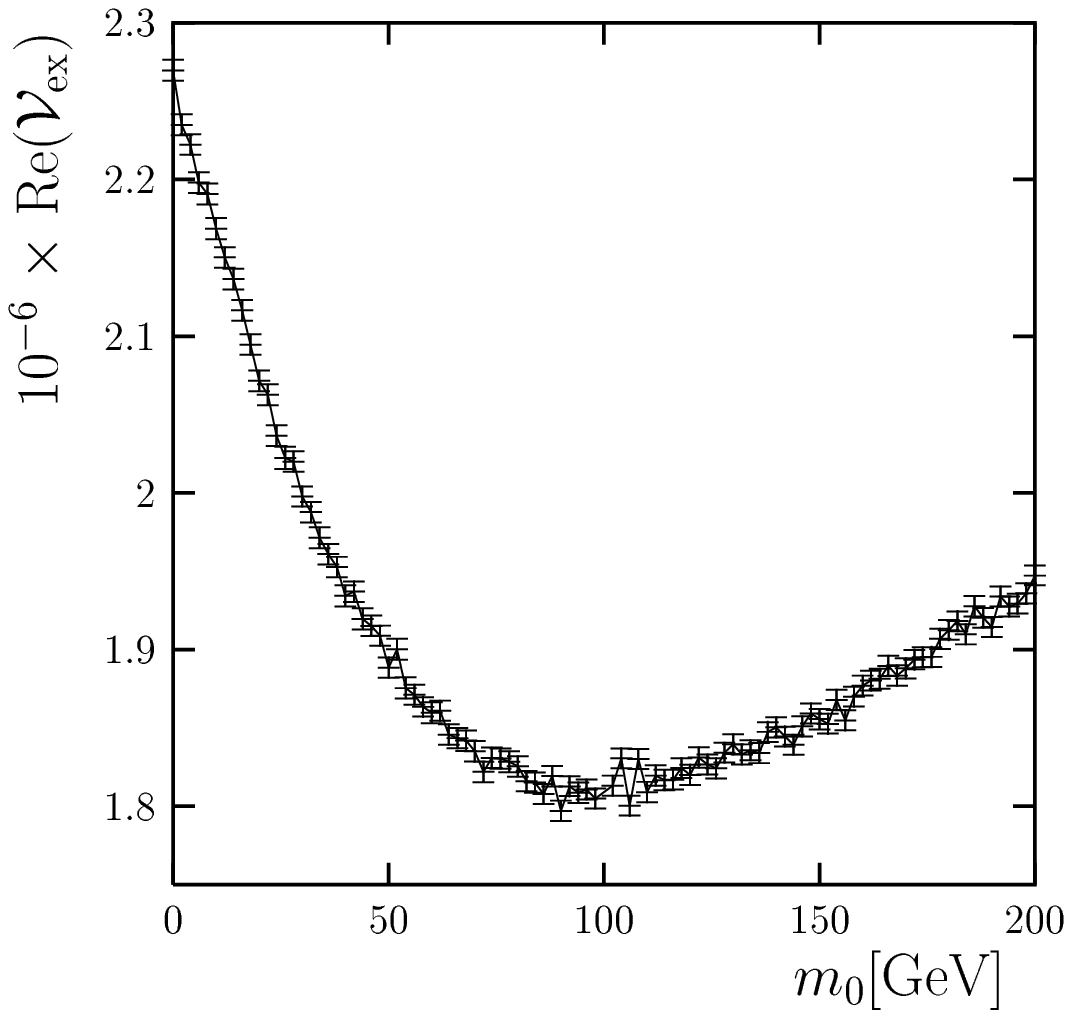, scale=0.6}
\epsfig{figure=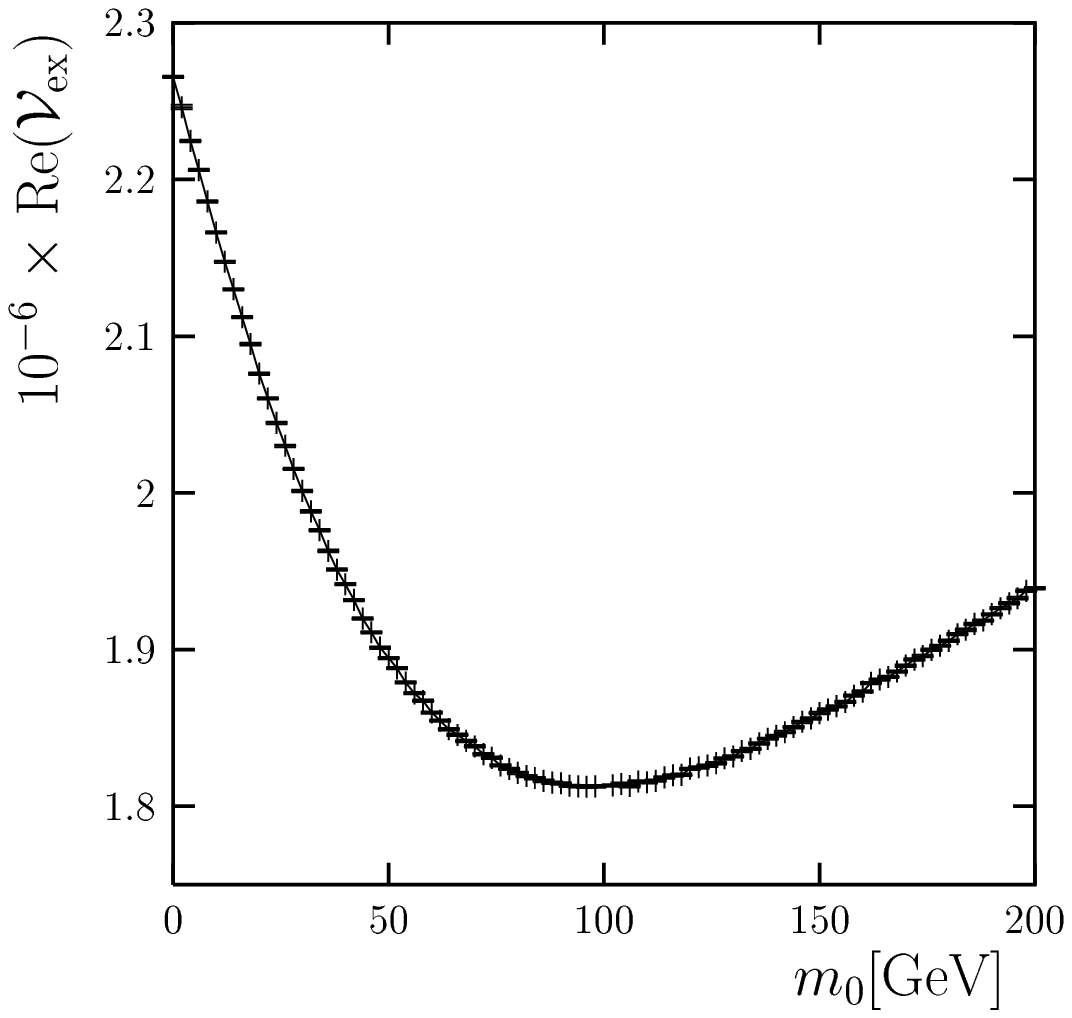, scale=0.6}
\epsfig{figure=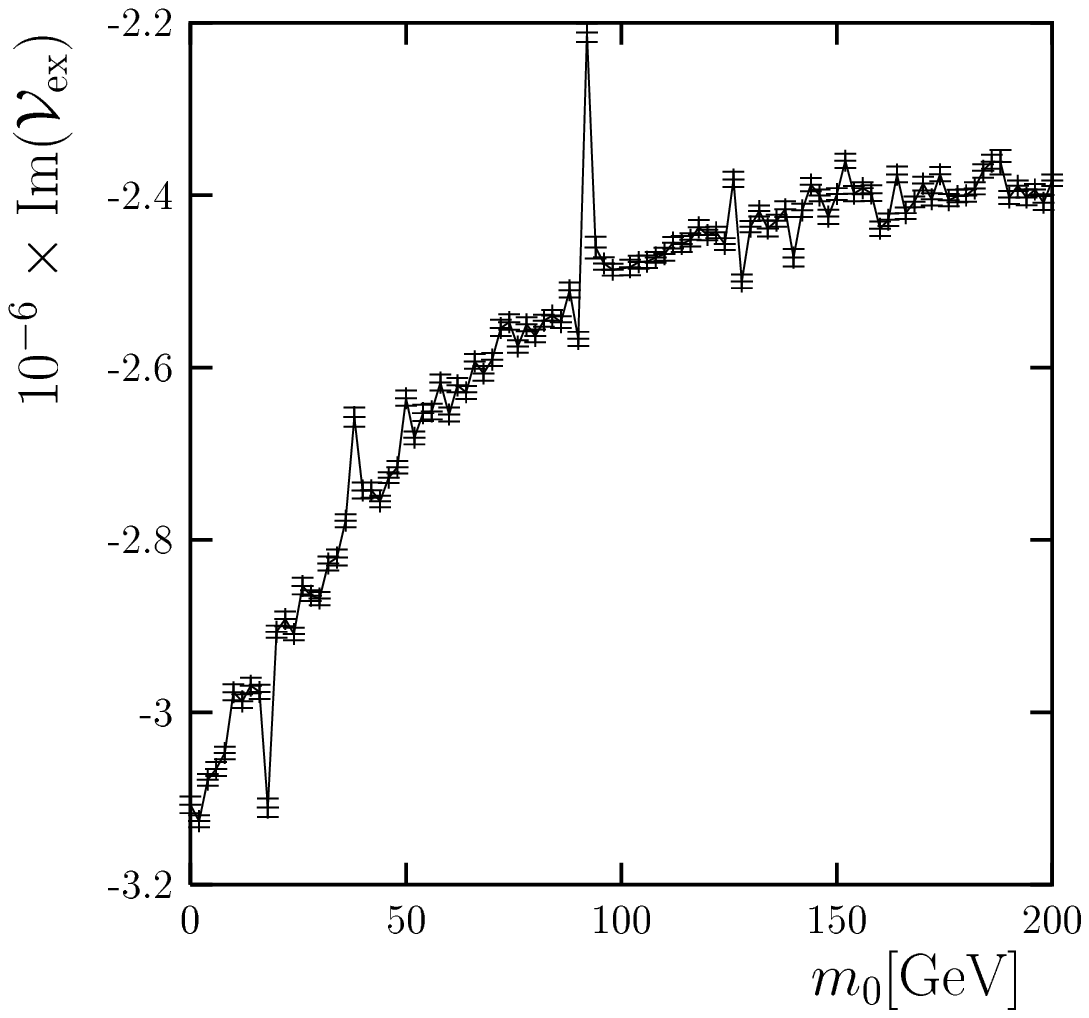, scale=0.6}
\epsfig{figure=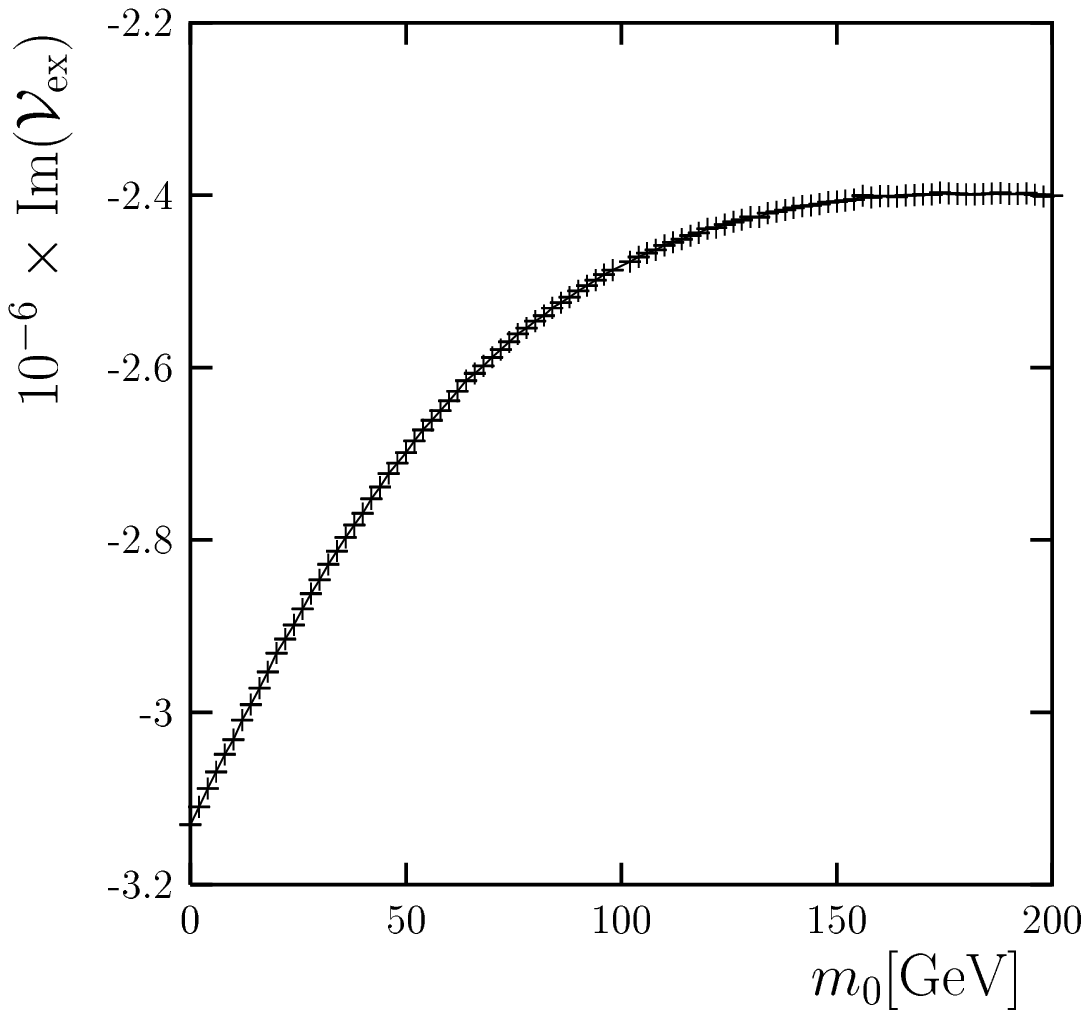, scale=0.6}
\caption{\label{Vexgra}Real part (top) and imaginary part (bottom) of the
integral ${\cal V}_{\rm ex}$ as a function of the mass parameter $m_0$ of the
subtraction masses for the decay mass value on the  two-particle threshold at
$M=\sqrt{p^2}=500\GeV$. The integral ${\cal V}_{\rm ex}$ is the sum of the
UV-finite parts of the master integrals corresponding to ${\cal T}_{\rm ex}$
in Eq.~(\ref{topex}) which after tensor reduction keep the planar and rotated
reduced planar topology. Shown are the results for separated (left hand side)
and combined numerical integration (right hand side) with VEGAS.}
\end{center}\end{figure}

For the physical masses and outgoing momenta we again use the values in
Eq.~(\ref{standmass}). The square of the decay mass $M=\sqrt{p^2}=500\GeV$ is
chosen to be equal to the two-particle threshold $p^2=(m_5+m_6)^2$. For the
subtraction masses we take values with an offset $m_0$ which in this example
runs from $0\GeV$ to $200\GeV$,\footnote{Note that ${\cal V}_{\rm ex}$ still
does not need to be independent of the subtraction masses because only in the
sum with the finite parts of the subtraction terms this dependence will
vanish.}
\begin{equation}
\matrix{m_{1,1}=5\GeV+m_0&m_{1,2}=25\GeV+m_0\cr
m_{2,1}=15\GeV+m_0&m_{2,2}=35\GeV+m_0.\cr}
\end{equation}
Subtraction terms and corresponding subtraction masses for the second loop
momentum $l$ are not necessary, since in this particular case the contributing
planar and rotated reduced planar topologies do not contain subdivergences in
$l$. The results are shown in Fig.~\ref{Vexgra}. The left hand side of
Fig.~\ref{Vexgra} shows real and imaginary parts for the separated integration,
the right hand side shows real and imaginary parts for the combined
integration. It is obvious that the combined integration is much more stable.

This result is easy to understand. Assuming that the relative variances are
the same for all $n$ partial integrals and for the complete integral, we can
use the same number of intersection points in the Monte Carlo integration for
each seperate integral and for the complete integral. In this case, the
standard deviation of the seperately integrated and summed integrals is bigger
by a factor of $\sqrt{n}$ than the standard deviation of the complete
integral~\cite{Kleiss:2005du,vanHameren:2001ng}.

\section{Conclusions}
Using the parallel/orthogonal space method, we have elaborated on a new method
to calculate the planar and rotated reduced planar ladder two-loop three-point
diagrams in the general massive case. Our method does not depend on a specific
physical decay process but is suitable to calculate all planar two-loop vertex
diagrams which may arise in the Standard Model of electroweak interactions
using Feynman gauge.

We have presented a tensor reduction which reduces the mentioned topologies
with arbitrary tensor structure to a set of master integrals with strongly
restricted, well-defined numerator structure. We have developed and
implemented an algorithm for the semi-analytical calculation of the UV-finite
part of those master integrals which still have the mentioned topologies after
tensor reduction. While six of the eight integrations are done analytically,
for the remaining two integrations we used different numerical methods.

The integration of the UV-finite part of the important master integrals was
checked exhaustively in the limit of vanishing subtraction parameters by
comparing the semi-analytical calculation with the results of a numerical
integration. As the final step of the semi-analytical calculation, the
numerical integration turns out to be stable, i.e.\ one obtains the same
result as for the above mentioned limit. This is valid especially in the
case where the numerical integration is not done for each master integral
separately but for the sum of all analytical results necessary for the given
process. 

For the remaining master integrals and subtraction terms with simpler
topologies there exist other methods. In part, these methods are already
implemented in \xloops. Therefore, in writing links to existing codes, these
cases can easily be implemented into our calculations. The still missing
topologies are the two-loop three-point topologies containing a two-point
subloop. These parts still have to be implemented~\cite{Bauberger:1994by,
Post:1997}. After having done this, we will be able to
evaluate semi-analytically all IR-finite NNLO corrections containing planar
two-loop vertex topologies as they arise in the Standard Model.

Finally, because the calculation of the scalar integrals for the non-planar
two-loop vertex function~\cite{Frink:1996} is quite similar to the calculation
of the scalar integrals for the planar two-loop vertex
function~\cite{Czarnecki:1994td}, our algorithm is expected to be easily
extendable to the non-planar two-loop vertex topology. Having extended our
algorithm in this direction, we are then able to calculate all NNLO vertex
corrections as they arise in the Standard Model.

As far as possible, the computer codes written in the course of this paper are
developed and implemented for most general decay processes. Because of this,
we plan to include our calculations in the \xloops-project in the near future.

\subsection*{Acknowledgements}
We thank H.S.~Do, J.G.~K\"orner, A.v.~Hameren, M.~Rogal, K.~Schilcher,
C.~Schwinn, H.~Spiesberger, and E.~Tuiran for helpful discussions. M.K.\ wants
to thank C.~Bauer and J.~Vollinga for computer-technical support. S.G.\
acknowledges support by the DFG as a guest scientist in Mainz, by the Estonian
target financed project No.~0182647s04, and by the Estonian Science Foundation
under grant No.~6216. M.K.\ was supported in parts by the BMBF and by the
Graduiertenkolleg ``Eichtheorien -- experimentelle Tests und theoretische
Grundlagen'', Mainz University.

\begin{appendix}

\section{Integral basis\label{appa}}
\setcounter{equation}{0}\def\theequation{A\arabic{equation}}
In this appendix we list the analytical results for the basic integrals
in Eq.~(\ref{basicint}) in terms of the parameters $r_s$, $r_t$, $r_0$,
$s_0$, and $t_0$. In order to calculate these integrals, we introduce cutoffs
$\Lambda_s$ and $\Lambda_t$. The results depending on these cutoffs are
extremely long and contain products of two logarithms as well as
dilogarithms~\cite{Knodel:2005}. However, one can show that these divergences
cancel in the sums in Eqs.~(\ref{k1l1sint0}) and~(\ref{k1l1sintx}). In order
to make the reader familiar with these features, we explain the calculation
for the simplest example.

\subsection{The integral ${\cal F}^{st}_{0,0}$}
The regularized integral is given by
\begin{eqnarray}
\lefteqn{{\cal F}^{st}_{0,0}\ =\ \int_0^{\Lambda_t}dt\int_0^{\Lambda_s}ds
  \frac1{\sqrts{(r_0-r_ss-r_tt+i\eta)^2-4st}}}\nonumber\\
  &=&-\frac{\Lambda_t}{r_s}\ln\left|\frac{r_sr_0+(2-r_sr_t)\Lambda_t
  -r_s^2\Lambda_s-r_s\Lambda}{2(r_sr_0+(1-r_sr_t)\Lambda_t)}\right|
  -\frac{\Lambda_s}{r_t}\ln\left|\frac{r_tr_0+(2-r_sr_t)\Lambda_s
  -r_t^2\Lambda_t-r_t\Lambda}{2(r_tr_0+(1-r_sr_t)\Lambda_s)}\right|
  \nonumber\\&&
  +\frac{r_0}{1-r_sr_t}\Bigg(\ln\left|\Lambda_t+\frac{r_sr_0}{1-r_sr_t}\right|
  +\ln\left|\Lambda_s+\frac{r_tr_0}{1-r_sr_t}\right|+i\pi\theta(r_0)
  \nonumber\\&&
  -\ln\left|r_0(1+r_sr_t)
  +(1-r_sr_t)(r_s\Lambda_s+r_t\Lambda_t+\Lambda)\right|
  -\ln\left|\frac{r_0}{2(1-r_sr_t)^2}\right|\Bigg)
\end{eqnarray}
where
\begin{equation}
\Lambda:=\sqrt{(r_0-r_s\Lambda_s-r_t\Lambda_t)^2-4\Lambda_s\Lambda_t}.
\end{equation}
What to do with the singularity? As a first step we can separate the
divergent part which depends on $\Lambda_s$, $\Lambda_t$ and $\Lambda$ from
the convergent part. But how does the divergent part cancel? If we look at
Eq.~(\ref{k1l1sintx}) where ${\cal F}^{st}_{0,0}$ appears, we see that it
actually appears as part of a sum. If we concentrate on the summation over $n$
and consider how ${\cal N}^k_{1,i,a}$ and ${\cal N}^k_{2,i,a}$ are defined in
Eqs.~(\ref{n1jadef}) and~(\ref{n2jadef}), and if we take into account that
$P_{2,j}\big|_{P_{1,j}}=\xi_{2,j}-\xi_{1,j}=-P_{1,j}\big|_{P_{2,j}}$, we see
that ${\cal N}^k_{1,i,a}=-{\cal N}^k_{2,i,a}$. The same is valid for
${\cal N}^l_{x,j,b}$ and ${\cal N}^l_{6,j,b}$, i.e.\
${\cal N}^l_{x,j,b}=-{\cal N}^l_{6,j,b}$. Therefore, an expression which is of
relevance for the subtracted integrals is the sum
\begin{equation}
\sum_{n=1}^2\sum_{m=x,6}(-1)^{n+m}{\cal F}^{st}_{0,0}(r_s,r_t,r_{n,i,m,j}).
\end{equation}
In replacing $\Lambda_s$ by $\Lambda_s^0/\epsilon_\Lambda$ and
$\Lambda_t$ by $\Lambda_t^0/\epsilon_\Lambda$ with $\epsilon_\Lambda\ll 1$
with arbitrary but positive and fixed values $\Lambda_s^0$ and $\Lambda_t^0$
we indeed can show that
\begin{equation}
\sum_{n=1}^2\sum_{m=x,6}(-1)^{n+m}
  {\cal F}^{st,{\rm div}}_{0,0}(r_s,r_t,r_{n,i,m,j})=O(\epsilon_\Lambda).
\end{equation}
For $\epsilon_\Lambda\to 0$, therefore, the divergent part will cancel in the
difference. We only have to consider the convergent part
\begin{equation}
{\cal F}^{st,{\rm conv}}_{0,0}=\frac{r_0}{1-r_sr_t}
  \left(i\pi\theta(r_0)-\ln\left|\frac{r_0}{2(1-r_sr_t)^2}\right|\right).
\end{equation}
Finally, we see that
\begin{equation}
\sum_{n=1}^2\sum_{m=x,6}(-1)^{n+m}r_{n,i,m,j}
  \to\frac{\pm 2}{\theta_0}\sum_{m=x,6}(\xi_{1,i}-\xi_{2,i})=0
\end{equation}
(cf.\ Eq.~(\ref{r0stdef}) including remarks given there). Therefore, we can
skip all terms which are linear in $r_0$. The result which we have to
implement is
\begin{equation}
{\cal F}^{st,{\rm impl}}_{0,0}=\frac{r_0}{1-r_sr_t}
  \left(i\pi\theta(r_0)-\ln|r_0|\right).
\end{equation}
In the following we will only give results which are to be implemented. For
the integrals with one index we will show only results for ${\cal F}^s_\alpha$.
Results for ${\cal F}^t_\beta$ can easily be obtained by replacing
$s_0\leftrightarrow t_0$ and $r_s\leftrightarrow r_t$.

The integral ${\cal F}^{st}_{0,0}$ is used for illustrative reasons only. It
does not appear in calculations because the integration region vanishes. The
same is valid for the integrals ${\cal F}^{st}_{i,j}$. Therefore, it turns out
that the integrals ${\cal F}^s_\alpha$ and ${\cal F}^t_\beta$ are sufficient
for calculations within the Standard Model of electroweak interactions using
Feynman gauge.

\subsection{The integrals ${\cal F}^s_\alpha$}
\begin{eqnarray}
r_s{\cal F}^{s,{\rm impl}}_0
  &=&-\real\left(\Li_2\left(1+\frac{r_sr_0}{t_0(1-r_sr_t)}\right)\right)
  \nonumber\\&&
  -i\pi\left\{\theta(r_0)\ln\left|1+\frac{r_sr_0}{t_0(1-r_sr_t)}\right|
  -\theta(t_0)\ln\left|t_0+\frac{r_sr_0}{1-r_sr_t}\right|\right\}\\
r_s^3{\cal F}^{s,{\rm impl}}_1
  &=&-(2-r_sr_t)\left\{\frac{r_sr_0}{1-r_sr_t}\ln|r_0|
  +\left(t_0+\frac{r_sr_0}{2-r_sr_t}\right)
  \real\left(\Li\left(1+\frac{r_sr_0}{t_0(1-r_sr_t)}\right)\right)\right\}
  \nonumber\\&&
  +i\pi\Bigg\{(2-r_sr_t)\theta(r_0)\left(\frac{r_sr_0}{1-r_sr_t}
  -\left(t_0+\frac{r_sr_0}{2-r_sr_t}\right)
  \ln\left|1+\frac{r_sr_0}{t_0(1-r_sr_t)}\right|\right)\nonumber\\&&\qquad
  +\left((2-r_sr_t)t_0+r_sr_0\right)\theta(t_0)
  \ln\left|t_0+\frac{r_sr_0}{1-r_sr_t}\right|\Bigg\}\\
r_s^5{\cal F}^{s,{\rm impl}}_2
  &=&\Bigg\{\frac12\pfrac{r_sr_0}{1-r_sr_t}^2(6-6r_sr_t+r_s^2r_t^2)
  \nonumber\\&&
  -\frac{r_sr_0}{1-r_sr_t}\left((6-6r_sr_t+r_s^2r_t^2)t_0+2(3-r_sr_t)r_sr_0
  \right)\Bigg\}\ln|r_0|\nonumber\\&&\kern-12pt
  -\left((6-6r_sr_t+r_s^2r_t^2)t_0^2+2(3-r_sr_t)r_sr_0t_0+r_s^2r_0^2\right)
  \real\left(\Li\left(1+\frac{r_sr_0}{t_0(1-r_sr_t)}\right)\right)
  \nonumber\\&&\kern-12pt
  -i\pi\Bigg\{\theta(r_0)\Bigg(\frac12\pfrac{r_sr_0}{1-r_sr_t}^2
  (6-6r_sr_t+r_s^2r_t^2)\nonumber\\&&\quad-\frac{r_sr_0}{1-r_sr_t}
  \left((6-6r_sr_t+r_s^2r_t^2)t_0+2(3-r_sr_t)r_sr_0\right)\nonumber\\&&\quad
  +\left((6-6r_sr_t+r_s^2r_t^2)t_0^2+2(3-r_sr_t)r_sr_0t_0+r_s^2r_0^2\right)
  \ln\left|1+\frac{r_sr_0}{t_0(1-r_sr_t)}\right|\Bigg)
  \nonumber\\&&
  -\theta(t_0)\left((6-6r_sr_t+r_s^2r_t^2)t_0^2+2(3-r_sr_t)r_sr_0t_0
  +r_s^2r_0^2\right)\ln\left|t_0+\frac{r_sr_0}{1-r_sr_t}\right|\Bigg\}
  \nonumber\\
\end{eqnarray}

\end{appendix}

\end{document}